 \documentstyle[11pt,aaspp4]{article}

\begin{document}

\title{ABUNDANCES AT HIGH REDSHIFTS: THE CHEMICAL ENRICHMENT HISTORY
OF DAMPED LYMAN-ALPHA GALAXIES}
\author{Limin Lu$^{1}$, Wallace L. W. Sargent, \& Thomas A. Barlow}
\affil{California Institute of Technology, 105-24, Pasadena, CA 91125}
\altaffiltext{1}{Hubble Fellow}

\begin{abstract}

   We study the elemental abundances of C, N, O, Al, Si, S, Cr, Mn, Fe, Ni, 
and Zn in a sample of 14 damped Ly$\alpha$ systems (galaxies) with
H I column density $N$(H I)$\geq10^{20}$ cm$^{-2}$, using high
quality spectra of quasars obtained with the 10m Keck telescope. 
To ensure accuracy, only weak, unsaturated absorption lines are used to
derive ion column densities and elemental abundances. 
Combining these abundance measurements with similar measurements
in the literature, we investigate the chemical evolution of
damped Ly$\alpha$ galaxies based on a sample of
23 systems in the redshift range
$0.7<z<4.4$. The main conclusions are as follows.

1. The damped Ly$\alpha$ galaxies have (Fe/H) in the range of 
1/10 to 1/300 solar, clearly indicating that these are young galaxies 
in the early stages of chemical evolution. The $N$(H I)-weighted
mean metallicity of the damped Ly$\alpha$ galaxies between $2<z<3$
is (Fe/H)=0.028 solar. There is a large scatter,
about a factor of 30, in (Fe/H) at $z<3$, which we 
argue probably results from the different formation 
histories of the absorbing galaxies or a mix of galaxy types.

2. Comparisons of the distribution of (Fe/H) vs redshift for the sample
of damped Ly$\alpha$ galaxies with the similar relation for the Milky
Way disk  indicate
that the damped Ly$\alpha$ galaxies are much less metal-enriched than
the Galactic disk in its past. Since there is evidence from our
analyses that
depletion of Fe by dust grains in the sample  galaxies is
relatively unimportant, the difference in the enrichment level between
the sample of damped Ly$\alpha$ galaxies and the Milky Way disk suggests
that damped Ly$\alpha$ galaxies are probably not high-redshift spiral disks
in the traditional sense. Rather, they could represent a thick disk 
phase of galaxies, or more likely the spheroidal component of galaxies,
or dwarf galaxies.

3. The mean metallicity of the damped Ly$\alpha$ galaxies is found to
increase with decreasing redshift, as is expected.
All four of the damped Ly$\alpha$ galaxies at $z>3$ in our sample
have (Fe/H) around 1/100 solar or less. In comparison, a large fraction
of the damped Ly$\alpha$ galaxies at $z<3$ have reached ten times
higher metallicity. This suggests that the time  around $z=3$
may be the epoch of galaxy formation in the
sense that galaxies are beginning to form the bulk of their stars.
Several other lines of evidence appear to point to the same conclusion,
including the evolution of the neutral baryon content of damped Ly$\alpha$
galaxies, the evolution in the quasar space density, and the morphology
of $z>3$ galaxies.

4. The relative abundance patterns of the elements studied here
clearly indicate that the bulk of heavy elements in these 
high-redshift galaxies were produced by Type II supernovae; there
is little evidence for significant contributions from
stellar mass loss of low-to-intermediate mass stars, or from 
Type Ia supernovae. 

5. Although earlier studies have attributed the overabundance of Zn relative
to Cr or Fe found in damped Ly$\alpha$ galaxies to
selective depletion of Cr and Fe by dust grains, 
such an interpretation is inconsistent with many of the other
elemental abundance ratios seen in these galaxies, most notably 
N/O and Mn/Fe. Several other tests also indicate that there is no
significant evidence for dust depletion in these galaxies.
We suggest that the overabundance of Zn relative to
Cr in damped Ly$\alpha$ galaxies may be intrinsic to their
stellar nucleosynthesis. If this interpretation is correct, it will provide 
important new information to the theory of stellar nucleosynthesis.

6. The absorption profiles of Al III in damped Ly$\alpha$ galaxies
are found to resemble those of the low-ionization lines. The profiles of
Si IV and C IV absorption, while resembling each other in general,
are almost always different from those of the low-ionization
absorption lines. These results suggest that Al III is probably produced
in the same physical region as the low-ionization species in the
absorbing galaxies, while the high-ionization species (Si IV and C IV)
mostly likely come from distinct physical regions. 

7. We discuss possible ways to obtain information on the history of star
formation (i.e., continuous or episodic) in damped Ly$\alpha$ galaxies, 
and on the shapes of the stellar initial mass functions. 

8. We review the  evidence for, and against, the hypothesis that damped
Ly$\alpha$ galaxies are disks or proto-disks at high redshifts,
and discuss the implications.

9. We determine upper limits on the temperature of the
 cosmic microwave background radiation
at several redshifts using absorption from the fine structure level of
the C II ion. These upper limits are consistent with the predicted
increase of $T_{CMB}$ with redshift.

\end{abstract}


\section{INTRODUCTION}

   A fundamental question in the study of galaxy formation and
evolution is: what is the chemical evolutionary history of galaxies?
While some information pertaining to the general question of galactic
chemical evolution has been obtained from studies of stellar populations and
the interstellar medium in the Milky Way and in nearby external galaxies, 
{\it direct} observations
of the chemical evolution of galaxies in the early universe has been
lacking. The situation  changed when it was recognized  that the narrow
absorption lines in the spectra of high-redshift quasars can be
used to probe conditions in the early universe. One of the most important
developments along this direction has been the recognition 
of ``damped Ly$\alpha$ absorption systems'' as a significant population
of quasar absorbers and the subsequent systematic studies of their
basic properties, which demonstrate that damped Ly$\alpha$ galaxies
are likely to be the progenitors of present-day galaxies (Wolfe et al. 1986;
Wolfe 1995; Lanzetta, Wolfe,
\& Turnshek 1995). 

   The pioneering work by Meyer, Welty, \& York (1989) and by 
Pettini, Boksenberg, \& Hunstead (1990) clearly demonstrated that 
one can extract significant information on the abundances
of chemical elements in damped Ly$\alpha$ galaxies through the analysis of
the absorption lines in quasar spectra. 
Pettini et al. (1994)
conducted the first systematic survey of elemental abundances in
damped Ly$\alpha$ galaxies,  though, for practical reasons, the study
was limited to the elements Zn and Cr. Nonetheless, their study 
revealed the important fact that damped Ly$\alpha$ galaxies are generally
metal poor (mean (Zn/H)$\sim 1/10$ solar), confirming that these are
young galaxies in the early stages of chemical enrichment. It was also
noted that the element of Cr is generally less abundant than Zn relative
to their solar proportion. Since Zn and Cr are thought to be made in
solar proportions throughout the Milky Way based on their abundances in
Galactic disk and halo stars (see discussion in \S4.3), the overabundance
of Zn relative to Cr found in damped Ly$\alpha$ galaxies was taken as evidence
that some Cr was depleted from the gas phase by dust grains.

   The large collecting power of the 10m Keck telescope has made it
feasible to study elemental abundances in damped Ly$\alpha$ systems
in much more detail (Wolfe et al. 1994; Lu et al.
1996a). The improvements come in three facets. The first is the higher
spectral resolution (FWHM$<10$ km s$^{-1}$), which allows one to see
more component structure in the absorption. The second is the much improved 
S/N of the spectra, which makes it considerably easier to study the
weak absorption lines. Weak absorption lines are crucial for the
determination of elemental abundances since they do not suffer nearly
as much as the strong lines from line saturation. The third
has to do with the large spectral coverage of the Keck High Resolution
Spectrometer (HIRES), which enables the study of many elements
simultaneously. This is important because the relative abundances of
elements contain valuable information about the nature of the stellar
nucleosynthesis that made these elements (cf. Wheeler, Sneden, \& Truran 1989).

   For a number of years, we have been collecting high quality spectra
of quasars using the Keck HIRES in order to carry out a comprehensive
analysis of the various absorption phenomena. One particular aspect
of this program is the study of elemental abundances in damped Ly$\alpha$
systems and their implications for galactic chemical evolution
and for stellar nucleosynthesis. Some preliminary results
of this nature have been reported in Lu, Sargent, \& Barlow (1995a).
The current paper presents the full analyses. The results described here
supersede the earlier results reported  in Lu et al. (1995a).

   In \S2 we describe the observations, data reduction, and the
methods used to estimate equivalent widths and column densities. 
Abundance determinations for the 14 damped Ly$\alpha$ systems studied 
here are discussed in \S3. These abundance measurements are combined 
with similar
measurements  in the literature to study the chemical evolution of damped
Ly$\alpha$ galaxies over the redshift range $0.7<z<4.4$ (\S4). 
In \S5 we discuss other observed properties
of damped Ly$\alpha$ galaxies and implications for their nature.
The redshift dependence of the temperature of the
cosmic microwave background as estimated from the C II* $\lambda$1335
fine structure line is discussed in \S6. 
\S7 summarizes the main conclusions.

   Throughout this study, we will assume 
$q_0=0.1$ and $H_{0}=50$ km s$^{-1}$ Mpc$^{-1}$.
We will use $z$ for general references of redshift,
$z_{em}$ for the emission redshift of quasars, $z_{damp}$
for the redshift of damped Ly$\alpha$ systems,
and $z_{abs}$ for any other absorption systems or for
absorption systems in general.

\section{OBSERVATIONS, DATA REDUCTION, AND MEASUREMENTS}

\subsection{Observations and Data Reduction}

    Table 1 gives the journal of the Keck HIRES observations. A 0.86" slit
width was used in all cases to yield a resolution of FWHM=6.6 km s$^{-1}$
with roughly 3 pixels per resolution element. The echelle format of HIRES
is such that one can only get complete coverage of the free spectral range
for $\lambda<5100$ \AA\ in a single setup with
the current TeK 2048x2048 CCD.
Except in a few cases, we used two partially overlapping setups for
 each wavelength region in order to get complete spectral coverage. 
Data reductions were done using a software written
by T. A. Barlow. After the echelle orders were optimally 
extracted and wavelength-
and flux-calibrated, they were resampled to 2 pixels per resolution element,
scaled to the same flux level, and added together weighted according 
to their S/N. Because some spectral
regions have more exposures than others due to the overlapping-setup
scheme, the S/N of the spectrum from even a single object can show quite
large variations from one region to another. 

    Most metal lines associated with damped Ly$\alpha$ systems are
blended with Ly$\alpha$ absorption lines if they occur in the Ly$\alpha$
forest. Thus our analyses will primarily
rely on measurements made redward of the Ly$\alpha$
emission line. We estimate the continuum level redward of the Ly$\alpha$ 
emission by fitting cubic splines to regions deemed free of
absorption lines using the IRAF task {\it continuum}. In rare
cases, we may also use measurements made in the Ly$\alpha$ forest if the
lines are reasonably clean. In such
cases, the continuum level is estimated by picking out small ``peaky''
regions in the spectral region of interest that are free of obvious
absorption lines and interpolating them with low-order cubic splines. 
The resulting
continuum appears to describe the data quite well.

\subsection{Equivalent Width and Column Density Measurements}

    For each damped Ly$\alpha$ system studied here, we give a table
listing the absorption lines detected, the measured equivalent widths,
and column densities (see \S3). 

    To carry out the measurements for a given damped Ly$\alpha$ system, 
we first plot the continuum-normalized profiles of the absorption 
lines in velocity space for an
adopted redshift for the system (which generally corresponds to the 
strongest component in the low ion absorption).  Equivalent width for each
line is then estimated by integrating the observed spectrum over the 
width of the absorption line defined by two limiting velocities:
$v_{-}$ and $v_{+}$. The 1$\sigma$ uncertainty associated with the measured
equivalent width is also calculated including both the effects of 
statistical noise and of uncertainties in the placement of the continuum
level (cf. Savage et al. 1993).  Only features over the 4$\sigma$
significance level will be considered a real detection.

To obtain column densities, we use the
apparent optical depth method (cf. Savage \& Sembach 1991). We first
convert the normalized absorption line profiles, $I(v)$, into measures of the
apparent optical depth per unit velocity, $\tau_{a}(v)$, through the
relation $\tau_a(v)=$ln[$1/I(v$)], which can then be converted into the
apparent column density per unit velocity, $N_a(v)$, in units of
atoms cm$^{-2}$ (km s$^{-1}$)$^{-1}$ through the relation
$$ N_a(v)=3.767\times 10^{14}{\tau_a(v) \over {f\lambda}}, \eqno(1)$$
where $\lambda$ is the rest wavelength in \AA\ and $f$ is the
oscillator strength. The total column density, obtained
by integrating $N_a(v)$ over the width of the absorption profile,
should represent the true column density to within the measurement 
uncertainty if the absorption line contains no unresolved, saturated
components\footnote{In principle, one can measure the column density
of a saturated absorption line if the absorption line is
fully resolved, {\it and} if the S/N is good enough to clearly indicate
that the flux in the center of the absorption line does not go to zero.
In practice, this rarely happens due to limited S/N in real data.}. 
In the case that the absorption line does contain unresolved,
saturated components, the integrated $N_a(v)$ only gives a lower limit
to the true column density.  For an ion species for which several 
absorption lines with different $f\lambda$ values are observed, comparisons
of their integrated $N_a(v)$ will generally allow one to infer if 
unresolved saturated components are present. This is because lines with larger
$f\lambda$ values will yield lower integrated column densities than
lines with smaller $f\lambda$ values in such cases. This technique 
was explored in details 
by Savage \& Sembach (1991). Examples of applications may be found in
Lu et al. (1995b).

Empirically, we find that, given the spectral resolution and S/N typical of
the data used in this work, absorption lines with peak  absorption
depth less than $\sim 80$\% (i.e., peak optical  depth $\tau_0<1.6$ if 
the lines are resolved)
show no evidence of unresolved saturation that would render invalid the
column densities derived from $N_a(v)$  integrations. For example,
the Fe II $\lambda\lambda$2249, 2260 lines differ in their $f\lambda$
values by a factor of 1.3, and they yield the same $N$(Fe II) for several
systems in Q 0449$-$1326, Q 0450$-$1312, and Q 0454+0356 even though
the intrinsically stronger $\lambda$2260 absorption lines are 60-75\%
deep. Similarly, the Si II $\lambda\lambda$1304, 1526 absorption lines,
which differ by a factor of 1.5 in their $f\lambda$ values, yield consistent
$N$(Si II)  in several systems in Q 1946+7658, Q 2212$-$1626, and
Q 1202$-$0725 (Lu et al. 1996a) even though the intrinsically stronger
$\lambda$1526 absorption lines are 70-85\% deep. 
Also, the S II $\lambda\lambda$1250, 1253
lines  (which differ in their $f\lambda$ values by a factor
of 2) in the $z_{damp}=2.8110$ system toward Q 0528$-$2505  yield consistent
column densities despite the fact that the intrinsically stronger
$\lambda$1253 absorption is 85\% deep. The most dramatic example
is offered by the $z_{damp}=2.8268$ system toward Q 1425+6039, where the
Fe II $\lambda\lambda$1608, 1611 absorption lines, which differ in their
$f\lambda$ values by a factor of 62, yield $N$(Fe II) values differing
by less than 0.1 dex, and the $\lambda$1608 absorption line is 80\% deep.
Similar examples may be found for the more ionized species such
as Al III, Si IV, and C IV, each having doublet absorption lines differing
in their $f\lambda$ values by a factor of 2.  In fact, we have yet to find
an example where an absorption line with peak absorption depth $<80$\% shows 
convincing evidence of being saturated and unresolved. Thus, for
ion species for which only a single absorption line is observed (i.e.,
for which we cannot check independently if there is any unresolved
saturation in the absorption), we will assume that the column density
derived from integrating its $N_a(v)$ profile is unbiased with respect
to unresolved saturation if its peak absorption depth is less than
80\%. In reality, most of the single absorption lines studied here
are $<50$\% deep, corresponding to  a peak optical depth of 0.7
if the lines are resolved. The only single absorption lines with peak   
absorption depth between 50 to 80\% for which
we have adopted their $N_a(v)$-integrated column density as
real measurements (rather than lower limits) are the Si II $\lambda$1808
lines at $z_{damp}=2.8110$ toward Q 0528$-$2505 and at $z_{damp}=4.0803$
toward Q 2237$-$0608, the Fe II $\lambda$1608 line at $z_{damp}=2.8268$
toward Q 1425+6039, and Al II $\lambda$1670 lines at $z_{damp}=2.8443$
toward Q 1946+7658 and at $z_{damp}=4.0803$ toward Q 2237$-$0608 (see \S3
for details).

We emphasize that, for unsaturated or mildly saturated lines,
the apparent optical depth method will yield column density measurements 
in agreement with the profile fitting  technique. For heavily saturated lines,
neither method will yield very reliable column densities in practice.
For this reason, the analyses in this paper will concentrate on column
density measurements made from unsaturated lines. Column density
{\it limits} from saturated lines may be used occasionally in the discussion.
For lines that are undetected or detected at significance level less
than 4$\sigma$, we will use the 4$\sigma$ upper limits of the equivalent 
width to estimate the upper limits of the column density assuming linear
part of the curve of growth.

Atomic data, including rest-frame vacuum wavelength, oscillator strength
($f$-value), and radiation damping constant, are taken from Morton (1991)
except for those which have revised $f$-values as given in the 
compilation of Tripp, Lu, \& Savage (1996; see Savage \& Sembach 1996
for additional information).

\section{DISCUSSION OF INDIVIDUAL SYSTEMS}

In this section, we derive column densities and elemental
abundances for each individual damped Ly$\alpha$ system studied here.
In general, absorption lines occurring in the Ly$\alpha$ forest will
not be discussed unless the line is relatively clean of 
contamination from forest absorption.
It will be assumed that the ionization state of the absorbing gas is
dominated by species appropriate for H I gas (e.g., O I, N I,
C II, Si II, Fe II, etc.)
so that abundance corrections for unobserved
ionization stages are negligible. This is believed to be a good assumption
given the large H I column density of the systems (see Viegas 1995).

Two damped Ly$\alpha$ systems not covered by the  spectra
listed in Table 1 will also be discussed. These are the $z_{damp}=1.1743$ 
system toward Q 0450$-$1312, and
the $z_{damp}=0.8598$  system toward Q 0454+0356. Keck HIRES observations
of these two quasars were made by Drs. Christopher Churchill and Steven Vogt,
who have kindly made their spectra available to us. Details of the observations
and reductions for these two objects are discussed in Churchill \& Vogt (1997).

\subsection{ Q 0000$-$2620 $z_{damp}=3.3901$ Damped System}
   
This damped Ly$\alpha$ system has been known ever since the quasar was
first discovered by C. Hazard (unpublished; but see Webb et al. 1988).
It is one of the four damped Ly$\alpha$ galaxies at $z>3$ for which
abundance information has been obtained.
     The HIRES spectrum used here for this object was already published by Lu 
et al. (1996b). We estimate  $N$(H I)=$(2.6\pm0.5)\times 10^{21}$ 
cm$^{-2}$ from fitting the Ly$\alpha$ damping profile (figure 1), 
which agrees quite
well with other determinations (Webb et al. 1988; Sargent, Steidel, \& 
Boksenberg 1989; Savaglio, D'Odorico, \& Moller 1994). 
Absorption lines detected redward of
the Ly$\alpha$ emission are listed in Table 2; the absorption profiles
are shown in figure 2. Several other lines
(Si II $\lambda\lambda$1190,1193,1260,1304, Si III $\lambda$1206, 
O I $\lambda$1302,
C II $\lambda$1334, Si IV $\lambda\lambda$1393,1402) are clearly detected
in the Ly$\alpha$ forest but are blended with Ly$\alpha$ forest absorption 
(see the spectrum in Lu et al. 1996b). These lines will not be discussed
here.

Except for the line tentatively identified as S II $\lambda$1253, 
all the low ion lines shown in figure 2 are clearly saturated, so
only lower limits to their column densities are given in Table 2, along
with the abundance limits. Similarly, only a lower limit to the C IV
column density is given. For Fe II, we give both the lower and upper limits
to the column density determined from the saturated Fe II $\lambda$1608
absorption and from the absence of the Fe II $\lambda$1611 absorption.

    In Lu  et al. (1996b) we tentatively identified a weak absorption feature
in the Ly$\alpha$ forest as the S II $\lambda$1253 line associated with this
damped Ly$\alpha$ system. This weak feature is also shown in figure 2,
and it shows excellent redshift agreement with the other low ion lines.
The corresponding S II $\lambda\lambda$1250,1259 lines are badly blended with
Ly$\alpha$ forest absorption.  If our identification of this S II line  
is correct, then the estimated column density,
log $N$(S II)=$14.70\pm0.03$ (Lu et al. 1996b), would imply an abundance of
[S/H]=$-1.98$. This  is consistent with the limits obtained for
the other elements. We note that Pettini et al. (1995b) estimated
[Zn/H]$\leq-1.76$ (3$\sigma$) and [Cr/H]$=-2.46$.

\subsection{ Q 0216$+$0803 $z_{damp}=2.2931$ System}
   
    This damped Ly$\alpha$ system was discovered by Lanzetta et al. (1991)
and confirmed by Lu \& Wolfe (1994), 
who estimated log $N$(H I)$=20.45\pm0.16$ for the system. Our Keck 
spectrum does not
cover the wavelength region containing the damped Ly$\alpha$ absorption.
We will therefore adopt the $N$(H I) value from Lu \& Wolfe (1994).
The absorption lines detected redward of the Ly$\alpha$ emission are 
listed in Table 3, and the absorption line profiles are shown in 
figure 3. Lines detected in the Ly$\alpha$ forest, which will not be 
used in the analysis,
include C II $\lambda$1334, O I $\lambda$1302, Si II $\lambda$1304, and
Si IV $\lambda\lambda$1393,1402.

    The Zn II $\lambda$2026 and Cr II $\lambda$2056 absorption lines
are detected at the 2.3$\sigma$ and 2.6$\sigma$ levels, respectively;
we will only use their 4$\sigma$ upper limits in the subsequent
analyses. The Fe II $\lambda$1608 absorption is probably saturated and we 
conservatively give  a lower limit of log$N$(Fe II)$>14.81$. 
The much weaker Fe II $\lambda$2260
absorption is detected at the 2.6$\sigma$ level, and the corresponding 
4$\sigma$ upper limit is log $N$(Fe II)$<14.97$. We will therefore adopt
log $N$(Fe II)$=14.89\pm0.08$.

\subsection{ Q 0216$+$0803 $z_{damp}=1.7688$ System}

    This system was identified as a candidate damped Ly$\alpha$ absorption
by Lanzetta et al. (1991) based on a low resolution spectrum. No 
higher-resolution confirmation of the damped Ly$\alpha$ absorption 
is available.
However, the fact that we detect intrinsically weak absorption lines 
like Fe II $\lambda$2260,
Mn II $\lambda$2576, and Ni II $\lambda$1751 suggests that most likely this
is a damped Ly$\alpha$ system. For example, even for solar metallicity, the
observed Fe II column density would imply $N$(H I)$\sim 10^{19}$
cm$^{-2}$. The Ly$\alpha$ line equivalent given by
Lanzetta et al. (1991) implies log $N$(H I)=20.
We tentatively adopt log$N$(H I)$=20\pm0.18$, assuming a 50\% uncertainty.
The absorption lines redward of Ly$\alpha$ emission for this 
system are listed in Table 4, and the profiles are shown in figure 4.
Note that none of the Zn II and Cr II absorption lines is detected at over
4$\sigma$ significance level so only upper limits to their column densities
are listed in Table 4. Metal lines detected in the Ly$\alpha$ forest include
C IV $\lambda\lambda$1548,1550, Fe II $\lambda$1608, and Al II $\lambda$1670.

\subsection{ Q 0449$-$1326 $z_{damp}=1.2667$ System}

    The redshift of this system is too low to see the corresponding 
Ly$\alpha$ absorption line ($\lambda_{obs}=2755$ \AA) from the ground.
The damped Ly$\alpha$ nature of the system is inferred from the detection 
of the intrinsically weak Mn II triplet lines and the 
Fe II $\lambda\lambda$2249,2260 lines (see Table 5), which even for
solar abundances, implies a $N$(H I)$\sim 4\times 10^{19}$ cm$^{-2}$. 
The metal lines detected
in this system are listed in Table 5. Figure 5 shows the absorption line
profiles.

    The derivation of the Mn II column density deserves some discussion.
The estimated values of $N$(Mn II) obtained by integrating $N_a(v)$ increases 
systematically  with decreasing $f\lambda$ value of the lines (Table 5). 
This seems to indicate that all three Mn II lines are moderately saturated. 
However, we argue that this is most likely not the case. 
The Mn II lines are weaker than or comparable to the Fe II $\lambda$2260
absorption in the system. The good agreement 
in the estimated $N$(Fe II) from the
Fe II $\lambda$2260 absorption and the weaker Fe II $\lambda$2249 absorption
indicates that the Fe II $\lambda$2260 absorption is not saturated.
Thus it is difficult to understand why the Mn II lines should be saturated.
The S/N near Mn II $\lambda$2576 is twice as good as the region near
Mn II $\lambda$2594, and three times as good as the region near
Mn II $\lambda$2606.  The values of $N$(Mn II) from the 
$\lambda$2576 and $\lambda$2594 lines agree well within 1$\sigma$.  It is 
only the $N$(Mn II) from the $\lambda$2606 line that differs significantly.
A careful look at the actual apparent column density profiles (figure 6) of
the three Mn II lines confirms that it is the Mn II $\lambda$2606 absorption
that shows significant deviation from the other two lines. It is likely 
that the Mn II $\lambda$2606 absorption is contaminated by an unidentified
weak absorption
line from another redshift system. We will therefore adopt the Mn II column 
density determined from the $\lambda$2576 and $\lambda$2594 lines.

    Since we do not have the $N$(H I) information, it is not
possible to derive absolute abundances for the system. However, it may be
interesting to get a crude idea of what the abundances might be for reasonable
guess of the $N$(H I).  For $N$(H I)=$2\times 10^{20}$ cm$^{-2}$, the column 
densities given in Table 5 implies: [Fe/H]=$-0.68$, [Mn/H]=$-0.96$, and
[Mg/H]$>-1.78$. As we will see later in \S4, the overabundance of Fe
relative to Mn in this system, [Fe/Mn]=+0.28, 
is quite representative of damped Ly$\alpha$ systems
in general and has important implications for the nucleosynthetic history
of the absorbing galaxies.

\subsection{ Q 0450$-$1312 $z_{damp}=1.1743$ System}

    The HIRES spectrum of this object came from Churchill \& Vogt (1997).
Figure 7 shows the metal line profiles, where the intrinsically weak
Fe II $\lambda\lambda$2249, 2260, Mn II $\lambda\lambda$2576, 2594, 2606, and
Cr II $\lambda$2056 lines are clearly detected. The equivalent widths and
column densities of the lines are given in Table 6. The $N$(H I) of the system
is unknown so it is not possible to derive absolute metal abundances. However,
the Fe II column density implies $N$(H I)=$4\times 10^{19}$ cm$^{-2}$ even for
solar metallicity. Hence we believe that this is a damped Ly$\alpha$ system.

\subsection{ Q 0454$+$0356 $z_{damp}=0.8598$ System}

    Abundances of Fe, Zn, and Cr for this damped Ly$\alpha$ system
have been determined by Steidel et al. (1995) using a medium-resolution
spectrum; they found [Fe/H]$=-1.04$, [Zn/H]$=-1.08$,
and [Cr/H]$=-1.07$ (note that these values have been corrected for the
$f$-values of Tripp  et al. 1996). Steidel  et al. also
estimated a $N$(H I)=$(5.7\pm0.3)\times 10^{20}$ cm$^{-2}$ from the
damped Ly$\alpha$ absorption in a HST FOS spectrum of the quasar.
The Keck HIRES spectrum used in this study came from Churchill \& Vogt
(1997). Figure 8 shows the metal lines, while Table 7 gives the measurements.
The stronger Zn II $\lambda$2026 line is not covered by the HIRES spectrum.
In any case, the S/N of the spectrum near the Zn II and Cr II lines is very
low, and none of the Zn II or Cr II lines is detected at
$\geq4\sigma$ significance level. The upper limits
on [Zn/H] and [Cr/H] (Table 7) are consistent with the Steidel et al.
(1995) measurements. The [Fe/H] estimated from the HIRES spectrum also
agrees with the Steidel et al. estimate. The only additional
information obtained for this system is the Mn abundance.

\subsection{ Q 0528$-$2505 $z_{damp}=2.8110$ System}

   We detect over 30 metal lines from 15 different species 
in this system, partially
because the damped Ly$\alpha$ system has a redshift slightly higher than 
the emission redshift of the quasar ($z_{em}=2.779$), so more metal
lines than usual are shifted out of the Ly$\alpha$ forest. 
Even though our spectrum covers most of the damped Ly$\alpha$ 
absorption line, the coverage to the blue side of the
damped Ly$\alpha$ absorption is too short to reliably assess the continuum
level before the damped Ly$\alpha$ absorption.
We therefore adopt log $N$(H I)$=21.2\pm0.1$ from Morton et al. (1980) and
Foltz, Chaffee, \& Black (1988).  The metal lines are presented 
in Table 8 and figure 9. We also derive upper limits to the abundance
of several other elements, including P, Cl, Ti, Co, Cu, Ga, and Ge, in 
order to compare with the results of Songaila \& Cowie (1996). 
Note that our abundance upper limits are 4$\sigma$ values. The corresponding
2$\sigma$ limits should be 0.3 dex lower than those given in the table.

   This system is unique in several ways. First, the redshift of the damped
Ly$\alpha$ absorption is higher than that of the quasar, suggesting that
the absorbing galaxy may be physically associated with the quasar in a group
or cluster. The presumed close proximity of the background quasar, 
which is an intense
source of ionizing photons,  may have significant effect on the physical
state of the absorbing gas in the damped Ly$\alpha$ galaxy. Secondly,
this is the only damped Ly$\alpha$ system for which molecular hydrogen
absorption has been detected (Foltz et al. 1988; Songaila \& Cowie 1996). 
Thirdly, the metal absorption
lines are unusually wide and complex, spreading over 600 km s$^{-1}$
in the strongest low ion lines and 500 km s$^{-1}$ in the high ion lines.
It appears that the absorption consists of two groups of components
at above and below 140 km s$^{-1}$, respectively (figure 9). 
The complexity of the metal
line absorption may be caused by two galaxies in superposition.

    The Si II $\lambda$1808 absorption is relatively strong (about 70\% deep)
and might be somewhat saturated. However, the comparably strong
S II $\lambda$1253 absorption does not appear to show any unresolved
saturation based on a comparison of the $N$(S II) column densities derived from
this line and from the weaker $\lambda$1250 line (Table 8). We will therefore
assume that the $N$(Si II) derived from the $\lambda$1808 absorption
is correct.

    Of particular interest is the C II* $\lambda$1335 absorption.
The ratio of $N$(C II*)/$N$(C II) is a density indicator (\S5). The ratio
can also be used to estimate the temperature of the cosmic microwave
background radiation at the redshift of the absorption system (\S6).
The C II* absorption in this system is clearly detected but 
is partially blended with the C II $\lambda$1334 absorption line (figure 9). 
However, we note that the part of the C II* absorption 
at $v>140$ km s$^{-1}$ is free of any obvious blending.
Hence it is possible to estimate $N$(C II*) for this part of the absorption
(see last section of Table 8). 

    Another ion of special interest is N I since the abundance ratio
of N/O can be used to infer the origin of the heavy elements
(\S4). We have clearly detected N I absorption from the triplet near 1200 \AA, 
but they are all blended with each other (figure 9). However, 
the part of the absorption at 
$v>150$ km s$^{-1}$ for the weakest N I line $\lambda$1200.710 appears to
be free of blending. Thus we have attempted to derive $N$(N I) for this part of
the absorption, and also for a few other elements (see last section
of Table 8) in order to examine the relative abundances. A direct 
integration of the apparent column density profile of the N I $\lambda$1200.710
line between $v=[+140, +300]$ km s$^{-1}$ yields log $N$(N I)$=14.86\pm0.02$.
It is possible that the N I $\lambda$1200.710 absorption in
this velocity interval is somewhat saturated. In order to be conservative,
we use the Fe II lines to estimate the maximum possible degree of saturation
in the N I absorption. The Fe II $\lambda\lambda$1608,2260 lines yield
log $N$(Fe II)$=14.65\pm0.02$ and log $N$(Fe II)$<15.12$ (4$\sigma$) 
respectively, over the interval $v=[+140, +300]$ km s$^{-1}$ (Table 8). 
The Fe II $\lambda$1608 absorption
might be saturated in this velocity interval, but the column density from 
the $\lambda$2260 line indicates that the value of log $N$(Fe II)=14.65 
may underestimate the true
$N$(Fe II) by at most 0.47 dex. Since the N I $\lambda$1200.710 absorption
has similar optical depth as the Fe II $\lambda$1608 absorption over the
velocity interval [+140, +300] km s$^{-1}$, it may be reasonable to 
assume that the true log $N$(N I) over this velocity interval should be
no more than 14.86+0.47=15.33. Thus we find conservatively 
log $14.65<N$(N I)$<15.33$, and [N/O]$<+0.65$ or $-1.21<$[N/Si]$<-0.53$ for
the absorption complex at $v>140$ km s$^{-1}$.

\subsection{ Q 0528$-$2505 $z_{damp}=2.1410$ System}

    Morton  et al. (1980) estimated $N$(H I)$=5\times 10^{20}$ 
cm$^{-2}$ for this system, which we will adopt and assume a 20\% 
uncertainty.
Figure 10 shows the metal absorption lines detected in this
system. The measurements are given in Table 9.
The Zn II $\lambda$2026 absorption is probably detected but
unfortunately blended with the Al II $\lambda$1670 line at $z_{damp}=2.8110$, 
and we only have
an upper limit to $N$(Zn II) from the weaker Zn II $\lambda$2062 
line. The Fe II $\lambda\lambda$2249,2260 lines, which normally provide
the best measure of $N$(Fe II) owing to their weak but detectable strength, 
are unfortunately blended with the Al III $\lambda$1854 and $\lambda$1862
lines in the $z_{damp}=2.8110$ damped system, so no measurements 
are possible. We can only derive a lower
limit to $N$(Fe II) from the moderately saturated Fe II $\lambda$1608
line. An upper limit of log $N$(Fe II)$<15.21$  can be obtained from the 
absence of Fe II $\lambda$1611 absorption. We thus adopt 
log $N$(Fe II)=$14.94\pm0.26$ in the analysis.

\subsection{ Q 1425$+$6039 $z_{damp}=2.8268$ System}

   This bright high redshift quasar ($z_{em}=3.2$, V=16.5) was discovered 
by Stepanian et al. (1991). Our Keck spectrum for this object is of 
exceptionally high
quality, with typical S/N reaching 120 per resolution element redward
of Ly$\alpha$ emission.
The damped Ly$\alpha$ absorption line in the Keck spectrum
is shown in figure 11. The metal absorption lines are shown in
figure 13. A number of lines are detected in the Ly$\alpha$ forest,
including Si II $\lambda\lambda$1190,1193,1260,1304, Si III $\lambda$1206,
O I $\lambda$1302, and several Fe II lines. These will not be used in
the analysis.

  The metal absorption lines in this damped system contain two groups 
of components, one between [$-100$, +100] km s$^{-1}$, and another one between 
[+200, +300] km s$^{-1}$ (figure 13). The low ion absorption lines clearly 
indicate that the bulk of
neutral gas occurs at zero velocity ($z=2.82680$). However,
damping profiles fitted to the Ly$\alpha$ and Ly$\beta$ absorption lines, 
with the redshift fixed at 2.82680, yield satisfactory fits to the blue sides
of the profiles for $N$(H I)=$2\times 10^{20}$ cm$^{-2}$ 
and $b=10$ km s$^{-1}$,
but leave a big chunk of absorption unaccounted for at the red side of the 
line profiles (figure 11). The ``extra'' absorption is clearly due to the 
group of components
occurring between +200 and +300 km s$^{-1}$. Figure 12 shows the same 
as figure 11, but with another component added at
$z_{abs}=2.83058$, corresponds to $v=300$ km s$^{-1}$ in figure 13. The extra
component has $N$(H I)=$1.0\times 10^{19}$ cm$^{-2}$ and $b=10$ km s$^{-1}$,
which fits the Ly$\alpha$ profile well, but not that of Ly$\beta$. 
This is because
the red side of the Ly$\alpha$ profile is dominated by the damping wing of
the component at $v=300$ km s$^{-1}$ 
(i.e., independent of the velocity dispersion
of the gas), while the Ly$\beta$ line, saturated
but not damped, depends on the relative distribution of the gas in the group
of components between [+200, +300] km s$^{-1}$ and on the velocity dispersion
of each component. Figure 12 suggests that 
the total $N$(H I) of the group of components between [+200, +300] km s$^{-1}$
is roughly $10^{19}$ cm$^{-2}$. Experiments indicate that the total $N$(H I)
of this group is definitely no higher than
$2\times 10^{19}$ cm$^{-2}$. Because of this relatively low $N$(H I), one
might worry about ionization corrections. For this reason, we will confine
the analysis of this system to the components at $v<120$ km s$^{-1}$,
and adopt a log $N$(H I)$=20.3\pm0.03$. Table 10 lists the metal lines and
the measurements. 

    The N I triplet near 1200 \AA\  is  clearly detected (see figure 13),
although the N I $\lambda$1200.710 line (not shown) 
is contaminated by Ly$\alpha$ forest
absorption. We are only able to derive a lower limit to $N$(N I). The
O I $\lambda$1302 absorption occurs in the Ly$\alpha$ forest and is strongly
saturated. Thus even deriving a limit to the N/O ratio is not possible. 
Our spectrum does not extend red enough to cover the 
weak Si II $\lambda$1808 absorption line. The Si II $\lambda$1526 
absorption is saturated, and the component near $-90$ km s$^{-1}$
is blended with the C IV $\lambda$1548 absorption at $z_{abs}=2.7727$. 
We are only
able to derive a lower limit to $N$(Si II) from the Si II $\lambda$1526
absorption at $v>-50$ 
km s$^{-1}$. The Fe II $\lambda$1608 absorption is relatively strong and might
be somewhat saturated. We estimate log $N$(Fe II)$=14.44$ from integrating
its $N_a(v)$ profile.
The much weaker Fe II $\lambda$1611 absorption is detected with 
$w_r=0.007\pm0.002$ \AA\ (3.5$\sigma$ detection), with a column density of
log $N$(Fe II)=14.48. This suggests that the Fe II $\lambda$1608 absorption
is unsaturated or nearly unsaturated. However, in order to be self-consistent,
we adopt log $N$(Fe II)$<14.52$ (4$\sigma$) for Fe II $\lambda$1611, and
adopt log $N$(Fe II)$=14.48\pm0.04$ as the final column density.
The S II lines are apparently blended with strong Ly$\alpha$ absorption lines
so no measurements are possible. Most (perhaps all) of the absorption in the 
C II* $\lambda$1335 panel is from the C II $\lambda$1334 line. 
The contamination
made it impossible to estimate the $N$(C II*) or its upper limit.

\subsection{ Q 1946$+$7658 $z_{damp}=2.8443$ System}

   This damped Ly$\alpha$ system and another one at $z_{damp}=1.7382$ in the
same spectrum (see \S3.11) have been studied in some detail 
by Lu  et al. (1995b; see also Fan \& Tytler 1994).  The Keck spectrum
obtained here is of higher quality than those used in previous studies
and extends further into the blue.
Thus it is worthwhile to repeat the analysis, particularly 
in light of the N I lines which are covered by the Keck spectrum.
The $N$(H I) of the system is estimated to be
log $N$(H I)$=20.3\pm0.1$ by Lu  et al., which agrees well with the
new determination based on the Keck spectrum: log $N$(H I)=$20.27\pm0.06$
(see figure 14).

    The metal lines in this systems are shown in figure 15. 
The Si II $\lambda$1190,1193 and Si III $\lambda$1206 lines are detected
in the Ly$\alpha$ forest.  The comparison between the C II $\lambda$1334 
absorption and the O I $\lambda$1302 absorption is very informative. 
While the O I absorption has only one obvious component centered at 
zero velocity, the C II absorption shows at least two extra components
at $-70$ and +80 km s$^{-1}$. This difference is almost certainly not 
due to the relative strength of the O I and C II absorption since the 
much weaker Al II $\lambda$1670 absorption (and possibly the 
Si II  $\lambda$1526 absorption) also shows the two extra components
as seen in C II. The difference is most likely due to ionization effects. 
The O I ion may be  considered  the best tracer of neutral gas 
(other than H I itself) since it has an ionization potential 
(13.61 eV) nearly identical to that of H I. On the other hand, 
ions like C II, Al II, Si II, etc. can exist in partially ionized 
regions because of their higher ionization
potentials. Thus essentially all of the neutral hydrogen in the system
is associated with the component at zero velocity. 
We will hence restrict the column density measurements of the low
ion lines to the component at zero velocity 
(which is reflected in the $v_-$ and
$v_+$ values used to estimate the equivalent width and column density).
This distinction makes an appreciable difference only for Al II.
The results are given in Table 11.

    The C II* $\lambda$1335 absorption is clearly absent, 
and the upper limit obtained here is twice as good as that 
given by Lu  et al. (1995b). By chance, all three N I lines
occur in clean regions of the Ly$\alpha$ forest (figure 15; for space 
considerations the weakest N I line is not shown); however, none is detected
at $\geq4\sigma$ significance level. The upper
limit on $N$(N I) and the lower limit on $N$(O I) from 
the saturated O I $\lambda$1302
absorption yields a firm upper limit of [N/O]$<-0.76$ dex. 
We are also able to put a not-very-informative upper 
limit on [S/H] from the absence of the S II $\lambda$1250
absorption, which happens to occur in a clean region of the forest. 
The two stronger S II lines are blended with Ly$\alpha$ absorption. 
The absorption in the N V $\lambda$1242
panel is most likely due to Ly$\alpha$ forest lines.

\subsection{ Q 1946$+$7658 $z_{damp}=1.7382$ System}

   The Ly$\alpha$ line at this redshift occurs at a wavelength blueward of the
Lyman limit absorption from the $z_{damp}=2.8443$ damped system 
in the same quasar spectrum and is thus not observable. 
The damped Ly$\alpha$ nature of the system is inferred from the detection
of the intrinsically weak Si II $\lambda$1808, Fe II $\lambda\lambda$2249,2260 
and the Cr II lines (figure 16), which implies 
$N$(H I)$=1.6\times 10^{19}$ cm$^{-2}$
even for solar abundances. The metal lines detected in the system are shown in
figure 16 and listed in Table 12. The Cr II $\lambda$2066 absorption is
blended with the C IV $\lambda$1548 absorption at $z_{abs}=2.6541$.
The C IV absorption lines in this damped system occur in the Ly$\alpha$ forest
and is probably somewhat contaminated.

   The column densities for Al III obtained  from 
the Al III $\lambda\lambda$1854,1862 absorption seem 
to indicate unresolved saturation in the Al III lines as the difference
between the two estimates, 0.12 dex, is significantly larger than 
the 1$\sigma$ errors.
This is surprising given the relative weakness of the absorption lines.
However, most of the difference occurs near $v=20$ km s$^{-1}$, 
where the intrinsically weaker
Al III $\lambda$1862 absorption is even deeper than the intrinsically stronger
$\lambda$1854 absorption. We believe the Al III $\lambda$1862 absorption 
is probably contaminated by an unidentified metal absorption line. 
We will assume this is the case and adopt the
$N$(Al III) from the $\lambda$1854 line.

   The lack of $N$(H I) information prevents an estimate of the absolute 
abundances in this system. For reference, if log $N$(H I)=20.3, 
then the following abundances would
be obtained: [Si/H]$=-1.09$, [Fe/H]$=-1.35$, [Zn/H]=$<-1.12$, [Cr/H]$=-1.20$,
and [Mn/H]$=-1.71$ (adopting log $N$(Mn II)=12.12 from Lu  et al. 1995b).

\subsection{ Q 2212$-$1626 $z_{damp}=3.6617$ System}

     This is the third highest redshift damped Ly$\alpha$ system for which 
detailed analysis has been carried out. 
Only the $z_{damp}=4.0803$ damped system
toward Q 2237$-$0608 (\S3.14) and the $z_{damp}=4.3829$ damped system toward
BR 1202$-$07 (Lu  et al. 1996a) are at higher redshifts. From 
our Keck spectrum (see figure 17), we
estimate log $N$(H I)=$20.20\pm0.08$.
The metal lines are shown in figure 18 and listed in Table 13.
A number of other metal lines (e.g., Si III $\lambda$1206, 
Si II $\lambda\lambda$1190,1193,1260)
appear to be present in the Ly$\alpha$ forest but are badly blended with forest absorption.

     One peculiarity about this system is that the O I $\lambda$1302 absorption 
clearly does not reach zero flux in its line center, which is unusual for this
normally-strong absorption line.
The O I absorption occurs on top of the strong Ly$\alpha$ emission
line of this quasar. Since heavily saturated Ly$\alpha$ 
absorption lines to the blue of the O I
absorption clearly goes to zero flux in their line centers, 
the residual flux in the O I absorption line center is not an 
artifact of the data reduction procedure. Indeed, visual inspections
of the raw images clearly confirm that the flux in the O I absorption 
line center is not zero.
We consider two possible causes of this residual flux: 
resolution smearing and ionization effects.

     If the O I absorption consists of many extremely narrow components that
are not resolvable at our resolution (FWHM=6.6 km s$^{-1}$), smearing
of the components due to the inadequate resolution
could lift the flux in the absorption line
center above the zero level even if the absorption components are intrinsically
saturated. This requires the components to have $b<4$ km s$^{-1}$ or
temperature $T<1.5\times 10^4$ K. Experiments show that, indeed, one can
explain the residual flux in the O I absorption line center if it contains
several appropriately spaced components narrower than $b=4$ km s$^{-1}$. 

     Another possible explanation of the residual flux in the O I absorption 
is that it is {\it real}, namely, the O I absorption line is unsaturated. 
This is unusual but not impossible. 
If this is the case, then the $N$(O I) from 
integrating the $N_a(v)$ profile (see table 13) should be an actual
measurement rather than lower limit. This in turn implies [O/H]=$-2.37$
for this system. This O abundance is nearly 0.5 dex below the Si abundance
([Si/H]=$-1.90$) in the same system, which is inconsistent with nucleosynthesis
considerations (see \S4.3). One can reconcile this difference if 
the Si abundance in this system has been overestimated 
because some of the Si II ions come from partially ionized 
gas rather than from neutral gas\footnote{Recall that it takes 16.3 eV to turn 
Si II into Si III, so Si II ions can exist in H II regions. 
In contrast, the ionization potential of O I is 13.61 eV,
making it the best tracer of neutral gas other than H I itself.}. 
If this is the case, it  would mean substantial ionization 
corrections for this system. However, we consider this explanation 
unlikely for the following reasons. Firstly, when there are significant
ionization corrections, usually there are notable 
differences in the absorption profiles between O I 
and other ions that can exist in H II regions (e.g., C II,
Si II, etc.; see the $z_{damp}=2.8443$ system toward Q 1946+7658 in \S3.10). 
In contrast, the O I absorption in the Q 2212$-$2616 system has identical 
velocity structure to the C II and Si II absorption in the system.
Secondly, the upper limit on the temperature of the microwave 
background radiation at the redshift of this absorption system 
as derived from the C II*/CII ratio (\S6) is extremely close to the 
Big Bang prediction, indicating that collisional excitation of the 
C II atoms by free electrons is negligible in the absorbing gas.
This is in contradiction with the large electron density expected in this
system if the absorbing gas is substantially ionized. These considerations
suggest that the resolution-smearing interpretation given in the 
preceding paragraph
is likely the correct explanation. In fact, the implied low electron density in
this system from the C II*/CII ratio (\S6) suggests that this system may be
{\it less} ionized than  typical damped Ly$\alpha$ systems, which in turn
may imply a lower temperature and hence narrower line widths (see the preceding
paragraph).

\subsection{ Q 2231$-$0015 $z_{damp}=2.0662$ System}

     Our Keck spectrum does not cover the damped Ly$\alpha$ line.
Lu \& Wolfe (1994) estimated log $N$(H I)=$20.56\pm0.10$ for this system, 
which we adopt. The metal lines are shown in figure 19 and listed in 
Table 14. Metal lines detected in the Ly$\alpha$ forest 
include Si II $\lambda$1526, C II $\lambda$1334, 
Si IV $\lambda\lambda$1393,1402, and C IV $\lambda\lambda$1548,1550,
but they all appear to be contaminated by forest absorption. 
The Fe II $\lambda$1608 absorption yields a
log $N$(Fe II)$=14.72\pm0.01$ from integrating the apparent
optical depth profile. Given its strength, it is probably saturated.
Thus we take the above value as lower limit. The absence
of the Fe II $\lambda$1611 absorption yields an 4$\sigma$ upper 
limit of log $N$(Fe II)$<15.08$. Hence we conservatively adopt
log $N$(Fe II)$=14.90\pm0.18$. The Cr II absorption is below the
4$\sigma$ significance level.

\subsection{ Q 2237$-$0608 $z_{damp}=4.0803$ System}

    This is the second highest redshift damped Ly$\alpha$ galaxy 
studied  in detail. 
    The damped Ly$\alpha$ absorption line in the Keck spectrum is shown
in figure 20. We estimate a $N$(H I)=$3\times 10^{20}$ cm$^{-2}$ with
about 30\% uncertainty. This $N$(H I) estimate
 is more uncertain than the estimates for
the other damped systems based on Keck spectra because both
wings of the damped Ly$\alpha$ line are affected by strong, dense
Ly$\alpha$ forest absorption due to its very high redshift. The $N$(H I)
may have been slightly underestimated judging from 
the base of the damped Ly$\alpha$
absorption, although increasing $N$(H I) will overfit the wings farther
away from the line center. 

    The metal lines in this system are shown in figure 21;
the measurements are given in Table 15. The C II $\lambda$1334
and C II* $\lambda$1335 lines (and to some degree the Ni II $\lambda$1370 line)
happen to occur in the red wing of the strong Ly$\alpha$ emission 
line so the S/N are exceptionally good.  
The Si II $\lambda$1526 and Al II $\lambda$1670 absorption
lines are about 80\% and 70\% deep, respectively. Based on the discussion
in \S2.2, they are likely to be unsaturated or nearly unsaturated. 
We will therefore adopt their column densities
as measurements rather than lower limits.

    Many metal lines in this system are expected in the Ly$\alpha$ forest, 
but trying to identify them is meaningless given the strong possibility of
contamination from the extremely dense Ly$\alpha$ forest.

\section{CHEMICAL EVOLUTION OF DAMPED LYMAN-ALPHA GALAXIES}

   In this section we combine the abundance information obtained for the
sample of damped Ly$\alpha$ galaxies discussed in \S3 with those from
previous studies  in order to carry out a statistical analysis of the chemical 
evolution of damped Ly$\alpha$ galaxies.  In particular, 
we will examine the metallicity distribution of damped Ly$\alpha$
galaxies and compare that with those for the various components (disk, halo)
in the Milky Way galaxy and with external galaxies. 
We will examine the age-metallicity relation of the
damped Ly$\alpha$ galaxies and compare that with the similar relation
determined for the Milky Way. We will also examine the
abundance ratios of various elements with different nucleosynthetic
origin in order to gain more insight into the chemical enrichment processes
in these damped Ly$\alpha$ galaxies.
The effects of dust on the derived  abundances through selective depletion
will also be discussed.

\subsection{Sample Construction}

\subsubsection{Damped Ly$\alpha$ Abundances}
 
Elemental abundances have been measured for a number of damped Ly$\alpha$ 
galaxies in the literature with varying degrees of accuracy. 
In constructing the sample of damped Ly$\alpha$ galaxies for further
statistical analysis, we will only consider those abundance measurements
that we believe to be largely free of saturation effects.
This means that, in general, measurements (but not lower limits) from 
heavily saturated lines will be discarded. Previous abundance
estimates or upper limits  that were based on lines with less than 
4$\sigma$ significance
level will be converted where possible into 4$\sigma$ {\it upper limits}
 to conform
to the standard adopted in \S3, or else they will be discarded. 
Where appropriate, we will 
also correct the published abundance measurements for the set of new
oscillator strengths compiled by Tripp  et al. (1996) so that
all the measurements will be on the same footing.  
Following tradition, we will use Fe as the metallicity indicator.
Hence, in general, only damped Ly$\alpha$ galaxies 
whose Fe abundance is available
will be included in the sample. However, as we will see in \S4.3, the Cr
abundance tracks Fe abundance very well in damped Ly$\alpha$ galaxies,
i.e., the Cr/Fe abundance ratios in damped Ly$\alpha$ galaxies (when
both are measured) are consistent with their solar ratio. In order to 
increase the sample size, we will sometimes substitute [Cr/H] for [Fe/H]
when the Fe abundance is not available (this happens in only 4 of the 23 
systems listed in Table 16). Hence damped Ly$\alpha$ galaxies without
[Fe/H] measurement, but with [Cr/H] measurement, will be included in the
sample. The final sample of damped Ly$\alpha$ galaxies, including those
discussed in \S3 and those from previous studies that satisfy the above
selection criteria, is given in Table 16.
 
\subsubsection{Galactic Disk and Halo Star Abundances}

    The intrinsic chemical compositions of Galactic disk and halo stars
of different age reflect the conditions of the local interstellar medium 
in the Galaxy when the stars were formed, which depend on the cumulative
effects of the rate of star formation, the stellar initial mass function,
stellar yields,  gas infall/outflow, etc. prior to that
epoch.  The chemical compositions of Galactic stars not only
allow us to study the past history of chemical evolution of the Milky
Way, but also provide a fundamental reference 
of what might be expected of the chemical compositions of interstellar
gas at different stages of a galaxy's enrichment history. 
In \S4.3 we will compare the relative abundances of various elements
found in damped Ly$\alpha$ galaxies
to those found in Galactic stars in order to gain some insight into 
the chemical enrichment processes occurring in damped Ly$\alpha$ galaxies.
In constructing the sample of stellar abundances, the following sources
are used: 
Gratton \& Sneden 1988, 1991 (Fe, Si, Cr, Ni), 
Edvardsson  et al. 1993 (Fe, Si, Ni),
Clegg, Lambert, \& Tomkin 1981 (Fe, S),
Francois 1987, 1988 (Fe, S).
Gratton 1989 (Fe, Mn), 
Beynon 1978 (Fe, Mn), 
Sneden \& Crocker 1988 (Fe, Zn),
Sneden, Gratton, Crocker 1991 (Fe, Zn), 
and Magain 1989 (Fe, Cr).

We note that stellar abundances of sulphur are rarely studied mainly
because the most useful S I lines occur in the near-infrared
($\sim$8700 \AA) and are weak and blended.
Of the three references for sulphur that we can find, Clegg  et al.
(1981) studied 20 stars mostly with [Fe/H]$>-0.5$, while Francois (1987, 1988)
contain 26 stars mostly with $-1.5<$[Fe/H]$<-0.5$. Although the Francois 
sample stars are more useful for comparison 
with the damped Ly$\alpha$ galaxies from 
the metallicity point of view, Lambert (1989) argued that the sulphur
abundances of Francois for halo stars, $\langle$[S/Fe]$\rangle=0.6$ dex, 
may be too high by 0.2 dex owing to the S I $f$-values used. 
Fortunately, this uncertainty does not affect any of the conclusions reached
later.

\subsubsection{Galactic ISM Cloud Abundances}

It is well known that the relative abundances of elements 
in Galactic ISM clouds differ significantly from the 
solar composition and vary widely from sight line to sight line,
i.e., from cloud to cloud. These deviations from solar relative abundances
stem from the fact that some elements (refractory elements) 
are preferentially incorporated
into solid forms or dust grains, while other elements are much less affected.
Assuming that the intrinsic abundances of the elements in the ISM clouds
are the same as the solar abundances, the measured gas-phase abundances
of the elements in ISM clouds then allow one to infer the fractions of 
different
elements that are locked up in grains, i.e., the amount of depletion.
Thus the abundance ratios of elements measured in diffuse ISM clouds
provide a fundamental reference to gauge the effects of dust depletion.
With the understanding that the properties of dust grains in damped 
Ly$\alpha$ galaxies may be significantly different from those of Galactic 
dust\footnote{Indeed, the extinction curves
found for the LMC and the SMC are known to be different from 
each other and from that found
in the Milky Way, indicating that the dust grains in all three galaxies
are different. Similarly, the reddening law has been found to depend on 
radius in M31 (Iye \& Richter 1985).}, we will compare in \S4.3 
the abundance ratios found in damped Ly$\alpha$ galaxies with those 
found in the ISM clouds in order to get a handle on the possible presence or
absence of dust in damped Ly$\alpha$ galaxies.

Abundances of Si, S, Cr, Mn, Fe, Ni, and Zn measured for 
ISM clouds based on recent high quality HST GHRS data are compiled
in Table 17. Only abundance measurements with
relatively small uncertainties (typically less than 0.1 dex) that are
judged to be free of significant saturation effects will be used.
As with the
damped Ly$\alpha$ abundances, we have corrected the GHRS measurements for
the set of new $f$-values compiled by Tripp  et al. (1996) where
appropriate.

\subsection{Age-metallicity Relation}

    Figure 22 shows the distribution of [Fe/H] as a function of
redshift for the damped Ly$\alpha$ galaxies in our sample
(solid circles). This is equivalent to the age-metallicity relation
commonly known in Galactic chemical evolution studies.
We have used [Fe/H] as the metallicity indicator, as is 
conventional. In a few cases, [Cr/H] is used in place of [Fe/H]
when the latter is not available (see \S4.3 for justification).
Also displayed in figure 22 is the age-metallicity
relation for the sample of Galactic disk stars (``+'' symbols) 
studied by Edvardsson  et al.  (1993). The conversion
between age (starting from the Big Bang)  and redshift assumes
$q_0=0.1$ and $H_0=50$ km s$^{-1}$ Mpc$^{-1}$,
which gives a current age of the universe of 16.6 Gyrs, consistent
with the stellar age determinations.

   As is clear from figure 22, the damped Ly$\alpha$ galaxies have 
Fe-metallicities in the
range of 1/10 to 1/300 solar, or [Fe/H] between $-1.0$ and $-2.5$. 
Hence they represent a population of young galaxies that are still in the
early stages of their chemical evolution. In fact, as we argue in 
\S4.3, the star formation process in most of these galaxies
probably started less than 1 billion years before the epoch corresponding
to their redshift. The $N$(H I)-weighted  mean metallicity of the 
damped Ly$\alpha$
galaxies between $2<z<3$ is [Fe/H]=$-1.56$, which may be regarded as
the cosmic metallicity at $\langle z\rangle=2.5$.

It is important to note that
the true Fe abundances of damped Ly$\alpha$ systems could
be systematically higher than those indicated in figure 22 if
significant depletion of Fe onto dust grains has occurred. However,
as we argue in \S4.3, the amount of Fe depletion in the damped 
Ly$\alpha$ galaxies in our sample is probably negligible, and
should be on average no more than 0.4 dex. Hence modest depletion
of Fe will not change the qualitative picture presented in this work.

\subsubsection{The Large Spread in [Fe/H]}

    As already noted by Pettini  et al. (1994), the chemical
enrichment processes in damped Ly$\alpha$ galaxies appear quite inhomogeneous.
For example, the [Zn/H] distribution in the sample of damped systems studied
by Pettini  et al. spans at least a factor of 20 or 1.3 dex.
The [Fe/H] in our sample at $z\sim2-3$ spans roughly the same range.
A metallicity gradient similar to those found in the Milky Way and 
in local spiral disks 
can only account for a factor of $\sim$2 spread in the observed metallicities
of damped Ly$\alpha$ galaxies (Pettini  et al. 1994). 
If all galaxies in the sample formed at roughly the same time, then the 
large spread in [Fe/H] may indicate widely different star formation histories
in these galaxies.  However, we consider it more likely that
the large metallicity spread is due to the different formation epoch of 
the sample galaxies and/or that the sample contains a heterogeneous mix of
galaxy types. Steidel (1995) found 
that galaxies responsible for the MgII absorption
systems at $z<1.6$ in quasar spectra come with a variety of 
morphological types, colors, and luminosities. This may explain the
large metallicity spread in damped Ly$\alpha$ galaxies if they sample
the same parent population of galaxies. 
We point out that varying amounts of Fe depletion
due to dust grains are unlikely to account for the bulk of metallicity spread 
since [Zn/H] shows similar spread (Pettini et al. 1994), 
and Zn is not expected to be depleted heavily by dust grains (see discussion
in \S4.3).

\subsubsection{Evidence for $z\geq3$ as the Epoch of Galaxy Formation}

   There is evidence in figure 22 that the mean metallicity of the 
sample increases
with time. This is perhaps not unexpected, but nonetheless reaffirms
the notion that heavy metals gradually build up with time
in galaxies through successive generations of star formation. 
There was a hint of such a trend in earlier studies (Pettini  et al. 1995b),
but the trend is made much clearer here 
by the addition of several new measurements at $z>3$ and by the fact 
that most of our data points are actual measurements rather than
upper limits (as is the case in Pettini et al. 1995).
In particular, all four of the damped Ly$\alpha$ galaxies at $z>3$ have 
[Fe/H]$<-1.7$ (or, [Fe/H]$<-2.0$ if we use the 2$\sigma$ upper limit for
the Q2212$-$1626 system). In comparison, many of 
the galaxies at $2<z<3$ have reached ten times higher metallicity.
Thus the time around $z\sim3$ could be the epoch of galaxy formation in
the sense that it signifies the onset of significant star formation
in galaxies. Several other lines of evidence appear to point to the
same conclusion,  as we discuss below.

The first such evidence comes from the space density distribution
of quasars. It is found that the space density of quasars peaks at
$2<z<3$ and starts to decline at $z>3$ (Schmidt, Schneider, \& Gunn 1995;
Warren, Hewett, \& Osmer 1994). By $z\sim4.3$, the space density of quasars
has dropped by a factor of 7 compared to their peak density 
at $z=2-3$ (Kennefick, Djorgovski, \& de Carvalho 1995). If this drop 
in quasar space density at $z>3$ signifies the early phase of structure
formation from the highest density peaks of the primordial fluctuation, 
it is perhaps ``natural'' to expect galaxy 
formation and subsequent star formation 
to occur at slightly lower redshifts. The second supporting evidence
comes from studies of the neutral gas content in damped Ly$\alpha$ systems.
Previous studies found that $\Omega_{damp}$, the cosmological density 
contained in the  H I gas in damped Ly$\alpha$ galaxies,  decreases from
$z\sim3$ to $z=0$, which was interpreted to indicate the transformation
of gas into stars (Lanzetta et al. 1995; Wolfe et al. 1996;
but also see Pei \& Fall 1995). Recent extension of such studies to 
even higher redshift reveals that $\Omega_{damp}$ declines rapidly
at $z>3$ (Storrie-Lombardi \& Wolfe 1996). This may indicate that damped
Ly$\alpha$ galaxies are still forming at $z>3$.
The third supporting evidence comes from the morphology of high redshift
galaxies. HST images of $z>3$ galaxies identified from their redshifted
Lyman continuum break (Steidel, Pettini, \& Hamilton 1996a; 
Steidel et al. 1996b; Giavalisco, Steidel, \& Macchetto 1996) 
reveal morphological
structures that could be interpreted to indicate star formation in the
spheroidal component of massive galaxies. 

Even if none of these arguments is convincing by itself, taken together,
they do consistently point to a picture that galaxies started forming at
$z$ around or slightly higher than 3.

Finally, it may be significant that the lowest Fe-metallicity we find for the
sample of damped Ly$\alpha$ galaxies is about $-2.4$, 
which, coincidentally, also represents
the lowest metallicity found for Galactic globular clusters. Recent 
observations of extremely low metallicity 
halo stars ([Fe/H]$<-2.5$) reveal that
the Galaxy underwent a distinct phase of nucleosynthesis before reaching
[Fe/H]$\sim -2.4$, characterized by some unusual elemental abundance
ratios and large variations 
(McWilliams  et al. 1995; Ryan, Norris, \& Beers 1996).
Possibly, this marks a time before which the chemical composition of
the gas is only affected by a few Type II supernovae (Audouze \& Silk 1995),
namely, the epoch of formation of the first generation stars.
More studies of $z>3$ damped Ly$\alpha$ galaxies may eventually reveal
such systems (i.e., [Fe/H]$<-2.5$), which would provide unambiguous
evidence for the initial formation of galaxies.

\subsubsection{Comparison with the Milky Way Galaxy and Other Nearby Galaxies}

   We note from figure 22 that the degree of chemical enrichment in
the sample of damped Ly$\alpha$ galaxies is, on average, 
considerably lower than the Milky
Way disk at any given time in the past. This may bear significantly on the
nature of the damped Ly$\alpha$ galaxies. It was initially
suggested (cf. Wolfe 1988) that the damped Ly$\alpha$ absorbers may
trace disks or proto-disks of high-redshift spirals. But the low
metallicities of the damped Ly$\alpha$ galaxies cast some doubts
on this interpretation. A thick disk similar to that of our own Galaxy
also appears to be inconsistent with the 
observed abundance distribution in damped Ly$\alpha$
galaxies, since the mean metallicity of stars in the thick disk is
around [Fe/H]=$-0.6$ with a range of $-1.4$ to 0 (cf. Pardi, Ferrini, \&
Matteucci 1995).  However, Beers \& Sommer-Larsen (1995) 
suggested that the [Fe/H] distribution of
thick disk stars probably extends to as low as $-2$. Since existing
studies of thick-disk stars are still limited to within a few kpc of the
disk, the known metallicity distribution may be biased toward higher values.
Thus a thick-disk interpretation of the damped Ly$\alpha$ abundances 
remains viable, especially in light of the (somewhat 
preliminary) evidence that damped Ly$\alpha$ galaxies exhibit kinematics
expected of rotating structures (Wolfe 1995).
Detailed accounts of the properties of the thick disk of the Milky Way
may be found in Majewski (1993).

 On the other hand, the damped Ly$\alpha$ galaxies
have metallicities resembling those found for Galactic halo stars
and globular clusters. Coincidentally, the range of [Fe/H] spanned
by the damped Ly$\alpha$ galaxies at $z\sim 2-3$ is nearly identical
to that spanned by Galactic globular clusters. This suggests  the 
interesting possibility that damped Ly$\alpha$
absorbers may represent a phase when the spheroidal component of high-redshift
galaxies is being formed. Such a possibility has, in fact, been suggested
previously based on different arguments (Lanzetta et al. 1995; Wolfe  1995).
The Fe-abundance distribution of damped systems is  also consistent
with that of old stellar populations in local group dwarf spheroidal galaxies
(cf. Da Costa 1992). Dwarf spheroidals contain old to intermediate age
stellar populations but with little gas, and the metallicity distributions
of their stellar populations indicate episodic star formation
in their history. Presumably, most or all of them contained 
interstellar gas in their remote past that would be sufficient
to produce damped Ly$\alpha$ absorption lines. The more gas-rich 
dwarf irregular galaxies (e.g., LMC and SMC) also appear
to contain old and sometimes intermediate age stellar populations, in addition
to young stellar populations indicating ongoing or recent star formation.
The metallicities of the old stellar populations in these 
galaxies are also similar to those seen in damped Ly$\alpha$ galaxies. 
These considerations suggest that damped Ly$\alpha$ galaxies may represent
gas-rich dwarf galaxies or sub-galactic fragments which are still
collecting themselves to form more massive galaxies. More discussion
about the nature of damped Ly$\alpha$ galaxies may be found
in \S5.

\subsection{Abundance Ratios: Nucleosynthesis vs Dust Depletion}

    In figure 23 we show the abundance ratios of Si, S, Zn, Cr, Mn, and Ni to
Fe plotted against [Fe/H]. 
Again,  except for the [Cr/Fe] panel, the Cr abundance 
has been used in place of the Fe abundance when the latter is not available. 
This is justified because all the
Ly$\alpha$ galaxies for which we have both Fe and Cr measurements 
indicate [Cr/H]$\sim$[Fe/H]: $\langle$[Cr/Fe]$\rangle=0.04\pm0.10$. The
dispersion in [Cr/Fe] is consistent with the typical measurement uncertainty
of $\sim$0.1 dex. Most of the abundance limits 
for C, N, O, and Al in Table 16 are not
strong enough to constrain the origin of these elements, so they will
only be considered in special cases.

  In principle, elemental abundance ratios allow one to infer 
what kind of nucleosynthetic processes may be responsible for the
enrichment of the interstellar medium. They are important for constraining
models of stellar nucleosynthesis and the shape of the IMF.
This is because different elements
have different nucleosynthetic origins and different enrichment time scales. 
For example, essentially all the oxygen in the Galaxy and much of 
the observable $\alpha$-process elements (including Si, S, Ca and Ti) 
are produced by massive
stars ($M>10M_{\odot}$) through explosive nuclear burning during Type II
SN explosions (cf. Wheeler  et al. 1989). 
The corresponding  time scale is very short: $<2\times 10^7$ yrs.
In contrast, most of the Fe-peak elements (including V, Cr, Mn, Fe, Co, Ni,
and possibly the lighter elements Sc and Ti, and the heavier elements
Cu and Zn) are believed to be produced by Type Ia 
SN\footnote{Using the Type II supernova models of Woosley \& Weaver (1995)
and a simple chemical evolution model, Timmes, Woosley, \& Weaver (1995)
found that only 1/3 of the solar Fe abundance comes from Type Ia supernovae.
However, they also found that better agreement with most of the 
observed abundance data can be achieved if the Type II supernova yields
of Woosley \& Weaver are systematically reduced by a factor of 2. }
 (cf. Wheeler et al. 
1989), which evolve from low to intermediate
mass stars ($M_{\odot}<M<8M_{\odot}$). Thus the time scale for producing 
the bulk of Fe-group elements should be at least as long as
the lifetime of the progenitor stars of Type Ia SN, 
which is $>10^8-10^9$ yrs, and
is possibly much longer depending on the uncertain time scale
between the formation of the white dwarf and the ignition of SN Ia explosion.
For example, Yoshii, Tsujimoto, \& Nomoto (1996) deduced an ``effective'' 
lifetime of $\sim 1.5$ Gyrs for Type Ia SN progenitors based on 
the analysis of Galactic stellar abundance data.
The same low-to-intermediate mass stars also make much of the solar
N (and possibly C) 
through mass losses on comparable time scales. Hence the production
of C, N, and Fe-group elements is expected to 
lag behind that of O and $\alpha$-group
elements following an episode of star formation.

    However, in the case of elemental abundances derived from gas phase
absorption, one also needs to worry about the 
additional effect of dust depletion
if a significant amount of dust exists in these damped Ly$\alpha$ galaxies.
Refractory elements such as Si, Mn, Fe, Cr, Ni, etc. 
are preferentially incorporated 
into dust grains, which depletes their abundances 
in the gas phase. That different elements deplete in different amounts
significantly modifies the relative abundances of 
these elements in the gas phase.

    In order to distinguish between the effects of nucleosynthetic processes 
and dust depletion, we compare
the abundance ratios observed in damped Ly$\alpha$ galaxies with those
seen in Galactic disk and halo stars, and with those determined for diffuse
Galactic ISM clouds. The former serves to illustrate the effects
of nucleosynthesis at different stages of chemical evolution of the Milky
Way, while the latter illustrates the effects of dust depletion.
Granted, there is no guarantee that any of the damped Ly$\alpha$ galaxies
would have the same chemical enrichment history as the Milky Way, or that
the dust grains in these high-redshift galaxies would have properties similar
to Galactic dust, but these comparisons should nevertheless be useful
in providing some general guidance.

\subsubsection{Can Dust Depletion Explain the Observed Abundance Ratios?}

  Figure 24 shows the abundance ratios in diffuse ISM clouds in both the 
Galactic disk and halo from Table 17. Reviews of this subject may be found in
Jenkins (1987) and Savage \& Sembach (1996). 
Since it is believed that the intrinsic 
elemental abundances of the local interstellar medium are solar, the spread 
in [Fe/H] (the horizontal axis in figure 24) 
represents different levels of dust depletion in the clouds 
rather than variations in the intrinsic metallicity of the clouds (as
is the case for figure 23). The
departure of the observed  elemental abundances in ISM clouds from the solar
mixture is generally considered to be caused by condensation of refractory 
elements into solid forms (grains). In the diffuse Galactic ISM,
Fe, Cr, and Ni are among the most heavily depleted elements; 
Si and Mn are somewhat less depleted; while Zn, S and C, N, O are nearly
unaffected by dust depletion (Jenkins 1987). 
These depletion characteristics show up clearly in figure 24. 
Comparing figure 23 and figure 24, one finds good agreement in the 
[Cr/Fe] and [Ni/Fe] ratios. The [Si/Fe], [S/Fe] and [Zn/Fe] 
ratios in damped Ly$\alpha$
galaxies are consistent with the effects of dust depletion (figure 24)
in the {\it direction} of the deviations from the solar ratios, 
but the {\it amounts} of
deviations are much less in the damped Ly$\alpha$ galaxies. This may   be 
explained if the dust-to-gas ratio in damped Ly$\alpha$
galaxies is lower than the Galactic value or the mean gas density in
damped Ly$\alpha$ galaxies is lower; the latter follows from the fact that
the degree of elemental depletion is observed to 
correlate with the mean gas density along the line of sight in the 
Milky Way (see the review in Jenkins 1987). However, the Mn/Fe ratio found
for damped Ly$\alpha$ galaxies is completely opposite to 
what is expected from the dust depletion effect. 
Another strong piece of evidence against a dust
depletion interpretation of the observed abundance ratios in damped
Ly$\alpha$ galaxies is their N/O ratio (see below).

The N I $\lambda\lambda$1199.55, 1200.22, 1200.71 triplet occurs in 
the Ly$\alpha$ forest so
it is generally not possible to make reliable measurements of their column
densities. However, in two damped Ly$\alpha$ systems (see Table 16), 
the clear absence of N I
absorption and the presence of saturated O I absorption provide firm upper
limits on their [N/O] ratio of $<-0.76$ and $<-0.31$ dex, respectively. 
Stronger limits can
be found for [N/Si]: $<-1.31$ and $<1.05$, respectively. In addition, we found
$-1.21<$[N/Si]$<-0.53$ for several components 
in the $z_{damp}=2.8110$ system toward
Q 0528$-$2505 (see \S3.7). Since both Si and O are
believed to come from massive stars, and indeed, are observed to
have solar abundance ratio in Galactic halo stars (cf. Wheeler  et al. 1989) 
and in metal-poor
dwarf galaxies (Thuan, Izotov, \& Lipovetsky 1995), it is probably safe
to assume that [Si/O]=0 in damped Ly$\alpha$ galaxies. We then find
N/O at least a factor of 3.3 to 20 times {\it lower} than the solar ratio in 
these damped Ly$\alpha$ galaxies.

The abundances of N and O in ISM clouds have been measured 
toward many sightlines
using the {\it Copernicus} satellite by York  et al. (1983), whose data were
reanalyzed by  Hibbert, Dufton, \& Keenan (1985) and by Keenan, 
Hibbert, \& Dufton (1985) using improved $f$-values.
Similar determinations have been made for a few clouds using the GHRS
on board the HST (Cardelli, Savage, \& Ebbets 1991).  Since these 
measurements were made using the extremely weak intersystem transition
lines of N I] $\lambda$1160 and O I] $\lambda$1335 rather than the 
strongly saturated N I $\lambda$1200 triplet and O I $\lambda$1302 
transition, they do not suffer significantly from the usual  
saturation problem associated with the strong lines.
Their results indicate that the depletion of N and O in diffuse ISM clouds is 
almost always less than 0.3 dex, and in addition, N is usually
less depleted than O.
Hence, it is extremely difficult to explain the observed N/O ratios in damped
Ly$\alpha$ galaxies with dust depletion if the intrinsic N/O ratio is solar. 

    Thus we conclude from the above discussion that dust depletion 
alone cannot explain the abundance ratios observed in damped Ly$\alpha$
galaxies.

\subsubsection{Can Nucleosynthesis Explain the Observed Abundance Ratios?}

    Figure 25 shows the abundance ratios found in 
Galactic disk and halo stars (see \S4.1.2 for references).
The overabundance of Si to Fe relative to the solar ratio in halo
stars ([Fe/H]$<-1$), [Si/Fe]$\sim0.3-0.4$,
is now well  established, and is thought
to reflect the yield of these elements from Type II supernovae.
A similar overabundance
is seen for S (figure 25) and other $\alpha$-elements
(cf. Wheeler  et al. 1989). All the other
elements displayed in figure 25 (Cr, Mn, Fe, Ni, Zn) are Fe-peak elements.
The fact that the Fe-peak elements are believed to be formed under
nuclear statistical equilibrium and thus share a common origin may explain 
why their abundance ratios in the halo stars are so similar to the 
solar values.  The only exception is Mn, with
$\langle$[Mn/Fe]$\rangle\sim -0.3$ for halo stars.
This ``odd-even'' effect, namely, the relative underabundance of odd-Z
elements (where Z is the atomic number) compared to the abundances 
of even-Z elements of the same nucleosynthetic origin, 
is fairly well established observationally.
One theory of why this occurs is that
the production of odd-Z elements depends on
the neutron excess in the nuclear fuel, which in turn depends on the
initial metallicity of the stellar composition (Truran \& Arnett 1971). 
Calculations of explosive nucleosynthesis in metal poor stars by Truran
\& Arnett indicate that the yield of both odd-Z nuclei and the neutron-rich
isotopes of even-Z nuclei are significantly reduced relative to more
metal-rich stars.

   We find remarkable agreement in the Si/Fe, S/Fe, Cr/Fe, and Mn/Cr ratios
between damped Ly$\alpha$ galaxies
and  those seen in Galactic stars in the same metallicity range.
The difference in Ni/Fe between figures 23 and 25 may not be real
and may have to do with the uncertain $f$-values of the Ni II transitions
(cf. Morton 1991).
Since the stellar abundances are not determined from the
same Ni II lines (the stellar abundances are from Ni I lines) as for the
damped Ly$\alpha$ galaxies, a
systematic error in the Ni II $f$-values could explain the  nearly constant
offset of [Ni/Fe] from the zero point in damped Ly$\alpha$ galaxies. 
This may also explain why the
Ni/Fe ratios in damped Ly$\alpha$ galaxies more closely resemble those
seen in diffuse ISM clouds (figure 24) as in both cases the abundances of Ni
are derived from 
the same Ni II transitions with the same $f$-values. We consider the
Ni/Fe ratio unsuitable for the purpose of discriminating a nucleosynthetic
origin of the damped Ly$\alpha$ abundances from a dust depletion origin.
More reliable determinations of the Ni II $f$-values in the future would make
the Ni measurements more useful. We also note that, although in most damped 
Ly$\alpha$ systems the Al II $\lambda$1670 absorption lines are saturated,
the Al II absorption in the $z_{damp}=2.8443$ system toward Q 1946+7658 
and in the $z_{damp}=4.0803$ system toward Q 2237$-$0608 appears
weak enough that the estimated  $N$(Al II)'s probably do not suffer from
significant saturation effects. The ratios, [Al/Si]$=-0.30$  and $-0.35$
(Tables 11 and 15), are consistent with observations of Galactic halo
stars (cf. Wheeler et al. 1989). Incidentally, Al and Si are both produced 
during hydrostatic carbon and neon burning in massive stars, 
but Al is an odd Z element. Accordingly,
the underabundance of Al relative to
Si is consistent with the odd-even effect.
 
   The only significant difference between figure 23 and figure 25 is 
in the Zn/Fe ratio. A Zn/Cr overabundance, or equivalently, 
a Zn/Fe overabundance (since [Cr/Fe]$\sim$0 in damped Ly$\alpha$ galaxies)  
was known since the first studies of damped Ly$\alpha$ abundances 
(Meyer, Welty, \& York 1989; Pettini, Boksenberg, \& 
Hunstead 1990), and was interpretated to indicate that some Cr (and Fe) is 
depleted by dust grains from the gas phase. We will have more discussion of
this in \S4.3.3.

   In general, we consider the good agreement in the Si/Fe, S/Fe, Cr/Fe, and
Mn/Fe ratios between damped Ly$\alpha$ galaxies and Galactic halo stars
strong evidence for a Type II SN enrichment 
origin of these elements without significant
modifications by dust grains. In particular, the observed Mn/Fe ratios 
{\it require} a Type II SN-enrichment explanation since they cannot be 
explained otherwise. Another strong argument for SN II enrichment in
damped Ly$\alpha$ galaxies is the N/O ratios (see below). 

In \S4.3.1, we argued that the observed low N/O ratios in several damped
Ly$\alpha$ galaxies are inconsistent with the interpretation of
dust depletion. On the other hand,
these low N/O ratios  are very similar to      those
found in Galactic halo stars.
There was some debate about whether [N/Fe] in halo stars is more close to
zero or less than zero (see discussion in Wheeler  et al. 1989).
 But since [O/Fe] is known to be about +0.4 dex
in halo stars, it is safe to say that [N/O]$<-0.4$ dex at [Fe/H]$<-1$.
Similarly, observations of metal-poor ([O/H]$<-1$) external galaxies
indicate [N/O]$<-0.6$ (Pagel  et al. 1992; Skillman \&
Kennicutt 1993; Thuan  et al. 1995).
Such low N/O ratios are commonly interpreted 
in terms of the different nucleosynthetic origin of these elements.
While O is almost exclusively produced by short-lived massive stars,
N is thought to have two sources. For pristine gas or gas with very low
metallicity, primary N production is made by intermediate mass
stars ($3M_{\odot}<M<8_{\odot}$) during the third dredge-up episodes,
which bring carbon-rich material from the He-burning shell
into the hydrogen-burning shell
(Renzini \& Voli 1981). The N produced this way is primary because it
does not depend on the initial metal content of the gas.
The long evolution time scale of the intermediate mass stars necessarily 
means that there should be a time delay of several
hundreds of million years between the injection of O into the ISM
and that of primary N. 
Secondary production of N can occur in stars of any mass during 
main-sequence burning,  when the initial metallicity in the star 
is high enough to provide the seed C and O nuclei. 
Although the above popular interpretation is challenged
by some recent observational results (cf. Thuan et al. 1995),
which suggest that significant amounts of {\it primary} N may be produced
by massive stars, it remains true that the {\it observed} N/O ratio
in halo stars and in metal-poor external galaxies is low and comparable
to that observed in damped Ly$\alpha$ galaxies.
We thus believe that the low N/O ratios (and the low Mn/Fe ratios)
found for damped Ly$\alpha$ galaxies are indications that massive 
stars are the primary pollutants
of the ISM in these galaxies. This probably happens because star formation
in these galaxies has
occurred shortly before they were observed, so that
only massive stars have had time to evolve to Type II SN.
Given this time scale, we expect to see $\alpha$-elements
overabundance relative to Fe-peak elements.
This is clearly seen in figure 23. In fact, the observed
overabundances of Si to Fe and S to Fe, and the underabundance of Al to Si 
and Mn to Fe can be entirely explained away 
with the above nucleosynthesis argument.  The only remaining 
question is: is there a need for dust depletion on top of the 
nucleosynthesis pattern?

\subsubsection{Dust Depletions On Top of Nucleosynthesis Pattern?}

The preceding discussion demonstrates that dust depletion alone cannot explain
the observed elemental abundances in damped Ly$\alpha$ galaxies.
Rather, with the exception of Zn, all the observed abundances are 
nicely explained if Type II supernovae are the only significant
sources of pollution in these galaxies. Could the Zn/Fe ratios, then,
be suggesting that there is some dust depletion effect on top of the
nucleosynthesis pattern?

In Galactic stars the Zn/Fe ratio is found to be solar at all metallicities
(see figure 25).
On the other hand, the observed Zn/Fe ratios in damped Ly$\alpha$ galaxies are
essentially all above the solar ratio.
Since Zn is almost unaffected by dust in warm diffuse ISM clouds, while 
Fe is among the most heavily depleted elements, a modest amount of Fe depletion
would explain the observed Zn/Fe ratios in damped Ly$\alpha$ systems.
This was indeed the argument that was put forward previously to explain
the observed Zn/Cr (equivalent to Zn/Fe) ratios (cf. Pettini et al. 1994).
However, this interpretation does not appear to be consistent with the other
observed ratios.  For the damped Ly$\alpha$ systems in which both Zn and Fe 
abundances are
measured, $\langle$[Zn/Fe]$\rangle=+0.39$. In other words, if the above
dust-depletion explanation for the observed Zn/Fe 
ratios is correct, the gas phase
Fe abundance in damped Ly$\alpha$ galaxies must have been on average
depleted by about 0.4 dex  (or more if Zn is also depleted). 
To explain the other ratios shown in figure 23,
this requires Si, S, Cr, Mn being depleted by the same amount in order
to maintain the expected Si/Fe, S/Fe, Cr/Fe, and Mn/Fe ratios from
halo star observations and from theories of Type II SN enrichment 
(recall that the observed Mn/Fe and N/O ratios
in damped Ly$\alpha$ galaxies {\it require} that the enrichment process
be dominated by SN II). This is quite 
difficult to understand given the observed depletion patterns in Galactic
ISM clouds. As discussed in Sembach \& Savage (1996), 
even in warm diffuse halo clouds
for which elemental depletion is the least, there is a differential depletion
factor of 0.44 dex between Si and Fe (i.e., Fe is more depleted than Si by
0.44 dex). The difference in the Si and Fe depletion widens for cooler, denser
clouds. The problem is especially acute for S since in diffuse ISM clouds
S is essentially undepleted (see Table 17). 
Given that Fe is among the most heavily
depleted elements in the Galactic ISM, it is difficult to understand how S and 
Fe could have the same depletion, unless the depletion property of dust in
the damped Ly$\alpha$ galaxies is drastically different from Galactic dust.
However, if the latter is true, then the basis for interpretating the Zn/Fe
ratio to indicate dust depletion may no longer be valid. 

     The degree to which an element  is depleted by dust grains in the 
Galactic ISM is known to roughly anticorrelate with its condensation
temperature (Field 1974; Jenkins 1987), 
which is the temperature at which half of the atoms condenses
into solid forms of some sort under thermal equilibrium
conditions. To further look 
into the issue of dust depletion, we plot the abundances of the
elements found in damped Ly$\alpha$ galaxies  as a function of their 
condensation temperature using the data given in Table 16. 
Rather than plotting the abundance of individual elements for each damped
Ly$\alpha$ galaxy, which
would make the figure incomprehensible because of the large scatter, 
we chose to plot the elemental abundances relative to the Fe abundance
in the damped Ly$\alpha$ galaxies in order to see the mean trend. 
The result, shown in figure 26, is both quantitatively and qualitatively
different from the same relation seen in ISM clouds 
(cf. Jenkins 1987; Savage \& 
Sembach 1996). The most notable difference is in the behaviour of Mn and N.
The implication is that there is no significant evidence for  dust
depletion to be the origin of the observed abundance patterns in 
damped Ly$\alpha$ galaxies.
This is in agreement with similar findings for two individual damped 
Ly$\alpha$ galaxies previously studied by Wolfe (1995) and by
Prochaska \& Wolfe (1996).

     Another way to look for the effect of dust depletion is to examine
the variations of elemental abundance ratios among the different velocity
components in a given system. It is well established that the degree
of dust depletion in ISM clouds generally varies with their
locations (environment) in the Galaxy, with cool, diffuse disk clouds
showing much larger depletion than the warm, diffuse halo clouds (cf.
Sembach \& Savage 1996). Such a dependence of depletion on environment 
probably has to do  with
the modification or destruction of dust grains by supernova shocks.
If the damped Ly$\alpha$ galaxies contain dust grains that are affected
by similar energetic processes,
then one might expect to see large variations in the
elemental abundance ratios from one absorbing cloud to another.
The best elements for this test are probably (S, Zn) and (Cr, Fe, Ni)
since the two groups of elements are at the two extremes of dust depletion
and because they are generally measurable in damped Ly$\alpha$ systems.
In practice, however, such a test is difficult to carry out since one 
must rely on the weak lines to derive accurate column densities, which
require high resolution, high S/N measurements. Prochaska \& Wolfe (1996) 
examined the relative abundances of Si, Fe, Ni, Cr, and Zn in several
components of the $z_{damp}=2.462$ damped Ly$\alpha$ absorption system 
toward Q 0201+3634 and concluded that there is little dust present
in the absorbing gas. Among all the damped Ly$\alpha$ systems
studied in \S3, we consider the $z_{damp}=2.8110$ system toward Q 0528$-$2505
the best suited for this purpose since the S II $\lambda$1250, 1253
lines are well measured and we have good measurements for six Ni II lines
(see Table 8). Equally important is the fact that 
these ions are well measured across
a $\sim300$ km s$^{-1}$ interval encompassing many components. The three panels
to the left in figure 27 show
the profiles of $N_a(v)$, the column density per unit velocity interval,
for S II and Ni II derived from the average of two S II lines and of six 
Ni II lines in the system. The profile for Si II $\lambda$1808 is
also shown for comparison. The three panels to the right in figure 27 shows the
$N_a(v)$ ratios of these ions. Amazingly, the S II/Si II, S II/Ni II,
and Si II/Ni II ratios are nearly constant over the $\Delta v\sim300 $ km/s
interval where these ratios are relatively well measured. Given that
these ratios can be affected by many factors, including intrinsic 
relative abundance variations, ionization effects, 
as well as dust depletion, we find the relative constancy of these ratios
quite remarkable. In comparison, the S II/Ni II ratio varies from $\sim6$
in warm diffuse halo clouds to $>200$ in cool disk clouds in the Milky Way
(see Sembach \& Savage 1996). The above result becomes even more remarkable
in light of the fact that the $z_{damp}=2.8110$ system toward Q 0528$-$2505
is the only damped Ly$\alpha$ system known to contain molecular hydrogen
(Foltz et al. 1988; Songaila \& Cowie 1996), 
which makes it more likely to contain dust.
However, in this damped Ly$\alpha$ system the fraction of 
molecular hydrogen relative to neutral hydrogen is well below the 
Galactic value for sightlines with similar values of $N$(H I).

   In conclusion, we find it difficult to interpret the observed Zn/Fe
ratios (or Zn/Cr ratios) in damped Ly$\alpha$ galaxies as being a consequence
of dust depletion.

\subsubsection{Concluding Remarks}

   To summarize the preceding  discussion,  we favor the interpretation 
that the abundance ratios in damped Ly$\alpha$ galaxies are mostly or 
entirely a consequence of Type II SN
nucleosynthesis without significant modification by dust depletion.
The evidence for this conclusion is the much-lower-than-solar N/O ratio, 
the $\alpha$-element overabundance relative to Fe peak elements, and
the underabundance of Mn relative to Fe and Al relative to Si  
(odd-even effect).  Such abundance patterns are, in some sense, expected
in the early stages of galactic chemical evolution based on what we
know about the Milky Way (cf. Timmes, Lauroesch, \& Truran 1995), and
were indeed noted previously in individual damped Ly$\alpha$ systems
(e.g., Lu et al. 1995b; Meyer, Lanzetta, \& Wolfe 1995; 
Steidel et al. 1995; Wolfe 1995; Pettini et al. 1995a; Lu et al. 1996a;
Prochaska \& Wolfe 1996). The present study, however, puts the above
conclusion on much firmer ground.

The Zn/Fe overabundance found for damped Ly$\alpha$
galaxies can be explained without invoking dust if 
there is an additional source of Zn 
(particularly $^{64}$Zn, see figure 5 of
Timmes et al. 1995) in SN II besides 
that synthesized under nuclear
statistical equilibrium.  Such an alternative production mechanism
for Zn might occur in the neutrino-driven wind discussed by 
Hoffman  et al. (1996). 
One difficulty with this interpretation is that  a Zn/Fe overabundance
is not observed during the early history of the Milky Way (figure 25).
In that regard, it would be extremely valuable to obtain
the Zn/Fe ratio for some metal-poor stars in external galaxies in 
order to see if Zn/Fe is
universally solar. It is perhaps significant that we do not yet have
a clear theoretical understanding of why Zn should track Fe so well 
in Galactic disk
and halo stars. For example, using the Type II SN yield calculations
of Woosley \& Weaver (1995), both Timmes et al. (1995) and
Malaney \& Chaboyer (1996) found a much lower-than-solar
Zn/Fe ratio (by a factor of 3-7) for plausible assumptions about the 
stellar initial mass function (see Arnett 1995 for a more in-depth
discussion).
If indeed the observed Zn/Fe ratios in damped systems are intrinsic 
to the nucleosynthetic processes, it will have significant implications
for the theory of stellar nucleosynthesis. At this point, we consider
Zn an anomalous element. It is important to note that, even if the
observed Zn/Fe ratios are due to dust depletion (which we consider unlikely),
the amount of Fe depletion should be no higher than 0.4 dex on average, 
assuming that Zn is undepleted. In other words, the amount of Fe locked
up in grains should be less than 60\% of the total. This is in contrast to 
the depletion of Fe in ISM clouds in which up to more than 99\% of the Fe may
be locked up in grains. It should also be noted that a 0.4 dex depletion of Fe
would not explain all the differences between the metallicity distribution
of damped Ly$\alpha$ galaxies and that of the Milky Way disk stars (figure 22).

Regardless of whether or not dust is present in the damped Ly$\alpha$ galaxies,
the observed N/O ratios and Mn/Fe ratios strongly indicate a nucleosynthetic
origin.  This suggests that the chemical enrichment process
in these damped Ly$\alpha$ galaxies must have occurred so recently
prior to  when they were observed that only Type II SN have 
made significant contributions to the pollution of their
interstellar material; low to intermediate mass stars could not have had enough
time to evolve and to make their presence known 
by dumping their nucleosynthetic
products into the interstellar space. Hence the star formation process should
not have proceeded more than several hundred million years from the
time corresponding to their redshifts. One way to 
avoid this conclusion is if the stellar initial mass function in these young
galaxies had a low mass cutoff so that there simply aren't
that many low mass stars to make Type Ia SN, 
even if they had enough time to do so.
Observations of elemental abundance ratios in 
lower redshift and more 
metal-rich damped Ly$\alpha$ galaxies should help to clarify this issue,
since we expect SN Ia to catch up eventually.

\subsection{Discussion}

    The issue of dust is critical for a complete and accurate 
understanding of damped Ly$\alpha$ galaxies.
    Fall and collaborators (Fall \& Pei 1989; Fall, Pei, \& McMahon 1989;
Pei, Fall, \& Bechtold 1991)
found marginally significant evidence that quasars with known
damped Ly$\alpha$ galaxies in the foreground are statistically redder
than those without. This was taken as evidence for the presence of a modest
amount of dust  in damped Ly$\alpha$ galaxies. The dust-to-gas ratio
inferred for the damped Ly$\alpha$ galaxies is roughly 1/10 of that
in the Galaxy. However, the observational
evidence for this reddening has been challenged recently (Foltz 1996, private
communication; Storrie-Lombardi 1996, private communication).
The only other independent evidence for the presence of dust in damped
Ly$\alpha$ galaxies was thought to be the Zn/Cr or Zn/Fe ratios
(cf. Pettini et al. 1994). But as
we argued earlier, this interpretation has problems explaining
many of the other observed abundance ratios.

Fall \& Pei (1995; also Pei \& Fall 1995) argued that 
most of the more dust-rich galaxies 
have been missed in existing surveys.
This is because previous surveys for damped Ly$\alpha$ galaxies were based
on quasars selected optically, hence quasars with dusty foreground 
galaxies would be preferentially missed owing to dust obscuration.
Fall \& Pei  (1995) estimated that up to 70\% of the 
bright quasars at $z\sim 3$ could be missing from optically selected samples.
In principle, such a selection effect could occur even if 
there is no significant
amount of dust in the {\it existing} sample of damped Ly$\alpha$ galaxies.
If this is the case,
then the damped Ly$\alpha$ galaxies studied here will not be representative
of all high-redshift galaxies. In particular, the mean 
metallicity as given by [Fe/H] would be biased towards lower values, 
making them appearing (on average) more metal poor
than they really are. The abundance ratios may also be affected. A critical
test of this hypothesis is to conduct a new survey for damped Ly$\alpha$
galaxies using a sample of high redshift quasars selected from a {\it complete}
sample of radio sources regardless of their optical brightness. Such a program
is being pursued.

\section{OTHER PROPERTIES AND THE NATURE OF DAMPED LYMAN-ALPHA GALAXIES}

\subsection{Electron Densities}

    The column density ratio of C II*/C II is a density indicator
(Bahcall \& Wolfe 1968). The lower level of the C II* $\lambda$1335
transition is a fine structure level of the ground state of the C II
atom. Under normal interstellar conditions, the
population balance between the ground level and the fine structure
level of the C II ground state is determined by collisional excitation
with electrons and by spontaneous radiative decay. 
Table 18 summarizes the C II*
and C II measurements in damped Ly$\alpha$ systems studied here and
in Lu et al. (1996a). Upper limits on the electron density are
derived using the formula and coefficients adopted in Tripp et al. (1996)
by assuming a temperature of 5000 K.
The derived electron densities are upper limits because only lower
limits to $N$(C II) are available in all systems considered, and 
because other mechanisms might also contribute significantly to 
the excitation (e.g., excitation by microwave background photons; see \S6).
These limits are consistent with the mean electron density of 
$\sim 0.07$ cm$^{-3}$ estimated for
the Galactic ISM  (cf. Reynolds 1991; Spitzer \& Fitzpatrick 1993).
For reference, the derived upper limits are on average factors of 2-3 
lower if we estimate $N$(C II) by assuming a solar C/Si ratio.

\subsection{Ionization}

     The relative strength and velocity distribution of absorption from 
ion species with
different ionization properties can  provide insight into the
ionization structure and the spatial distribution of the absorbing gas. 
Table 19 summarizes such information for some ions of interest based
on data presented in this paper and other similar studies with Keck HIRES
(see references in the table).  We make the following comments:

(1) In all cases where such information is available, the Al III absorption
profiles always have similar appearance as the low ionization lines.
Additionally, the Al II lines are always much stronger than the Al III
lines, indicating $N$(Al II)$>>N$(Al III). 
These results suggest that Al III is likely to be produced in the same
physical region as the low ion gas, and that most of the gas is  neutral
rather than ionized (as expected from the large H I column densities
of the systems). This could happen if Al III comes from an ionized
shell surrounding the neutral gas which produces the bulk of low ions.

(2) The Si IV and C IV absorption line profiles almost always resemble
each other, and they usually have a very different appearance from the low
ion absorption lines. This suggests that the bulk of the high ions are probably
produced in physical regions distinct from that containing the low ions. 
The high ions could arise from low-density ionized gas (halo clouds?) 
surrounding the main structure which gives
rise to the low ionization species.

The profile differences between low-ion and high-ion absorption lines
have been noted previously in individual damped Ly$\alpha$ systems
(eg. Wolfe et al. 1994; Lu et al. 1995b). The current discussion makes
these trends much clearer. The results are reassuring in that, in
general, the metal
abundances derived from low ionization species without corrections for
ionization effects should be substantially correct. This is consistent
with theoretical expectations (Viegas 1995)

\subsection{Kinematics}

The profiles of metal absorption line in damped Ly$\alpha$
systems contain information about the kinematics of the gas.
The total velocity spread of the absorption lines  may
allow one to estimate the mass of the objects if they are
caused by random motions of the absorbing clouds confined by 
gravitational potentials. In practice, the velocity spread may
be affected by ordered motions (e.g., rotation, infall/outflow) and
by energetic events such as supernova explosions. One needs also
to take into account the fact that often the quasar sightlines
do not probe the full gravitational potential of the absorbing objects,
 depending
on the impact parameters. These considerations make it difficult
to infer the mass of the absorbing galaxies.

On the other hand, analyses of the profiles of the metal absorption lines have
yielded a potentially very exciting result. Wolfe (1995; see also
Lanzetta \& Bowen 1992) found preliminary evidence that the weak 
low-ionization metal absorption lines in damped Ly$\alpha$
systems often show an edge-leading asymmetry that is characteristic 
of absorption from rotating gaseous disks. 
It should be pointed out that only
weak, unsaturated low-ionization absorption lines can be used to trace 
reliably the velocity structure of the neutral gas in galactic disks. 
Strong, saturated 
absorption lines are not appropriate for this purpose for two reasons:
(1) one loses the information on the velocity structure of the absorption
in the saturated part; (2) strong absorption lines are more sensitive
to trace amounts of gas so they are more easily contaminated by
diffuse clouds that are not associated with galactic disks (e.g., halo
clouds). The implication of Wolfe's finding is 
that damped Ly$\alpha$ galaxies may be high-redshift spiral disks.
If this conclusion is confirmed with a much larger data set,
it will not only provide strong evidence for the physical nature
of the damped Ly$\alpha$ galaxies, but will also become one of the most 
significant constraints on cosmological 
models of structure formation in the early universe.
Data contained in this paper are currently being analyzed by Wolfe 
and collaborators for the purpose discussed above.

\subsection{Star Formation and Stellar Initial Mass Function}

  The rate and form of star formation and the initial mass function 
(IMF) of the stars formed are fundamental parameters for
understanding galactic chemical evolution. 
The failure to detect Ly$\alpha$ emission from damped Ly$\alpha$ galaxies
may suggest that damped Ly$\alpha$ galaxies 
generally do not have high star formation rates. 
For example, Lowenthal et al. (1995)
surveyed seven quasar fields with known foreground damped Ly$\alpha$
absorbers and failed to detect any of the galaxies in Ly$\alpha$
emission down to quite stringent limits. The implied star formation rates
are $<1M_{\odot}$ yr$^{-1}$ for normal IMF, assuming negligible
extinction by dust.  Of course, if a significant amount of dust is
present in these galaxies, the true star formation rate could be
much higher. However, the Hu et al.  (1993) observations of H$\alpha$
emission from three damped Ly$\alpha$ galaxies limit the star formation
rates to less than 20$M_{\odot}$ yr$^{-1}$ in those galaxies.

   On the other hand, the lack of Ly$\alpha$ emission from a damped
Ly$\alpha$ galaxy does not necessarily mean low star-formation
activity, even if there is no significant extinction by dust. 
As demonstrated by Valls-Gabaud (1993), the equivalent width of
Ly$\alpha$ emission from a galaxy is a strong function of the age
of the stellar population. In the case of a star burst, the strength
of the Ly$\alpha$ emission decreases rapidly after the burst,
essentially because the massive stars (O and early B stars) responsible
for the Ly$\alpha$ emission are short-lived. His model calculations
suggest that the Ly$\alpha$ photons  disappear from the galaxy in  
$\sim 10^7$ years after the burst and the galaxy spectrum will even show
stellar absorption in Ly$\alpha$ at later times.
The case of continuous star formation is qualitatively different, with the 
galaxy showing relatively strong Ly$\alpha$ emission 
(equivalent width $>100$ \AA)
at all times because O, B stars are continuously being formed. 
{\it If} the lack of strong Ly$\alpha$ emission in damped Ly$\alpha$
galaxies is indicative of the bursting nature of their star formation
process, it may have significant implications for their nature.
There is now strong evidence that normal spiral disks tend to
have relatively constant star formation during their lifetimes while
dwarf galaxies (dwarf irregulars and dwarf spheroidals) tend to have
bursts of star formation (cf. Kennicutt 1995). 
A bursting nature of star formation for
damped Ly$\alpha$ galaxies, {\it if established}, would favor them
being dwarf galaxies or subgalactic structures.

  The form of star formation in 
damped Ly$\alpha$ galaxies may also be inferred from the elemental
abundance ratios. In particular, the relative abundance of O and
$\alpha$-process elements to those of the Fe-group elements is
fairly sensitive to the intervals between episodes of star 
formation, essentially because of  the $\sim 1$ Gyr delay between the
almost-instantaneous enrichment from SN II and that from SN Ia.
If some of the damped Ly$\alpha$ galaxies in our sample
began with a strong burst of star formation which was followed by a long 
quiescent period, one would expect its Si/Fe
ratio to drop below the SN II expectation (i.e., $\sim$2-3 times
the solar value) after $\Delta t>>1$ Gyrs when Type Ia SN start to inject
Fe into the ISM. Thus a uniform distribution of Si/Fe ratios may well indicate
that the star formation process in these galaxies is  relatively uniform.
The existing data (figure 23) does show a fairly uniform
Si/Fe ratio: $\langle$[Si/Fe]$\rangle=0.36\pm0.11$ based on 
12 measurements (note that the scatter of 0.11 dex is comparable to 
the typical measurement uncertainty of 0.1 dex). However, it is premature
to make any firm conclusions because of the 
relatively small number  of data points, and because we don't know when
each galaxy was formed. The age span of the existing sample 
galaxies with Si/Fe measurements, $\sim3$ Gyrs, is probably too small 
to even make a statistical inference about when these galaxies were formed.
If the uniformity of the Si/Fe ratio persists to much lower redshifts
in a much larger sample, then it may seriously constrain the history
of star formation in these galaxies since at least some of the very
low-redshift galaxies should have formed much longer than 1 Gyrs ago.
A smooth star formation process would be more typical of disk galaxies 
than dwarfs.  Alternatively, a constant Si/Fe ratio may be maintained 
by a top-heavy IMF. Dust depletion, which is expected to become
significant at low enough redshift,  may further complicate the issue.

The relative constancy in the Si/Fe ratio among the damped Ly$\alpha$ galaxies 
in the sample also suggests that the IMF integrated stellar yields of these
elements are fairly uniform. If the relative yield of $\alpha$-elements 
and Fe from Type II SN varies strongly with the progenitor mass,
as was suggested in some theoretical calculations (e.g., Thielemann,
Nomoto, \& Hashimoto 1996), the scatter in the measured Si/Fe ratio
may be used to put significant constraints on the variations of the
stellar IMF in these galaxies over the range of 10-40 $M_{\odot}$ 
(responsible for SN II).
We did not attempt a quantitative calculation of the
IMF constraint because all existing models of SN II suffer from the
uncertain nature of the explosion mechanism, which significantly
affects the Fe yield.

\subsection{Are Damped Ly$\alpha$ Galaxies Spirals?}

   There is  direct evidence  that at least some damped Ly$\alpha$ absorption
systems may be associated with large, disk-like structures.
Briggs et al. (1989) found that the 21-cm absorption from the
$z_{damp}=2.04$ damped Ly$\alpha$ galaxy toward  PKS 0458$-$020 
shows striking similarity in the measurements made with the Arecibo 
single dish and with the VLBI. This  requires
the absorbing galaxy to cover the entire radio source, which is an 
extended structure of about 10 kpc size in the absorber's reference frame. 
Kinematic arguments further suggest
that the actual size of the absorber may be several times bigger.
More recently, Djorgovski et al. (1996) identified a galaxy at
$z=3.150$ responsible for a damped Ly$\alpha$ absorption system in the
spectrum of a background quasar.
The projected separation between the detected galaxy and
the quasar then indicates that the absorbing galaxy must be at least as
large as 17 kpc, comparable to the sizes of normal galaxy disks.

   Another evidence for high-redshift damped Ly$\alpha$ galaxies to
be spiral disks comes from the work of Wolfe (1995), who presented
evidence that the absorption profiles of the metal absorption lines
in damped Ly$\alpha$ galaxies generally show an edge-leading asymmetry
that is characteristic of absorption from rotating gaseous disks. If this
result is confirmed with more detailed modeling using a large sample,
it will constitute the strongest evidence that damped Ly$\alpha$
galaxies are spiral disks or proto-disks.

   Evidence regarding the nature of damped Ly$\alpha$ galaxies also
comes from direct imaging identification of their low-redshift counterparts.
Steidel et al. (1994) presented ground-based images of two quasar fields,
3C 286 (Q 1328+3045) and PKS 1229$-$021, containing foreground
low-redshift ($z<1$) damped Ly$\alpha$ absorbers, and identified
{\it candidate} damped Ly$\alpha$ absorbing galaxies (i.e., no spectroscopic
redshifts were obtained). The galaxy responsible
for the damped Ly$\alpha$ absorption toward 3C 286 appears to be a 
low surface brightness galaxy, while the one responsible for the
damped Ly$\alpha$ absorption toward PKS 1229$-$021 appears to be a normal
galaxy somewhat fainter than $L*$.  In another study (Cohen et al. 1996),
the damped Ly$\alpha$ absorber at $z_{damp}=0.437$ toward the quasar
3C 196 was identified with a spiral galaxy of luminosity roughly $L*$ in an HST
image, although no spectroscopic redshift confirmation was available.

   More broadly, Steidel and collaborators (see Steidel 1995 and references
therein; see also Bergeron and Boisse 1991) have been very successful
in identifying galaxies responsible for the Mg II absorption systems seen
in quasar spectra at $z_{abs}<1.6$. The colors and luminosities of the
identified galaxies are consistent with them being normal Hubble-sequence
galaxies, spanning the range of E to Irr but mostly early to late type spirals.
The implication is that normal galaxies are already in place by a
redshift of 1.6 with properties not that different from the local
population. This is consistent with results from deep surveys of field galaxies
(cf. Driver et al. 1995).

    Direct searches for high-redshift galaxies are also starting to produce
interesting results, some relevant to the nature of damped Ly$\alpha$
galaxies. Identifications of $z>3$ galaxies in quasar fields based
on the presence of a redshifted Lyman continuum break (Steidel et al. 1996a,b;
Giavalisco et al. 1996) have
produced a population of what appears to the progenitors
of normal, massive galaxies that are forming stars of their spheroidal
component, although
this interpretation is not unique. One such ``Lyman break''galaxy is
confirmed to be a damped Ly$\alpha$ galaxy (Steidel et al. 1996a;
Djorgovski et al. 1996), and
another is a candidate for a damped Ly$\alpha$ galaxy (Giavalisco et al.
1996). 

    These results suggest that it may not be unreasonable to think
that damped Ly$\alpha$ galaxies are high-redshift spirals or their
precursors. In fact, if one takes the age estimates of the disk stars
in the sample of Edvardsson et al. (1993) literally\footnote{Private
communications with Dr. Edvardsson suggest that, due to observational
uncertainties and uncertainties in stellar evolution models,  
the oldest disk stars in
the Edvardsson et al. (1993) sample are not necessarily as old as were
suggested in the study.
Some of the oldest stars in the sample may also be halo stars
that happen to have orbits very close to the Galactic plane. A detailed
discussion of the age of the Galactic disk is beyond the scope of
this paper. Interested readers are referred to the nice discussion
by Majewski (1993).}  (see figure 23),
then it is clear that the Milky Way disk was already in place 10-15 Gyrs
ago or at $z=2-4$ for the cosmology adopted here ($q_0=0.1$,
$H_0=50$). Unless our Milky Way is
among the first galaxies formed in the universe, one may be forced
to conclude that (some) spiral disks must exist by the redshift of
2-4 in some form. If we insert a sightline through such a galaxy,
a damped Ly$\alpha$ absorption will clearly result. The key question
is: what dominates the absorption cross-section at $N$(HI)=$10^{20}-10^{22}$
cm$^{-2}$ at the relevant redshifts, spirals or subgalactic structures?

    If the damped Ly$\alpha$ galaxies in our sample are indeed 
disks or proto-disks,
we must explain why their Fe metallicities are so much lower than the Milky 
Way disk in its past. One possibility is that  this is a selection bias
due to dust obscuration (Pei \& Fall 1995; see discussion in \S4.4).  
In principle, disks or proto-disk
could be present at $z>2$ without being included in our sample if the existing
sample of damped Ly$\alpha$ galaxies is dominated by a large
population of smaller structures. Such subgalactic structures could
be the building blocks of large, normal galaxies that are still
falling into each other, or they might
survive to the present epoch. If the damped Ly$\alpha$ galaxies
at $z=2-4$ are predominantly sub-galactic structures, then the number 
density of damped Ly$\alpha$ galaxies,
$dN/dz=0.3$ at $z\sim3$ (Lanzetta et al. 1991; Wolfe et al. 1996),
requires a space density of 322$h_{50}$/$R_0^2$ Mpc$^{-3}$, where $R_0$ 
is the characteristic size (radius in units of kpc) of the subgalactic
structures with $N$(H I)$>2\times 10^{20}$ cm$^{-2}$ for a spherical
geometry.  This may provide significant constraints
on cosmological models of structure formation.

\section{TEMPERATURE OF THE COSMIC MICROWAVE BACKGROUND}

    Big Bang cosmology predicts that the temperature $T_{CMB}$ of the
cosmic microwave background (CMB) radiation 
should behave like $T_{CMB}\propto (1+z)$. 
As suggested by Bahcall \& Wolf (1968), this relation is testable
since one can estimate the CMB temperature from the relative populations
of ground-state fine-structure levels of suitably selected atoms in 
high-redshift quasar absorption systems. This technique has previously been
applied to the C I atom  (Meyer et al. 1986; Songaila et al. 1994a)
and to the C II ion (Songaila et al. 1994b; Lu et al. 1996a). 
In \S3, we have estimated column densities (or limits) of C II 
and C II* for several damped Ly$\alpha$ absorption systems.
Since in most cases we only have lower limits to $N$(C II) and/or upper
limits to $N$(C II*), the resulting upper limits on $N$(C II*)/$N$(C II)
yield strict upper limits on the 
temperature of the CMB
since other mechanisms may contribute significantly to the excitation of
the C II ion (e.g., collisional excitations by electrons or hydrogen atoms,
direct UV pumping, etc.).  

Table 20 summarizes the available measurements 
of the temperature of the cosmic microwave background at high redshifts, 
including four based on measurements made in \S3, one from Lu et al. (1996a), 
and three from Songaila et al. (1994a,b). 
In the case of Q 1331+1704 (Songaila et al. 1994a), both the C I and C I*
absorption are detected so it was possible to actually measure the excitation
temperature rather than providing upper limits.  These results are 
displayed in figure 28, where the dotted line is the prediction
of $T_{CMB}=2.73(1+z)$ from Big Bang cosmology. All the existing
measurements or upper limits are consistent with the Big Bang prediction. 

We note that, other than the two measurements at $z=1.776$ (which are based
on C I*/C I), the two upper limits at $z>4$ are particularly tight.
This may be partially attributed to the good S/N of the data, but is also
due to the fact that at $z>4$, the peak of the microwave background radiation
is shifted to a wavelength region very close to the energy separation 
(158 $\mu$m) between the fine structure levels of the  C II ion ground
state. Thus the
C II ion  is a particularly good probe of the microwave background temperature
at very high redshifts. It is also significant that one of the measurements
at $z=1.7660$ and the upper limit at $z=4.0803$ are very close to the Big Bang
predictions, suggesting that contributions from 
other excitation mechanisms are negligible in these systems.

\section{SUMMARY}

    We have obtained high quality spectra of quasars using the High Resolution
Spectrometer (HIRES) on the 10m Keck telescope in order to study the elemental
abundances in 14 high-redshift ($0.9<z<4.0$) damped Ly$\alpha$ galaxies. 
The HIRES spectra have a resolution of FWHM=6.6 km s$^{-1}$, 
with a typical S/N$\sim 40$ per resolution element.
The chemical elements being studied include C, N, 
O, Al, Si, S, Cr, Mn, Fe, Ni, and Zn, although not every element listed
is measured for every damped Ly$\alpha$  galaxy. 
In order to ensure accuracy, we measure column densities and abundances using 
only those absorption lines with $\geq4\sigma$ detection
significance that are believed to be unsaturated. This allows us to circumvent
the difficulty associated with measuring reliable column densities using
saturated absorption lines. In all other cases, 
we derive abundance limits using either the $4\sigma$ upper 
limits (for weaker or undetected lines) or the lower limits (for
saturated lines) on their column densities. We also use the improved
oscillator strengths ($f$-values) for a number of important transitions
as given in the compilation of Tripp et al. (1996).
These new abundance measurements are combined with similar measurements
in the literature that are believed to be 
largely free of biases resulting from line 
saturations in order to investigate the chemical evolution of
damped Ly$\alpha$ galaxies. Where appropriate, 
we have corrected the earlier measurements
with the set of improved $f$-values (Tripp et al. 1996) so that all 
the measurements will be on the same footing. The total sample contains 
23 damped Ly$\alpha$ absorption systems in the redshift range
$0.7<z<4.4$. Our main results are the following.

1. The damped Ly$\alpha$ galaxies have [Fe/H] in the range of $-2.5$ to
$-1$, or (Fe/H) between 1/10 to 1/300 solar,
clearly indicating that these are young galaxies in the early
stages of chemical evolution. The $N$(H I)-weighted mean metallicity
of the damped Ly$\alpha$ galaxies with $2<z<3$ is [Fe/H]=$-1.56$, which
may be regarded as the cosmic metallicity at $\langle z\rangle=2.5$. 
There is a large scatter (about a factor of 30) in (Fe/H) at $z<3$, 
which we argue is real and which probably results from 
the different formation histories of the absorbing galaxies or a mix
of galaxy types.

2. Comparisons of the distribution of [Fe/H] vs redshift for the sample
of damped Ly$\alpha$ galaxies with the similar relation for the Milky
Way disk (i.e., the Galactic age-metallicity relation) as defined
by the stars in the Edvardsson et al. (1993) sample indicate
that the damped Ly$\alpha$ galaxies are much less metal-enriched than
the Galactic disk in its past. Since there is evidence from our
analyses that
depletion of Fe by dust grains in the damped Ly$\alpha$ galaxies is
unimportant, the difference in the enrichment level between
the sample of damped Ly$\alpha$ galaxies and the Milky Way disk suggests
that the damped Ly$\alpha$ galaxies are probably not high-redshift spiral 
disks
in the traditional sense. Rather, they could represent a thick disk 
phase of galaxies, or more likely the spheroidal component of galaxies,
or dwarf galaxies.

3. There is evidence that the mean metallicity in damped Ly$\alpha$
galaxies rises sharply at $z\leq3$, possibly indicating the onset of
significant star formation and chemical enrichment in early galaxies.
Such an epoch may be identified as the epoch of galaxy formation in the
sense that galaxies are beginning to form the bulk of their stars.
This conclusion is consistent with the drop in the neutral baryon content
of damped Ly$\alpha$ galaxies at $z>3$ (Storrie-Lombardi \& Wolfe 1996), with
the sharp decrease in quasar space 
density at $z>3$ (cf. Kennefick et al. 1995),
and with the morphology of $z\geq 3$ galaxies identified from their 
Lyman continuum  break (Steidel et al. 1996a,b; Giavalisco et al. 1996).
It would  be extremely important  to confirm the above results
using a much larger statistical sample.

4. The relative abundance patterns of the elements being studied here
closely resemble those seen in Galactic halo stars. This provides
(in our opinion) unambiguous evidence that the bulk of heavy elements in these 
high-redshift galaxies were produced by Type II supernovae, with little
contribution from stellar mass loss of low-to-intermediate mass
stars or from Type Ia supernovae. 
The evidence for Type II supernova enrichment includes the overabundance
of O relative to N, the overabundance of $\alpha$-elements (Si, S)
relative to the iron peak elements (Cr, Mn, Fe, Ni), and the underabundance
of Mn relative to Fe and Al relative to Si (i.e., the odd-even effect).

5. In agreement with earlier studies, we find Zn to be overabundant relative 
to Cr or Fe by an average of $\sim 0.4$ dex compared to their solar values.
Although earlier studies have attributed this effect to 
selective depletion of Cr and Fe by dust grains, 
such an interpretation is inconsistent with many of the other
elemental abundance ratios seen in these galaxies, most notably 
N/O and Mn/Fe. We find no dependence of the measured elemental abundances
on their condensation temperature in a way that would suggest the
depletion by dust
grains (cf. Jenkins 1987). The lack of significant variations
in the S/Ni ratio across a large velocity interval (several hundreds of km 
s$^{-1}$) in one damped Ly$\alpha$ system also argues against the presence of
dust in this galaxy.  Therefore, 
we suggest that the overabundance of Zn relative to
Cr in damped Ly$\alpha$ galaxies may be intrinsic to their stellar 
nucleosynthesis. If this interpretation is correct, it will provide important
new information to the theory of stellar nucleosynthesis.

6. We discuss the physical conditions in damped Ly$\alpha$ galaxies.
The upper limits on the electron density 
in  the absorbing gas as inferred from the C II*/C II ratios are
consistent with the value found in the Galactic ISM. It is found that
the absorption profiles of Al III in damped Ly$\alpha$ galaxies
always resemble those of the low-ionization lines. The profiles of
Si IV and C IV absorption, while resembling each other in general,
are almost always different from those of the low-ionization
absorption lines. These results suggest that Al III is probably produced
in the same physical region as the low-ionization species in the
absorbing galaxies, while the high-ionization species (Si IV and C IV)
mostly likely come from different physical regions. 

7. We discuss possible ways to get information on the history of star
formation (i.e., continuous or episodic) in damped Ly$\alpha$ galaxies, 
and on the shape of their stellar initial mass functions. 

8. We review the  evidence for and against the hypothesis that damped
Ly$\alpha$ galaxies are disks or proto-disks at high redshifts (Wolfe
et al. 1986), and discuss the implications.

9. We determine upper limits on the temperature of the
 cosmic microwave background radiation
at several redshifts using absorption from the fine structure level of
the C II ion (i.e., C II*). These upper limits are 
consistent with the predicted increase of $T_{CMB}$ with redshift.

\acknowledgements
 It is a great pleasure to acknowledge helpful discussions with 
Bengt Edvardsson, Max Pettini, Yongzhong Qian, Blair Savage, 
Chuck Steidel, David Valls-Gabaud, Art Wolfe, and Stan Woosley
on various topics related to this work.
We are very grateful to Steven Vogt and Chris Churchill for
making available their Keck HIRES spectra of 
Q 0450$-$1312 and Q 0454+0356 in advance of publication.
We thank Craig Foltz and Lisa Storrie-Lombardi for communicating
their separate, unpublished results on reddening by dust, and Lisa 
Storrie-Lombardi and Art Wolfe for their unpublished result on the 
evolution of $\Omega_{damp}$.  We also thank the referee, Frank
Timmes, for many helpful comments and suggestions, which 
led to significant improvement in the presentation of the results 
and discussion.
The W. M. Keck Observatory is operated as a scientific partnership
between the California Institute of Technology and the University
of California; it was made possible by the generous financial
support of the W. M. Keck Foundation. We especially thank Steven
Vogt and the HIRES team for building a superb spectrograph, and
the observatory staff for assisting with the observations.
Support for this work was provided by NASA through grant number 
HF1062.01-94A (LL) from the
Space Telescope Science Institute, which is operated by the Association
of Universities for Research in Astronomy, Inc., for NASA under contract
NAS5-26555. WWS acknowledges support from NSF grant AST92-21365.

\begin{planotable}{cccccr}
\tablewidth{0pc}
\tablecaption{JOURNAL OF OBSERVATIONS}
\tablehead{
\colhead{Object}           & \colhead{$z_{em}^a$}    
 &\colhead{Date}           & \colhead{Coverage}  
 &\colhead{Exposure}    } 
\startdata
Q 0000$-$2620  &4.108  &13 Nov 95   &5107-7608  &6000s  \nl
               &       &13 Nov 95   &5146-7660  &6000s  \nl
Q 0216+0803    &2.992  &11 Oct 94   &4203-6633  &6000s  \nl
               &       &27 Dec 94   &4223-6662  &6000s  \nl
               &       &22 Aug 95   &5153-7670  &6600s  \nl
Q 0449$-$1325  &3.097  &26 Dec 94   &4203-6632  &8000s  \nl
               &       &26 Dec 94   &4224-6662  &8000s  \nl
Q 0528$-$2505  &2.779  &27 Dec 94   &4541-7044  &6000s  \nl
               &       &27 Dec 94   &6274-8697  &9000s  \nl
               &       &23 Feb 95   &4540-7042  &6000s  \nl
               &       &23 Feb 95   &6386-8695  &3000s  \nl
Q 1425+6039    &3.173  &18 May 95   &4077-6540  &6000s  \nl
               &       &18 May 95   &4059-6513  &8400s  \nl
               &       &18 May 95   &3719-6179  &2400s  \nl
               &       &19 May 95   &3719-6179  &9000s  \nl
               &       &19 May 95   &3732-6199  &9000s  \nl
               &       &19 May 95   &4077-6540  &2400s  \nl
Q 1946+7658    &2.994  &12 Jun 94   &4013-6514  &4700s  \nl
               &       &12 Jun 94   &4031-6541  &4000s  \nl
Q 2212$-$1626  &3.992  &14 Nov 93   &5107-7608  &9000s  \nl
               &       &15 Nov 93   &5146-7660  &9000s  \nl
               &       &22 Aug 95   &4468-6887  &6000s  \nl
               &       &22 Aug 95   &4492-6920  &6000s  \nl
               &       &22 Aug 95   &5341-7775  &3000s  \nl
Q 2231$-$0015  &3.018  &16 Jun 94   &4060-6515  &7600s  \nl
               &       &17 Jun 94   &4032-6425  &6000s  \nl
Q 2237$-$0608  &4.559  &10 Oct 94   &6401-8713  &12000s \nl
               &       &11 Oct 94   &6473-8801  &18000s \nl
               &       &20 Aug 95   &4899-7455  &12000s \nl
               &       &20 Aug 95   &4933-7350  &3000s  \nl
               &       &21 Aug 95   &4933-7350  &9000s  \nl

\enddata
\tablenotetext{a}{ Redshift of the quasar as estimated from the Ly$\alpha$ 
emission line in the Keck spectrum except for Q 0528$-$2505, which is from
Hewitt \& Burbidge (1993).}

\end{planotable}

\begin{planotable}{lccrrrrrr}

\tablewidth{0pc}
\tablecaption{Measurements for the $z_{damp}=3.3901$ System Toward Q 0000$-$2620}
\tablehead{
\colhead{Ion}                   &\colhead{$\lambda^a$}
 &\colhead{$f^a$}               &\colhead{$v_{-}^b$}
 &\colhead{$v_{+}^b$}           &\colhead{$w_r\pm\sigma^c$}
 &\colhead{$N_a(v)\pm\sigma^d$} &\colhead{$N_{adopt}^e$}
 &\colhead{[Z/H]$^f$}            }
\startdata

H I   &1215.670 &0.4164 &\nodata &\nodata &\nodata         &\nodata        &21.41$\pm$0.08 &\nodata \nl
Al II &1670.787 &1.833  &$-60$   &+100    &0.286$\pm$0.020 &$>13.15$       &$>13.15$       &$>-2.74$\nl
Si II &1526.707 &0.11   &$-60$   &+100    &0.327$\pm$0.008 &$>14.56$       &$>14.56$       &$>-2.40$\nl
Fe II &1608.451 &0.0619 &$-60$   &+40     &0.178$\pm$0.013 &$>14.48$       &$>14.48$       &$>-2.44$\nl
      &1611.201 &0.00102&$-60$   &+40     &$0.003\pm0.017$ &$<15.45$       &$<15.45$       &$<-1.47$\nl
C IV  &1548.195 &0.1908 &$-170$  &+130    &0.966$\pm$0.020 &$>14.73$       &$>14.73$       &\nodata \nl
      &1550.770 &0.09522&$-170$  &+130    &0.814$\pm$0.022$^g$ &$>14.86^g$     &\nodata        &\nodata \nl

\enddata
\tablenotetext{a}{From Morton 1991 unless revised values appear
in the compilation of Tripp {\it et al.\ } 1996.}

\tablenotetext{b}{The velocity region over which the
equivalent width and column density are measured.}

\tablenotetext{c}{Rest-frame equivalent width and associated error.}

\tablenotetext{d}{Column density from integrating the $N_a(v)$ profile.
Upper limits are 4$\sigma$.}
\tablenotetext{e}{Final adopted column density.}
\tablenotetext{f}{Abundance relative to the solar value of Anders \& Grevesse 1989.}

\tablenotetext{g}{The measurement is somewhat contaminated due to blending with the
                  Mg II $\lambda$2796 absorption at $z_{abs}=1.433$.}

\end{planotable}

\begin{planotable}{lccrrrrrr}

\tablewidth{0pc}
\tablecaption{Measurements for the $z_{damp}=2.2931$ System Toward Q 0216+0803}
\tablehead{
\colhead{Ion}                   &\colhead{$\lambda^a$}
 &\colhead{$f^a$}               &\colhead{$v_{-}^b$}
 &\colhead{$v_{+}^b$}           &\colhead{$w_r\pm\sigma^c$}
 &\colhead{$N_a(v)\pm\sigma^d$} &\colhead{$N_{adopt}^e$}
 &\colhead{[Z/H]$^f$}            }
\startdata

H I   &1215.670 &0.4164 &\nodata &\nodata &\nodata &\nodata &$20.45\pm0.16$ &\nodata \nl
Si II &1526.707 &0.11   &$-340$ &+140 &0.920$\pm$0.034 &$>15.06$       &15.45$\pm$0.04 &$-0.55\pm0.16$\nl
      &1808.013 &0.00218&$-40$  &+100 &0.153$\pm$0.012 &$15.45\pm0.04$ &\nodata        &\nodata       \nl
Zn II &2026.136 &0.489  &$-40$  &+100 &0.067$\pm$0.029 &$<12.81$       &$<12.81$       &$<-0.29$      \nl
      &2062.664 &0.256  &$-40$  &+100 &0.036$\pm$0.031 &$<13.11$       &\nodata        &\nodata       \nl
Cr II &2056.254 &0.105  &$-40$  &+100 &0.064$\pm$0.024 &$<13.39$       &$<13.39$       &$<-0.74$      \nl
      &2062.234 &0.078  &$-40$  &+100 &0.055$\pm$0.030 &$<13.61$       &\nodata        &\nodata       \nl
Fe II &1608.451 &0.0619 &$-80$  &+140 &0.482$\pm$0.022 &$>14.81$       &14.89$\pm$0.08 &$-1.07\pm0.18$\nl
      &2249.877 &0.00182&$-40$  &+100 &0.027$\pm$0.028 &$<15.14$       &\nodata        &\nodata       \nl
      &2260.781 &0.00244&$-40$  &+100 &0.067$\pm$0.026 &$<14.97$       &\nodata        &\nodata       \nl
Ni II &1709.600 &0.0689 &$-40$  &+100 &0.064$\pm$0.024 &$<13.73$       &13.60$\pm$0.07 &$-1.10\pm0.17$\nl
      &1741.549 &0.1035 &$-40$  &+100 &0.101$\pm$0.016 &$13.60\pm0.07$ &\nodata        &\nodata       \nl
      &1751.910 &0.06375&\nodata$^g$ &\nodata$^g$ &\nodata$^g$ &\nodata$^g$ &\nodata   &\nodata     \nl
Al II &1670.787 &1.833  &$-340$ &+140 &1.091$\pm$0.030 &$>13.88$       &$>13.88$       &$>-1.05$      \nl
Al III&1854.716 &0.5602 &$-340$ &+140 &0.580$\pm$0.052 &$13.75\pm0.03$ &$13.74\pm0.02$ &\nodata       \nl
      &1862.790 &0.2789 &$-340$ &+140 &0.359$\pm$0.033 &$13.73\pm0.02$ &\nodata        &\nodata       \nl
C IV  &1548.195 &0.1908 &$-380$ &+130 &1.460$\pm$0.032 &$>14.93$       &$>15.06$       &\nodata       \nl
      &1550.770 &0.09522&$-380$ &+130 &1.150$\pm$0.040 &$>15.06$       &\nodata        &\nodata       \nl

\enddata
\tablenotetext{a}{From Morton 1991 unless revised values appear
in the compilation of Tripp {\it et al.\ } 1996.}

\tablenotetext{b}{The velocity region over which the
equivalent width and column density are measured.}

\tablenotetext{c}{Rest-frame equivalent width and associated error.}

\tablenotetext{d}{Column density from integrating the $N_a(v)$ profile. Upper limits are 4$\sigma$.}

\tablenotetext{e}{Final adopted column density.}

\tablenotetext{f}{Abundance relative to the solar value of Anders \& Grevesse 1989. The uncertainty includes
 that from the $N$(H I).}

\tablenotetext{g}{Blended with the C IV $\lambda$1550 absorption at $z_{abs}=2.7205$.}

\end{planotable}

\begin{planotable}{lccrrrrrr}

\tablewidth{0pc}
\tablecaption{Measurements for the $z_{damp}=1.7688$ System Toward Q 0216+0803}
\tablehead{
\colhead{Ion}                   &\colhead{$\lambda^a$}
 &\colhead{$f^a$}               &\colhead{$v_{-}^b$}
 &\colhead{$v_{+}^b$}           &\colhead{$w_r\pm\sigma^c$}
 &\colhead{$N_a(v)\pm\sigma^d$} &\colhead{$N_{adopt}^e$}
 &\colhead{[Z/H]$^f$}            }
\startdata

H I   &1215.670 &0.4164 &\nodata &\nodata &\nodata    &\nodata        &$20.00\pm0.18$ &\nodata       \nl
Si II &1808.013 &0.00218&$-30$  &+40 &0.044$\pm$0.011 &14.89$\pm$0.11 &14.89$\pm$0.11 &$-0.66\pm0.21$\nl
Ni II &1751.910 &0.06375&$-30$  &+40 &0.020$\pm$0.005 &13.09$\pm$0.10 &13.09$\pm$0.10 &$-1.16\pm0.21$\nl
Mn II &2576.877 &0.3508 &$-30$  &+40 &0.059$\pm$0.015 &12.48$\pm$0.11 &12.48$\pm$0.11 &$-1.05\pm0.21$\nl
      &2594.499 &0.2710 &$-30$  &+40 &0.052$\pm$0.017 &$<12.63$       &\nodata        &\nodata       \nl
Zn II &2026.136 &0.489  &$-30$  &+40 &0.030$\pm$0.010 &$<12.34$       &$<12.34$       &$<-0.31$      \nl
      &2062.664 &0.256  &\nodata$^g$ &\nodata$^g$ &\nodata$^g$ &\nodata$^g$ &\nodata  &\nodata       \nl
Cr II &2056.254 &0.105  &$-30$  &+40 &0.012$\pm$0.011 &$<13.06$       &$<13.06$       &$<-0.62$      \nl
      &2062.234 &0.078  &$-30$  &+40 &0.033$\pm$0.014 &$<13.29$       &\nodata        &\nodata       \nl
      &2066.161 &0.0515 &$-30$  &+40 &0.011$\pm$0.020 &$<13.62$       &\nodata        &\nodata       \nl
Fe II &2249.877 &0.00182&$-30$  &+40 &0.013$\pm$0.010 &$<14.68$       &14.53$\pm$0.09 &$-0.98\pm0.20$\nl
      &2260.781 &0.00244&$-30$  &+40 &0.035$\pm$0.007 &14.53$\pm$0.09 &\nodata        &\nodata       \nl
      &2344.214 &0.1097 &$-110$ &+40 &0.395$\pm$0.013 &$>14.34$       &\nodata        &\nodata       \nl
      &2374.461 &0.0326 &$-40$  &+40 &0.253$\pm$0.011 &$>14.45$       &\nodata        &\nodata       \nl
      &2382.765 &0.3006 &$-110$ &+40 &0.533$\pm$0.011 &$>14.02$       &\nodata        &\nodata       \nl
      &2586.650 &0.0684 &$-110$ &+40 &0.407$\pm$0.024 &$>14.46$       &\nodata        &\nodata       \nl
      &2600.173 &0.2239 &$-110$ &+40 &0.534$\pm$0.028 &$>14.03$       &\nodata        &\nodata       \nl
Al III&1854.716 &0.5602 &$-150$ &+40 &0.225$\pm$0.043 &13.23$\pm$0.08 &13.20$\pm$0.07 &\nodata       \nl
      &1862.790 &0.2789 &$-150$ &+40 &0.102$\pm$0.022 &13.17$\pm$0.12 &\nodata        &\nodata       \nl

\enddata
\tablenotetext{a}{From Morton 1991 unless revised values appear
in the compilation of Tripp {\it et al.\ } 1996.}

\tablenotetext{b}{The velocity region over which the
equivalent width and column density are measured.}

\tablenotetext{c}{Rest-frame equivalent width and associated error.}

\tablenotetext{d}{Column density from integrating the $N_a(v)$ profile. Upper
limits are 4$\sigma$.}
\tablenotetext{e}{Final adopted column density.}
\tablenotetext{f}{Abundance relative to the solar value of Anders \& Grevesse 1989. The uncertainty
includes that from the $N$(H I).}
\tablenotetext{g}{Blended with an unidentified absorption feature.}

\end{planotable}

\begin{planotable}{lccrrrrrr}

\tablewidth{0pc}
\tablecaption{Measurements for the $z_{damp}=1.2667$ System Toward Q 0449$-$1326}
\tablehead{
\colhead{Ion}                   &\colhead{$\lambda^a$}
 &\colhead{$f^a$}               &\colhead{$v_{-}^b$}
 &\colhead{$v_{+}^b$}           &\colhead{$w_r\pm\sigma^c$}
 &\colhead{$N_a(v)\pm\sigma^d$} &\colhead{$N_{adopt}^e$}
 &\colhead{[Z/H]$^f$}            }
\startdata

Fe II &2249.877 &0.00182&$-20$  &+40  &0.082$\pm$0.005 &15.11$\pm$0.03 &15.13$\pm$0.02 &\nodata       \nl
      &2260.781 &0.00244&$-20$  &+40  &0.103$\pm$0.009 &15.14$\pm$0.03 &\nodata        &\nodata       \nl
      &2344.214 &0.1097 &$-60$  &+50  &0.420$\pm$0.011 &$>14.27$       &\nodata        &\nodata       \nl
      &2374.461 &0.0326 &$-40$  &+50  &0.299$\pm$0.010 &$>14.67$       &\nodata        &\nodata       \nl
      &2382.765 &0.3006 &$-60$  &+140 &0.597$\pm$0.019 &$>13.98$       &\nodata        &\nodata       \nl
      &2586.650 &0.0684 &$-60$  &+50  &0.388$\pm$0.011 &$>14.39$       &\nodata        &\nodata       \nl
      &2600.173 &0.2239 &$-60$  &+50  &0.606$\pm$0.015 &$>14.02$       &\nodata        &\nodata       \nl
Mn II &2576.877 &0.3508 &$-20$  &+40  &0.112$\pm$0.008 &12.86$\pm$0.03 &12.87$\pm$0.03 &\nodata       \nl
      &2594.499 &0.2710 &$-20$  &+40  &0.097$\pm$0.015 &12.91$\pm$0.06 &\nodata        &\nodata       \nl
      &2606.462 &0.1927 &$-20$  &+40  &0.099$\pm$0.021 &13.07$\pm$0.09 &\nodata        &\nodata       \nl
Mg II &2796.352 &0.6123 &$-110$ &+140 &1.080$\pm$0.015 &$>13.86$       &$>14.11$       &\nodata       \nl
      &2803.531 &0.3054 &$-110$ &+140 &0.979$\pm$0.015 &$>14.11$       &\nodata        &\nodata       \nl
Mg I  &2852.964 &1.830  &$-40$  &+40  &0.267$\pm$0.009 &$>12.62$       &$>12.62$       &\nodata       \nl

\enddata
\tablenotetext{a}{From Morton 1991 unless revised values appear
in the compilation of Tripp {\it et al.\ } 1996.}

\tablenotetext{b}{The velocity region over which the
equivalent width and column density are measured.}

\tablenotetext{c}{Rest-frame equivalent width and associated error.}

\tablenotetext{d}{Column density from integrating the $N_a(v)$ profile.}
\tablenotetext{e}{Final adopted column density.}
\tablenotetext{f}{No absolute abundance estimates are possible owing to the lack
of $N$(H I) information.}

\end{planotable}

\begin{planotable}{lccrrrrrr}

\tablewidth{0pc}
\tablecaption{Measurements for the $z_{damp}=1.1743$ System Toward Q 0450$-$1312}
\tablehead{
\colhead{Ion}                   &\colhead{$\lambda^a$}    
 &\colhead{$f^a$}               &\colhead{$v_{-}^b$}  
 &\colhead{$v_{+}^b$}           &\colhead{$w_r\pm\sigma^c$}
 &\colhead{$N_a(v)\pm\sigma^d$} &\colhead{$N_{adopt}^e$}
 &\colhead{[Z/H]$^f$}            } 
\startdata
Mg II &2796.352 &0.6123  &$-50$  &+260 &$1.840\pm0.038$ &$>14.10$       &$>14.10$       &\nodata     \nl
      &2803.531 &0.3054  &\nodata$^g$ &\nodata$^g$ &\nodata$^g$ &\nodata$^g$ &\nodata   &\nodata     \nl
Cr II &2056.254 &0.105   &$-20$  &+130 &$0.086\pm0.022$ &$13.44\pm0.12$ &$13.44\pm0.12$ &\nodata     \nl
      &2062.234 &0.078   &$-20$  &+130 &$0.103\pm0.032$ &$<13.64$       &\nodata        &\nodata     \nl
      &2066.161 &0.0515  &$-20$  &+130 &$0.044\pm0.033$ &$<13.83$       &\nodata        &\nodata     \nl
Mn II &2576.877 &0.3508  &$-20$  &+130 &$0.159\pm0.019$ &$12.96\pm0.08$ &$12.92\pm0.07$ &\nodata     \nl
      &2594.499 &0.2710  &$-20$  &+130 &$0.100\pm0.020$ &$12.84\pm0.11$ &\nodata        &\nodata     \nl
      &2606.462 &0.1927  &$-20$  &+130 &$0.069\pm0.019$ &$<13.02$       &\nodata        &\nodata     \nl
Fe II &2249.877 &0.00182 &$-20$  &+130 &$0.105\pm0.018$ &$15.06\pm0.10$ &$15.12\pm0.07$ &\nodata     \nl
      &2260.781 &0.00244 &$-20$  &+130 &$0.129\pm0.023$ &$15.17\pm0.09$ &\nodata        &\nodata     \nl
      &2344.214 &0.1097  &$-50$  &+240 &$0.785\pm0.042$ &$>14.55$       &\nodata        &\nodata     \nl
      &2374.461 &0.0326  &$-50$  &+240 &$0.494\pm0.050$ &$>14.86$       &\nodata        &\nodata     \nl
      &2382.765 &0.2006  &$-50$  &+240 &$1.113\pm0.038$ &$>14.29$       &\nodata        &\nodata     \nl
      &2600.173 &0.2239  &$-50$  &+240 &$1.114\pm0.044$ &$>14.33$       &\nodata        &\nodata     \nl
Zn II &2026.136 &0.489   &$-20$  &+130 &$0.067\pm0.028$ &$<12.80$       &$<12.80$       &\nodata     \nl
      &2062.664 &0.256   &$-20$  &+130 &$0.043\pm0.035$ &$<13.16$       &\nodata        &\nodata     \nl

\enddata
\tablenotetext{a}{From Morton 1991 unless revised values appear 
in the compilation of Tripp {\it et al.\ } 1996.}

\tablenotetext{b}{The velocity region over which the
equivalent width and column density are measured.}

\tablenotetext{c}{Rest-frame equivalent width and associated error.}

\tablenotetext{d}{Column density from integrating the $N_a(v)$ profile. Upper limits are 4$\sigma$.}
\tablenotetext{e}{Final adopted column density.}
\tablenotetext{f}{No absolute abundance estimates are possible owing to the lack of $N$(H I) information.}
\tablenotetext{g}{Only partially covered.}

\end{planotable}

\begin{planotable}{lccrrrrrr}

\tablewidth{0pc}
\tablecaption{Measurements for the $z_{damp}=0.8598$ System Toward Q 0454$+$0356}
\tablehead{
\colhead{Ion}                   &\colhead{$\lambda^a$}    
 &\colhead{$f^a$}               &\colhead{$v_{-}^b$}  
 &\colhead{$v_{+}^b$}           &\colhead{$w_r\pm\sigma^c$}
 &\colhead{$N_a(v)\pm\sigma^d$} &\colhead{$N_{adopt}^e$}
 &\colhead{[Z/H]$^f$}            } 
\startdata

H I   &1215.67  &0.4164  &\nodata &\nodata &\nodata     &\nodata        &$20.76\pm0.03$ &\nodata     \nl
Mg II &2803.531 &0.3054  &$-110$ &+100 &$1.450\pm0.017$ &$>14.38$       &$>14.38$       &$>-1.97$    \nl
Mg I  &2852.964 &1.830   &$-90$  &+60  &$0.303\pm0.027$ &$12.44\pm0.04$ &$12.44\pm0.04$ &\nodata     \nl
Cr II &2056.254 &0.105   &$-90$  &+40  &$0.093\pm0.108$ &$<14.04$       &$<14.04$       &$<-0.40$    \nl
      &2062.234 &0.078   &$-90$  &+40  &$0.070\pm0.098$ &$<14.13$       &\nodata        &\nodata     \nl
      &2066.161 &0.0515  &$-90$  &+40  &$0.043\pm0.106$ &$<14.34$       &\nodata        &\nodata     \nl
Mn II &2576.877 &0.3508  &$-90$  &+40  &$0.158\pm0.031$ &$12.96\pm0.09$ &$12.93\pm0.06$ &$-1.36\pm0.07$\nl
      &2594.499 &0.2710  &$-90$  &+40  &$0.113\pm0.025$ &$12.90\pm0.09$ &\nodata        &\nodata     \nl
      &2606.462 &0.1927  &$-90$  &+40  &$0.079\pm0.033$ &$<13.06$       &\nodata        &\nodata     \nl
Fe II &2249.877 &0.00182 &$-90$  &+40  &$0.074\pm0.044$ &$<15.33$       &$15.27\pm0.09$ &$-1.00\pm0.09$\nl
      &2260.781 &0.00244 &$-90$  &+40  &$0.170\pm0.034$ &$15.27\pm0.09$ &\nodata        &\nodata     \nl
      &2344.214 &0.1097  &$-100$ &+80  &$0.977\pm0.018$ &$>14.74$       &\nodata        &\nodata     \nl
      &2374.461 &0.0326  &$-100$ &+80  &$0.718\pm0.024$ &$>15.00$       &\nodata        &\nodata     \nl
      &2382.765 &0.2006  &$-100$ &+80  &$1.141\pm0.016$ &$>14.34$       &\nodata        &\nodata     \nl
      &2586.650 &0.0684  &$-100$ &+80  &$1.014\pm0.016$ &$>14.83$       &\nodata        &\nodata     \nl
      &2600.173 &0.2239  &$-100$ &+80  &$1.228\pm0.011$ &$>14.49$       &\nodata        &\nodata     \nl
Zn II &2062.664 &0.256   &$-90$  &+40  &$0.096\pm0.095$ &$<13.60$       &$<13.60$       &$<+0.19$    \nl

\enddata
\tablenotetext{a}{From Morton 1991 unless revised values appear 
in the compilation of Tripp {\it et al.\ } 1996.}

\tablenotetext{b}{The velocity region over which the
equivalent width and column density are measured.}

\tablenotetext{c}{Rest-frame equivalent width and associated error.}

\tablenotetext{d}{Column density from integrating the $N_a(v)$ profile. Upper limits are 4$\sigma$.}
\tablenotetext{e}{Final adopted column density.}
\tablenotetext{f}{Abundance relative to the solar value of Anders \& Grevesse 1989.}

\end{planotable}

\begin{planotable}{lccrrrrrr}

\tablewidth{0pc}
\tablecaption{Measurements for the $z_{damp}=2.8110$ System Toward Q 0528$-$2505}
\tablehead{
\colhead{Ion}                   &\colhead{$\lambda^a$}    
 &\colhead{$f^a$}               &\colhead{$v_{-}^b$}  
 &\colhead{$v_{+}^b$}           &\colhead{$w_r\pm\sigma^c$}
 &\colhead{$N_a(v)\pm\sigma^d$} &\colhead{$N_{adopt}^e$}
 &\colhead{[Z/H]$^f$}            } 
\startdata

H I   &1215.670 &0.4164  &\nodata &\nodata &\nodata     &\nodata        &$21.20\pm0.10$ &\nodata     \nl
C II  &1334.532 &0.1278  &$-260$ &+380 &$2.234\pm0.014^g$ &$>15.58^g$   &$>15.58$       &$>-2.18$      \nl
C IV  &1548.195 &0.1908  &\nodata$^h$ &\nodata$^h$ &\nodata$^h$ &\nodata$^h$  &$>15.00$       &\nodata     \nl
      &1550.770 &0.09522 &$-180$ &+320 &$0.887\pm0.033$ &$>15.00$       &\nodata        &\nodata     \nl
N I   &1199.550 &0.1327  &\nodata$^i$ &\nodata$^i$ &\nodata$^i$ &\nodata$^i$  &\nodata  &\nodata     \nl
      &1200.223 &0.0885  &\nodata$^i$ &\nodata$^i$ &\nodata$^i$ &\nodata$^i$  &\nodata  &\nodata     \nl
      &1200.710 &0.0442  &\nodata$^i$ &\nodata$^i$ &\nodata$^i$ &\nodata$^i$  &\nodata  &\nodata     \nl
N V   &1238.821 &0.157   &$-60$  &+100 &$0.163\pm0.015$ &$14.00\pm0.04$ &$13.99\pm0.04$ &\nodata     \nl
      &1242.804 &0.07823 &$-60$  &+100 &$0.080\pm0.014$ &$13.94\pm0.07$ &\nodata        &\nodata     \nl
O I   &1302.169 &0.04887 &$-260$ &+380 &$1.945\pm0.018^j$ &$>15.94^j$   &$>15.94$       &$>-2.19$    \nl
Mg II &1239.925 &0.00125 &$-40$  &+300 &$0.080\pm0.033$ &$<15.88$       &$<15.88$       &$<-0.91$     \nl
      &1240.395 &0.000625&$-40$  &+300 &$0.040\pm0.033$ &$<16.19$       &               &            \nl
Al II &1670.787 &1.833   &$-260$ &+380 &$2.174\pm0.018$ &$>14.20$       &$>14.20$       &$>-1.48$    \nl
Al III&1854.716 &0.5602  &$-110$ &+320 &$0.966\pm0.026$ &$>14.04$       &$>14.07$       &\nodata     \nl
      &1862.790 &0.2789  &$-110$ &+320 &$0.649\pm0.025$ &$>14.07$       &\nodata        &\nodata     \nl
Si II &1260.422 &1.007   &$-260$ &+380 &$2.276\pm0.012^k$ &$>14.70^k$   &$16.00\pm0.04$ &$-0.75\pm0.11$\nl
      &1304.370 &0.086   &$-120$ &+380 &$1.526\pm0.016$ &$>15.55$       &\nodata        &\nodata     \nl
      &1526.707 &0.11    &$-260$ &+380 &$1.998\pm0.022^l$ &$>15.44^l$   &\nodata        &\nodata     \nl
      &1808.013 &0.00218 &$-80$  &+300 &$0.495\pm0.027$ &$16.00\pm0.04$ &\nodata        &\nodata     \nl
Si IV &1393.755 &0.514   &$-180$ &+320 &$1.006\pm0.023$ &$>14.41$       &$>14.52$       &\nodata     \nl
      &1402.770 &0.2553  &$-180$ &+320 &$0.744\pm0.031$ &$>14.52$       &\nodata        &\nodata     \nl
S II  &1250.584 &0.005453&$-80$  &+300 &$0.261\pm0.028$ &$15.60\pm0.05$ &$15.59\pm0.03$ &$-0.88\pm0.10$\nl
      &1253.811 &0.01088 &$-80$  &+300 &$0.429\pm0.030$ &$15.58\pm0.03$ &\nodata        &\nodata     \nl
      &1259.519 &0.01624 &\nodata$^m$ &\nodata$^m$ &\nodata$^m$ &\nodata$^m$ &\nodata  &\nodata     \nl
Cr II &2056.254 &0.105   &$-40$  &+300 &$0.159\pm0.040$ &$13.65\pm0.12$ &$13.65\pm0.12$ &$-1.23\pm0.16$\nl
      &2062.234 &0.078   &\nodata$^n$ &\nodata$^n$ &\nodata$^n$ &\nodata$^n$  &\nodata  &\nodata     \nl
      &2066.161 &0.0515  &$-40$  &+300 &$0.066\pm0.037$ &$<13.89$       &\nodata        &\nodata     \nl
Fe II &1608.451 &0.0619  &$-120$ &+300 &$1.264\pm0.018$ &$>15.32$       &$15.45\pm0.11$ &$-1.26\pm0.15$\nl
      &1611.201 &0.00102 &$-40$  &+300 &$0.072\pm0.026$ &$<15.64$       &\nodata        &\nodata     \nl
      &2249.877 &0.00182 &$-40$  &+300 &$0.194\pm0.091$ &$<15.65$       &\nodata        &\nodata     \nl
      &2260.781 &0.00244 &$-40$  &+300 &$0.274\pm0.066$ &$15.45\pm0.11$ &\nodata        &\nodata     \nl
Ni II &1317.217 &0.146   &$-40$  &+300 &$0.136\pm0.025$ &$13.83\pm0.08$ &$13.89\pm0.03$ &$-1.56\pm0.10$\nl
      &1370.132 &0.131   &$-40$  &+300 &$0.159\pm0.026$ &$13.92\pm0.07$ &\nodata        &\nodata     \nl
      &1454.842 &0.0596  &$-40$  &+300 &$0.089\pm0.029$ &$<14.02$       &\nodata        &\nodata     \nl
      &1709.600 &0.0689  &$-40$  &+300 &$0.141\pm0.023$ &$13.94\pm0.07$ &\nodata        &\nodata     \nl
      &1741.549 &0.1035  &$-40$  &+300 &$0.182\pm0.024$ &$13.87\pm0.06$ &\nodata        &\nodata     \nl
      &1751.910 &0.06375 &$-40$  &+300 &$0.124\pm0.025$ &$13.88\pm0.09$ &\nodata        &\nodata     \nl
Zn II &2026.136 &0.489   &$-40$  &+300 &$0.201\pm0.030$ &$13.09\pm0.07$ &$13.09\pm0.07$ &$-0.76\pm0.12$\nl
      &2062.664 &0.256   &\nodata$^o$ &\nodata$^o$ &\nodata$^o$ &\nodata$^o$  &\nodata  &\nodata     \nl
\cutinhead{Undetected Elements}
P II  &1532.533 &0.00761 &$-80$  &+300 &$0.012\pm0.027$ &$<14.84$       &$<14.84$       &$<+0.07$    \nl
Cl II &1347.240 &0.118   &$-80$  &+300 &$-0.012\pm0.044$&$<13.97$       &$<13.97$       &$<-0.50$    \nl
Ti II &1298.697 &0.0952  &$-80$  &+300 &$0.015\pm0.031$ &$<13.95$       &$<13.95$       &$<-0.18$    \nl
Co II &1466.203 &0.140   &$-80$  &+300 &$0.022\pm0.031$ &$<13.67$       &$<13.67$       &$<-0.44$    \nl
Cu II &1358.773 &0.380   &$-80$  &+300 &$0.000\pm0.030$ &$<13.29$       &$<13.29$       &$<-0.18$    \nl
Ga II &1414.402 &1.801   &$-80$  &+300 &$0.009\pm0.032$ &$<12.61$       &$<12.61$       &$<+0.28$    \nl
Ge II &1237.059 &0.876$^p$ &$-80$&+300 &$0.015\pm0.040$ &$<13.13$       &$<13.13$       &$<+0.30$    \nl
\cutinhead{For the absorption complext at $v>140$ km s$^{-1}$}
C II  &1334.532 &0.1278  &+140   &+380 &$0.904\pm0.006$ &$>15.18$       &$>15.18$       &\nodata     \nl
C II* &1335.708 &0.1149  &+140   &+300 &$0.226\pm0.010$ &$14.22\pm0.02$ &$14.22\pm0.02$ &\nodata     \nl 
N I   &1200.710 &0.0442  &+150$^q$ &+300 &$0.277\pm0.009$ &$>14.86$     &$<15.33^r$     &\nodata     \nl
O I   &1302.169 &0.04887 &+140   &+380 &$0.807\pm0.007$ &$>15.56$       &$>15.56$       &\nodata     \nl
Si II &1808.013 &0.00218 &+140   &+300 &$0.129\pm0.012$ &$15.36\pm0.06$ &$15.36\pm0.06$ &\nodata     \nl
S II  &1250.584 &0.005453&+140   &+300 &$0.089\pm0.013$ &$15.10\pm0.07$ &$15.04\pm0.04$ &\nodata     \nl
      &1253.811 &0.01088 &+140   &+300 &$0.129\pm0.015$ &$15.00\pm0.05$ &\nodata        &\nodata     \nl
Fe II &1608.451 &0.0619  &+140   &+300 &$0.406\pm0.008$ &$>14.65$       &$>14.65$       &\nodata     \nl
      &1611.201 &0.00102 &+140   &+300 &$0.016\pm0.014$ &$<15.37$       &\nodata        &\nodata     \nl
      &2260.781 &0.00244 &+140   &+300 &$0.070\pm0.036$ &$<15.12$       &\nodata        &\nodata     \nl
\enddata
\end{planotable}

\clearpage
\noindent Note to Table 8:

$^a${From Morton 1991 unless revised values appear 
in the compilation of Tripp  et al. 1996.}

$^b${The velocity region over which the
equivalent width and column density are measured.}

$^c${Rest-frame equivalent width and associated error.}

$^d${Column density from integrating the $N_a(v)$ profile. 
  Upper limits are 4$\sigma$.}

$^e${Final adopted column density.}

$^f${Abundance relative to the solar value of Anders \& Grevesse 1989. The
uncertainty includes that from the $N$(H I).}

$^g${May be slightly contaminated at the red 
edge by the C II* $\lambda$1335 absorption.
The effect on the column density limit should be negligible.}

$^h${Only partially covered by our spectrum.}

$^i${The three N I lines are blended with each other.}

$^j${Blended with the Si IV $\lambda$1402 absorption 
at $z_{abs}=2.5381$, but the
effect on the column density limit should be negligible.}

$^k${Contaminated at the blue edge by the S II $\lambda$1259 
absorption in the same system. The equivalent width is thus somewhat 
overestimated, but the effect on the column density
limit should be negligible.}

$^l${Slightly contaminated by the Al III $\lambda$1854 
absorption at $z_{abs}=2.1410$, but the effects on the 
measured equivalent width and the column density limit should be negligible.}

$^m${Blended with the Si II $\lambda$1260 absorption in the same system.}

$^n${Blended with the Zn II $\lambda$2062 absorption in the same system.}

$^o${Blended with the Cr II $\lambda$2062 absorption in the same system.}

$^p${The $f$-value is from Cardelli  et al. 1991a. }

$^q${The $v_-$ is adjusted to avoid the contaminating absorption at the 
left end. This should not have any significant effect 
on the estimated column density.}

$^r${see text in \S3.5.}

\clearpage

\begin{planotable}{lccrrrrrr}

\tablewidth{0pc}
\tablecaption{Measurements for the $z_{damp}=2.1410$ System Toward Q 0528$-$2505}
\tablehead{
\colhead{Ion}                   &\colhead{$\lambda^a$}    
 &\colhead{$f^a$}               &\colhead{$v_{-}^b$}  
 &\colhead{$v_{+}^b$}           &\colhead{$w_r\pm\sigma^c$}
 &\colhead{$N_a(v)\pm\sigma^d$} &\colhead{$N_{adopt}^e$}
 &\colhead{[Z/H]$^f$}            } 
\startdata

H I   &1215.670 &0.4164  &\nodata &\nodata &\nodata     &\nodata        &$20.70\pm0.08$ &\nodata     \nl
C IV  &1548.195 &0.1908  &$-140$ &+60  &$0.619\pm0.012$ &$>14.57$       &$>14.71$       &\nodata     \nl
      &1550.770 &0.09522 &$-140$ &+60  &$0.499\pm0.014$ &$>14.71$       &\nodata        &\nodata     \nl
Al II &1670.787 &1.833   &$-120$ &+120 &$0.612\pm0.013$ &$>13.46$       &$>13.46$       &$>-1.72$    \nl
Al III&1854.716 &0.5602  &$-50$  &+30  &$0.104\pm0.007^g$ &$12.85\pm0.03^g$ &$12.77\pm0.08$ &\nodata     \nl
      &1862.790 &0.2789  &$-40$  &+30  &$0.049\pm0.009$ &$12.77\pm0.08$ &\nodata        &\nodata     \nl
Si II &1526.707 &0.11    &$-120$ &+120 &$0.596\pm0.017$ &$>14.79$       &$15.26\pm0.04$ &$-0.99\pm0.09$\nl
      &1808.013 &0.00218 &$-60$  &+30  &$0.098\pm0.008$ &$15.26\pm0.04$ &\nodata        &\nodata     \nl
Cr II &2056.254 &0.105   &$-30$  &+30  &$0.044\pm0.005$ &$13.09\pm0.05$ &$13.10\pm0.04$ &$-1.28\pm0.09$\nl
      &2062.234 &0.078   &$-30$  &+30  &$0.035\pm0.005$ &$13.10\pm0.06$ &\nodata        &\nodata     \nl
      &2066.161 &0.0515  &$-30$  &+30  &$0.028\pm0.007$ &$13.16\pm0.11$ &\nodata        &\nodata     \nl
Mn II &2576.877 &0.3508  &$-30$  &+30  &$0.045\pm0.011$ &$12.38\pm0.10$ &$12.38\pm0.10$ &$-1.85\pm0.13$\nl
Fe II &1608.451 &0.0619  &$-120$ &+120 &$0.320\pm0.024$ &$>14.68$       &$14.94\pm0.26$ &$-1.27\pm0.27$\nl
      &1611.201 &0.00102 &$-30$  &+30  &$0.017\pm0.010$ &$<15.21$       &\nodata        &\nodata     \nl
      &2249.877 &0.00182 &\nodata$^h$ &\nodata$^h$ &\nodata$^h$ &\nodata$^h$ &\nodata   &\nodata     \nl
      &2260.781 &0.00244 &\nodata$^h$ &\nodata$^h$ &\nodata$^h$ &\nodata$^i$ &\nodata   &\nodata     \nl
      &2344.214 &0.1097  &$-120$ &+120 &$0.778\pm0.019$ &$>14.56$       &\nodata        &\nodata     \nl
      &2382.765 &0.2006  &$-120$ &+120 &$1.091\pm0.024$ &$>14.27$       &\nodata        &\nodata     \nl
Ni II &1709.600 &0.0689  &$-30$  &+30  &$0.019\pm0.009$ &$<13.31$       &$13.22\pm0.06$ &$-1.73\pm0.10$\nl
      &1741.549 &0.1035  &$-30$  &+30  &$0.040\pm0.007$ &$13.18\pm0.08$ &\nodata        &\nodata     \nl
      &1751.910 &0.06375 &$-30$  &+30  &$0.030\pm0.007$ &$13.27\pm0.10$ &\nodata        &\nodata     \nl
Zn II &2026.136 &0.489   &\nodata$^i$ &\nodata$^i$ &\nodata$^i$ &\nodata$^i$ &$<12.28$  &$<-1.07$    \nl
      &2062.664 &0.256   &$-30$  &+30  &$0.008\pm0.005$ &$<12.28$       &\nodata        &\nodata     \nl

\enddata
\tablenotetext{a}{From Morton 1991 unless revised values appear 
in the compilation of Tripp {\it et al.\ } 1996.}

\tablenotetext{b}{The velocity region over which the
equivalent width and column density are measured.}

\tablenotetext{c}{Rest-frame equivalent width and associated error.}

\tablenotetext{d}{Column density from integrating the $N_a(v)$ profile.
Upper limits are 4$\sigma$.}
\tablenotetext{e}{Final adopted column density. }
\tablenotetext{f}{Abundance relative to the solar value of Anders \& Grevesse 1989.
The uncertainty includes that from the $N$(H I).}

\tablenotetext{g}{This line occurs in the red wing of the strong
Si II $\lambda$1526 line at $z_{abs}=2.8110$ so it may be contaminated.}

\tablenotetext{h}{Blended with the Al III $\lambda\lambda$1854, 1962 lines at $z_{abs}=2.8110$, respectively.}
\tablenotetext{i}{Blended with the Al II $\lambda$1670 line at $z_{abs}=2.8110$.}
\end{planotable}

\begin{planotable}{lccrrrrrr}

\tablewidth{0pc}
\tablecaption{Measurements for the $z_{damp}=2.8268$ System Toward Q 1425$+$6039}
\tablehead{
\colhead{Ion}                   &\colhead{$\lambda^a$}    
 &\colhead{$f^a$}               &\colhead{$v_{-}^b$}  
 &\colhead{$v_{+}^b$}           &\colhead{$w_r\pm\sigma^c$}
 &\colhead{$N_a(v)\pm\sigma^d$} &\colhead{$N_{adopt}^e$}
 &\colhead{[Z/H]$^f$}            } 
\startdata

H I   &1215.670 &0.4164  &\nodata &\nodata &\nodata     &\nodata        &$20.30\pm0.04$  &\nodata     \nl
C II  &1334.532 &0.1278  &$-160$ &+120 &$0.842\pm0.001$ &$>15.19$       &$>15.19$       &$>-1.67$      \nl
C IV  &1548.195 &0.1908  &$-140$ &+120 &$0.545\pm0.003$ &$14.32\pm0.01$ &$14.32\pm0.01$ &\nodata     \nl
      &1550.770 &0.09522 &$-140$ &+120 &$0.322\pm0.004$ &$14.31\pm0.01$ &\nodata        &\nodata     \nl
N I   &1199.550 &0.1327  &$-120$ &+60  &$0.291\pm0.004$ &$>14.62$       &$>14.62$       &$>-1.73$    \nl
      &1200.223 &0.0885  &$-120$ &+60  &$0.202\pm0.004$ &$>14.54$       &\nodata        &\nodata     \nl
      &1200.710 &0.0442  &\nodata$^g$ &\nodata$^g$ &\nodata$^g$ &\nodata$^g$ &\nodata   &\nodata     \nl
Al II &1670.787 &1.833   &$-120$ &+100 &$0.634\pm0.003$ &$>13.55$       &$>13.55$       &$>-1.23$    \nl
Si II &1526.707 &0.11    &$-50^h$&+100 &$>0.471^h$      &$>14.78^h$       &$>14.78$       &$>-1.07$    \nl
Si IV &1393.755 &0.514   &$-140$ &+120 &$0.288\pm0.003$ &$13.62\pm0.01$ &$13.62\pm0.01$ &\nodata     \nl
      &1402.770 &0.2553  &$-140$ &+120 &$0.164\pm0.006$ &$13.62\pm0.01$ &\nodata        &\nodata     \nl
Fe II &1608.451 &0.0619  &$-120$ &+100 &$0.266\pm0.004$ &$14.44\pm0.01$ &$14.48\pm0.04$ &$-1.33\pm0.06$\nl
      &1611.201 &0.00102 &$-40$  &+40  &$0.007\pm0.002$ &$<14.52$       &\nodata        &\nodata     \nl
Ni II &1370.132 &0.131   &$-40$  &+40  &$0.017\pm0.001$ &$12.90\pm0.03$ &$12.90\pm0.03$ &$-1.65\pm0.00$\nl
      &1454.842 &0.0596  &$-40$  &+40  &$0.009\pm0.002$ &$12.91\pm0.07$ &\nodata        &\nodata     \nl

\enddata
\tablenotetext{a}{From Morton 1991 unless revised values appear 
in the compilation of Tripp {\it et al.\ } 1996.}

\tablenotetext{b}{The velocity region over which the
equivalent width and column density are measured.}

\tablenotetext{c}{Rest-frame equivalent width and associated error.}

\tablenotetext{d}{Column density from integrating the $N_a(v)$ profile.
Upper limits are 4$\sigma$.}
\tablenotetext{e}{Final adopted column density.}
\tablenotetext{f}{Abundance relative to the solar value of Anders \& Grevesse 1989.
The uncertainty includes that from the $N$(H I).}
\tablenotetext{g}{Contaminated by Ly$\alpha$ forest absorption.}

\tablenotetext{h}{The component at $v\sim -90$ km s$^{-1}$ is blended with the
C IV $\lambda$1548 absorption at $z_{abs}=2.7727$. Thus we restrict the integration
to $v>-50$ km s$^{-1}$ to yield lower limits of the equivalent width and column density.}
\end{planotable}

\begin{planotable}{lccrrrrrr}

\tablewidth{0pc}
\tablecaption{Measurements for the $z_{damp}=2.8443$ System Toward Q 1946$+$7658}
\tablehead{
\colhead{Ion}                   &\colhead{$\lambda^a$}    
 &\colhead{$f^a$}               &\colhead{$v_{-}^b$}  
 &\colhead{$v_{+}^b$}           &\colhead{$w_r\pm\sigma^c$}
 &\colhead{$N_a(v)\pm\sigma^d$} &\colhead{$N_{adopt}^e$}
 &\colhead{[Z/H]$^f$}            } 
\startdata

H I   &1215.670 &0.4164  &\nodata &\nodata &\nodata     &\nodata        &$20.27\pm0.06$ &\nodata     \nl
C II  &1334.532 &0.1278  &$-40$  &+20  &$0.119\pm0.003$ &$>14.14$       &$>14.14$       &$>-2.69$      \nl
C II* &1335.708 &0.1149  &$-20$  &+20  &$0.001\pm0.003$ &$<12.76$       &$<12.76$       &\nodata     \nl 
C IV  &1548.195 &0.1908  &$-100$ &+110 &$0.482\pm0.006$ &$14.36\pm0.01$ &$14.36\pm0.01$ &\nodata     \nl
      &1550.770 &0.09522 &$-100$ &+110 &$0.319\pm0.006$ &$14.37\pm0.01$ &\nodata        &\nodata     \nl
N I   &1199.550 &0.1327  &$-20$  &+20  &$0.011\pm0.003$ &$<12.82$       &$<12.82$       &$<-3.50$    \nl
      &1200.223 &0.0885  &$-20$  &+20  &$-0.002\pm0.003$&$<13.01$       &\nodata        &\nodata     \nl
      &1200.710 &0.0442  &$-20$  &+20  &$0.008\pm0.003$ &$<13.28$       &               &            \nl
O I   &1302.169 &0.04887 &$-20$  &+20  &$0.081\pm0.002$ &$>14.46$       &$>14.46$       &$>-2.74$    \nl
Al II &1670.787 &1.833   &$-40$  &+20  &$0.068\pm0.005$ &$12.26\pm0.03$ &$12.26\pm0.03$ &$-2.49\pm0.07$\nl
Si II &1304.370 &0.086   &$-20$  &+20  &$0.040\pm0.002$ &$13.64\pm0.02$ &$13.63\pm0.02$ &$-2.19\pm0.06$\nl
      &1526.707 &0.11    &$-20$  &+20  &$0.061\pm0.002$ &$13.61\pm0.02$ &\nodata        &\nodata     \nl
Si IV &1393.755 &0.514   &$-100$ &+110 &$0.374\pm0.008$ &$13.86\pm0.01$ &$13.86\pm0.01$ &\nodata     \nl
      &1402.770 &0.2553  &$-100$ &+110 &$0.241\pm0.007$ &$13.87\pm0.01$ &\nodata        &\nodata     \nl
S II  &1250.584 &0.005453&$-20$  &+20  &$0.003\pm0.002$ &$<14.00$       &$<14.00$       &$<-1.54$    \nl
Fe II &1608.451 &0.0619  &$-20$  &+20  &$0.029\pm0.003$ &$13.38\pm0.04$ &$13.38\pm0.04$ &$-2.40\pm0.07$\nl
Ni II &1370.132 &0.131   &$-20$  &+20  &$0.003\pm0.002$ &$<12.57$       &$<12.57$       &$<-1.95$    \nl

\enddata
\tablenotetext{a}{From Morton 1991 unless revised values appear 
in the compilation of Tripp {\it et al.\ } 1996.}

\tablenotetext{b}{The velocity region over which the
equivalent width and column density are measured.}

\tablenotetext{c}{Rest-frame equivalent width and associated error.}

\tablenotetext{d}{Column density from integrating the $N_a(v)$ profile. Upper limits
are 4$\sigma$.}
\tablenotetext{e}{Final adopted column density.}
\tablenotetext{f}{Abundance relative to the solar value of Anders \& Grevesse 1989.
The uncertainty includes that from the $N$(H I).}

\end{planotable}

\begin{planotable}{lccrrrrrr}

\tablewidth{0pc}
\tablecaption{Measurements for the $z_{damp}=1.7382$ System Toward Q 1946$+$7658}
\tablehead{
\colhead{Ion}                   &\colhead{$\lambda^a$}    
 &\colhead{$f^a$}               &\colhead{$v_{-}^b$}  
 &\colhead{$v_{+}^b$}           &\colhead{$w_r\pm\sigma^c$}
 &\colhead{$N_a(v)\pm\sigma^d$} &\colhead{$N_{adopt}^e$}
 &\colhead{[Z/H]$^f$}            } 
\startdata

Al III&1854.716 &0.5602  &$-20$ &+40 &$0.053\pm0.004$ &$12.58\pm0.03$ &$12.58\pm0.03$ &\nodata     \nl
      &1862.790 &0.2789  &$-20$ &+40 &$0.039\pm0.004$ &$12.70\pm0.04$ &\nodata        &\nodata     \nl
Si II &1808.013 &0.00218 &$-20$ &+30 &$0.033\pm0.003$ &$14.76\pm0.03$ &$14.76\pm0.03$ &\nodata     \nl
Cr II &2056.254 &0.105   &$-20$ &+20 &$0.025\pm0.004$ &$12.84\pm0.07$ &$12.78\pm0.06$ &\nodata     \nl
      &2062.234 &0.078   &$-20$ &+20 &$0.015\pm0.003$ &$12.72\pm0.08$ &\nodata        &\nodata     \nl
      &2066.161 &0.0515  &\nodata$^g$ &\nodata$^g$ &\nodata$^g$ &\nodata$^g$ &\nodata &\nodata     \nl
Fe II &2249.877 &0.00182 &$-20$ &+20 &$0.023\pm0.003$ &$14.49\pm0.05$ &$14.46\pm0.04$ &\nodata     \nl
      &2260.781 &0.00244 &$-20$ &+20 &$0.026\pm0.005$ &$14.41\pm0.07$ &\nodata        &\nodata     \nl
      &2344.214 &0.1097  &$-30$ &+40 &$0.254\pm0.006$ &$>14.17$       &\nodata        &\nodata     \nl
Zn II &2026.136 &0.489   &$-20$ &+20 &$0.006\pm0.003$ &$<11.83$       &$<11.83$       &\nodata     \nl
      &2062.664 &0.256   &$-20$ &+20 &$0.004\pm0.003$ &$<12.10$       &\nodata        &\nodata     \nl

\enddata
\tablenotetext{a}{From Morton 1991 unless revised values appear 
in the compilation of Tripp {\it et al.\ } 1996.}

\tablenotetext{b}{The velocity region over which the
equivalent width and column density are measured.}

\tablenotetext{c}{Rest-frame equivalent width and associated error.}

\tablenotetext{d}{Column density from integrating the $N_a(v)$ profile.
Upper limits are 4$\sigma$.}
\tablenotetext{e}{Final adopted column density.}
\tablenotetext{f}{Abundance relative to the solar value of Anders \& Grevesse 1989.
The uncertainty includes that from the $N$(H I).}

\tablenotetext{g}{Blended with the C IV $\lambda$1548 absorption at $z_{abs}=2.6541$.}
\end{planotable}

\begin{planotable}{lccrrrrrr}

\tablewidth{0pc}
\tablecaption{Measurements for the $z_{damp}=3.6617$ System Toward Q 2212$-$1626}
\tablehead{
\colhead{Ion}                   &\colhead{$\lambda^a$}    
 &\colhead{$f^a$}               &\colhead{$v_{-}^b$}  
 &\colhead{$v_{+}^b$}           &\colhead{$w_r\pm\sigma^c$}
 &\colhead{$N_a(v)\pm\sigma^d$} &\colhead{$N_{adopt}^e$}
 &\colhead{[Z/H]$^f$}            } 
\startdata
H I   &1215.670 &0.4164  &\nodata &\nodata &\nodata   &\nodata        &$20.20\pm0.08$ &\nodata     \nl
C II  &1334.532 &0.1278  &$-50$ &+50 &$0.276\pm0.005$ &$>14.53$       &$>14.53$       &$>-2.29$    \nl
C II* &1335.708 &0.1149  &$-50$ &+50 &$0.017\pm0.007$ &$<13.20$       &$<13.20$       &\nodata     \nl 
C IV  &1548.195 &0.1908  &$-40$ &+50 &$0.140\pm0.018$ &$13.73\pm0.06$ &$13.78\pm0.06$ &\nodata     \nl
      &1550.770 &0.09522 &$-40$ &+50 &$0.101\pm0.016$ &$13.84\pm0.07$ &\nodata        &\nodata     \nl
O I   &1302.169 &0.04887 &$-50$ &+50 &$0.242\pm0.001$ &$>14.82$       &$>14.82$       &$>-2.37$    \nl
Si II &1304.370 &0.086   &$-50$ &+50 &$0.087\pm0.003$ &$13.91\pm0.01$ &$13.91\pm0.01$ &$-1.90\pm0.08$\nl
      &1526.707 &0.11    &$-50$ &+50 &$0.139\pm0.013$ &$13.92\pm0.04$ &\nodata        &\nodata     \nl
Si IV &1393.755 &0.514   &\nodata$^g$ &\nodata$^g$ &\nodata$^g$ &\nodata$^g$  &$13.56\pm0.04$ &\nodata     \nl
      &1402.770 &0.2553  &$-40$ &+50 &$0.111\pm0.009$ &$13.56\pm0.04$ &\nodata        &\nodata     \nl
Fe II &1608.451 &0.0619  &$-50$ &+50 &$0.038\pm0.034$ &$<13.99$       &$<13.99$       &$<-1.78$    \nl
Ni II &1370.132 &0.131   &$-50$ &+50 &$-0.012\pm0.012$&$<12.48$       &$<12.48$       &$<-2.01$    \nl

\enddata
\tablenotetext{a}{From Morton 1991 unless revised values appear 
in the compilation of Tripp {\it et al.\ } 1996.}

\tablenotetext{b}{The velocity region over which the
equivalent width and column density are measured.}

\tablenotetext{c}{Rest-frame equivalent width and associated error.}

\tablenotetext{d}{Column density from integrating the $N_a(v)$ profile. Upper limits
are 4$\sigma$.}
\tablenotetext{e}{Final adopted column density.}
\tablenotetext{f}{Abundance relative to the solar value of Anders \& Grevesse 1989.
The uncertainty includes that from the $N$(H I).}

\tablenotetext{g}{Blended with the Mg II $\lambda$2796 absorption line at $z_{abs}=1.3234$.}
\end{planotable}

\begin{planotable}{lccrrrrrr}

\tablewidth{0pc}
\tablecaption{Measurements for the $z_{damp}=2.0662$ System Toward Q 2231$-$0015}
\tablehead{
\colhead{Ion}                   &\colhead{$\lambda^a$}    
 &\colhead{$f^a$}               &\colhead{$v_{-}^b$}  
 &\colhead{$v_{+}^b$}           &\colhead{$w_r\pm\sigma^c$}
 &\colhead{$N_a(v)\pm\sigma^d$} &\colhead{$N_{adopt}^e$}
 &\colhead{[Z/H]$^f$}            } 
\startdata

H I   &1215.67  &0.4164  &\nodata &\nodata &\nodata     &\nodata        &$20.56\pm0.10$ &\nodata     \nl
Al II &1670.787 &1.833   &\nodata$^g$ &\nodata$^g$ &\nodata$^g$ &\nodata$^g$ &\nodata   &\nodata     \nl
Al III&1854.716 &0.5602  &$-160$ &+30  &$0.173\pm0.012$ &$13.07\pm0.03$ &$13.07\pm0.03$ &\nodata     \nl
      &1862.790 &0.2789  &$-160$ &+30  &$0.084\pm0.017$ &$13.03\pm0.09$ &\nodata        &\nodata     \nl
Si II &1808.013 &0.00218 &$-110$ &+20  &$0.091\pm0.012$ &$15.23\pm0.06$ &$15.23\pm0.06$ &$-0.88\pm0.12$\nl
Cr II &2056.254 &0.105   &\nodata$^h$ &\nodata$^h$ &\nodata$^h$ &\nodata$^h$ &$<13.26$  &$<-0.98$    \nl
      &2062.234 &0.078   &$-110$ &+20  &$0.028\pm0.013$ &$<13.26$       &\nodata        &\nodata     \nl
      &2066.161 &0.0515  &$-110$ &+20  &$0.021\pm0.012$ &$<13.40$       &\nodata        &\nodata     \nl
Fe II &1608.451 &0.0619  &$-160$ &+30  &$0.422\pm0.006$ &$>14.72$       &$14.90\pm0.18$ &$-1.17\pm0.20$\nl
      &1611.201 &0.00102 &$-110$ &+20  &$0.004\pm0.007$ &$<15.08$       &\nodata        &\nodata     \nl
Ni II &1709.600 &0.0689  &$-110$ &+20  &$0.034\pm0.007$ &$13.30\pm0.10$ &$13.28\pm0.07$ &$-1.53\pm0.12$\nl
      &1741.549 &0.1035  &$-110$ &+20  &$0.048\pm0.009$ &$13.26\pm0.08$ &\nodata        &\nodata     \nl
      &1751.910 &0.06375 &$-110$ &+20  &$0.023\pm0.012$ &$<13.45$       &\nodata        &\nodata     \nl
Zn II &2026.136 &0.489   &$-110$ &+20  &$0.035\pm0.008$ &$12.33\pm0.12$ &$12.33\pm0.12$ &$-0.88\pm0.16$\nl
      &2062.664 &0.256   &$-110$ &+20  &$0.029\pm0.013$ &$<12.74$       &\nodata        &\nodata     \nl

\enddata
\tablenotetext{a}{From Morton 1991 unless revised values appear 
in the compilation of Tripp {\it et al.\ } 1996.}

\tablenotetext{b}{The velocity region over which the
equivalent width and column density are measured.}

\tablenotetext{c}{Rest-frame equivalent width and associated error.}

\tablenotetext{d}{Column density from integrating the $N_a(v)$ profile.
Upper limits are 4$\sigma$.}
\tablenotetext{e}{Final adopted column density.}
\tablenotetext{f}{Abundance relative to the solar value of Anders \& Grevesse 1989.
The uncertainty includes that from the $N$(H I).}

\tablenotetext{g}{Blended with the Si IV $\lambda$1402 absorption at $z_{abs}=2.6523$.}
\tablenotetext{h}{Badly affected by a night sky line.}
\end{planotable}

\begin{planotable}{lccrrrrrr}

\tablewidth{0pc}
\tablecaption{Measurements for the $z_{damp}=4.0803$ System Toward Q 2237$-$0608}
\tablehead{
\colhead{Ion}                   &\colhead{$\lambda^a$}    
 &\colhead{$f^a$}               &\colhead{$v_{-}^b$}  
 &\colhead{$v_{+}^b$}           &\colhead{$w_r\pm\sigma^c$}
 &\colhead{$N_a(v)\pm\sigma^d$} &\colhead{$N_{adopt}^e$}
 &\colhead{[Z/H]$^f$}            } 
\startdata

H I   &1215.670 &0.4164  &\nodata &\nodata &\nodata     &\nodata        &$20.52\pm0.11$ &\nodata     \nl
C II  &1334.532 &0.1278  &$-140$ &+80  &$0.579\pm0.002$ &$>14.89$       &$>14.89$       &$>-2.19$      \nl
C II* &1335.708 &0.1149  &$-130$ &+60  &$-0.001\pm0.003$&$<12.83$       &\nodata        &\nodata     \nl 
C IV  &1548.195 &0.1908  &$-120$ &+80  &$0.198\pm0.019$ &$13.81\pm0.04$ &$13.81\pm0.04$ &\nodata     \nl
      &1550.770 &0.09522 &$-120$ &+80  &$0.109\pm0.019$ &$13.81\pm0.08$ &\nodata        &\nodata     \nl
Al II &1670.787 &1.833   &$-130$ &+60  &$0.243\pm0.013$ &$12.85\pm0.02$ &$12.85\pm0.02$ &$-2.15\pm0.11$\nl
Si II &1526.707 &0.11    &$-130$ &+60  &$0.291\pm0.007$ &$14.27\pm0.02$ &$14.27\pm0.02$ &$-1.80\pm0.11$\nl
Si IV &1393.755 &0.514   &$-120$ &+80  &$0.183\pm0.010$ &$13.41\pm0.03$ &$13.44\pm0.03$ &\nodata     \nl
      &1402.770 &0.2553  &$-120$ &+80  &$0.119\pm0.006$ &$13.47\pm0.03$ &\nodata        &\nodata     \nl
Fe II &1608.451 &0.0619  &$-130$ &+60  &$0.087\pm0.022$ &$13.85\pm0.11$ &$13.85\pm0.11$ &$-2.18\pm0.15$\nl
Ni II &1370.132 &0.131   &$-100$ &+300 &$-0.003\pm0.008$&$<13.17$       &$<13.17$       &$<-1.6$     \nl

\enddata
\tablenotetext{a}{From Morton 1991 unless revised values appear 
in the compilation of Tripp {\it et al.\ } 1996.}

\tablenotetext{b}{The velocity region over which the
equivalent width and column density are measured.}

\tablenotetext{c}{Rest-frame equivalent width and associated error.}

\tablenotetext{d}{Column density from integrating the $N_a(v)$ profile.
Upper limits are 4$\sigma$.}
\tablenotetext{e}{Final adopted column density.}
\tablenotetext{f}{Abundance relative to the solar value of Anders \& Grevesse 1989.
The uncertainty includes that from the $N$(H I).}

\end{planotable}

\begin{planotable}{ccrrlc}
\tablewidth{38pc}
\tablecaption{Dummy Table as Placeholder}
\tablehead{
\colhead{Object}            &\colhead{$z_{damp}$}    
 &\colhead{$N$(C II)}       &\colhead{$N$(C II*)$^a$}  
 &\colhead{$n_e$ (cm$^{-3}$)} &\colhead{reference}   } 
\startdata
\nodata        & \nodata    &\nodata     &\nodata     &\nodata  &1      \nl
\nodata        & \nodata    &\nodata     &\nodata     &\nodata  &1      \nl
\nodata        & \nodata    &\nodata     &\nodata     &\nodata  &1      \nl
\nodata        & \nodata    &\nodata     &\nodata     &\nodata  &1      \nl
\nodata        & \nodata    &\nodata     &\nodata     &\nodata  &1      \nl
\nodata        & \nodata    &\nodata     &\nodata     &\nodata  &1      \nl
\nodata        & \nodata    &\nodata     &\nodata     &\nodata  &1      \nl
\nodata        & \nodata    &\nodata     &\nodata     &\nodata  &1      \nl
\nodata        & \nodata    &\nodata     &\nodata     &\nodata  &1      \nl
\nodata        & \nodata    &\nodata     &\nodata     &\nodata  &1      \nl
\enddata

\end{planotable}

\begin{planotable}{ccrrlc}
\tablewidth{38pc}
\tablecaption{Dummy Table as Placeholder}
\tablehead{
\colhead{Object}            &\colhead{$z_{damp}$}    
 &\colhead{$N$(C II)}       &\colhead{$N$(C II*)$^a$}  
 &\colhead{$n_e$ (cm$^{-3}$)} &\colhead{reference}   } 
\startdata
\nodata        & \nodata    &\nodata     &\nodata     &\nodata  &1      \nl
\nodata        & \nodata    &\nodata     &\nodata     &\nodata  &1      \nl
\nodata        & \nodata    &\nodata     &\nodata     &\nodata  &1      \nl
\nodata        & \nodata    &\nodata     &\nodata     &\nodata  &1      \nl
\nodata        & \nodata    &\nodata     &\nodata     &\nodata  &1      \nl
\nodata        & \nodata    &\nodata     &\nodata     &\nodata  &1      \nl
\nodata        & \nodata    &\nodata     &\nodata     &\nodata  &1      \nl
\nodata        & \nodata    &\nodata     &\nodata     &\nodata  &1      \nl
\nodata        & \nodata    &\nodata     &\nodata     &\nodata  &1      \nl
\nodata        & \nodata    &\nodata     &\nodata     &\nodata  &1      \nl
\enddata

\end{planotable}

\begin{planotable}{ccrrlc}
\tablewidth{38pc}
\tablecaption{Electron Densities}
\tablehead{
\colhead{Object}            &\colhead{$z_{damp}$}    
 &\colhead{$N$(C II)}       &\colhead{$N$(C II*)$^a$}  
 &\colhead{$n_e$ (cm$^{-3}$)} &\colhead{reference}   } 
\startdata
0528$-$2505    & 2.8110$^b$ & $>15.18$   & 14.22      & $<1.6$  &1      \nl
1946+7658      & 2.8443 & $>14.14$   & $<12.46$   & $<0.30$ &1 \nl
2212$-$1626    & 3.6617 & $>14.53$   & $<12.90$   & $<0.34$ &1      \nl
2237$-$0608    & 4.0803 & $>14.89$   & $<12.53$   & $<0.06$ &1      \nl
1202$-$0725    & 4.3829 & $>14.96$   & $<13.06$   & $<0.18$ &2      \nl
\enddata

\tablerefs{(1) this paper; (2) Lu {\it et al.\ } 1996a.}

\tablenotetext{a}{ Upper limits are 2$\sigma$.}

\tablenotetext{b}{ For the components at $v>140$ km s$^{-1}$.}

\end{planotable}

\begin{planotable}{ccccccc}
\tablewidth{43pc}
\tablecaption{Comparisons of Different Ionization Species}
\tablehead{
\colhead{Object}            &\colhead{$z_{damp}$}    
 &\colhead{Al II/Al III$^a$}    &\colhead{Low Ion vs Al III$^b$}  
 &\colhead{Si IV vs C IV$^b$}   &\colhead{Low Ion vs High Ion$^b$}
 &\colhead{Ref}   } 
\startdata
0000$-$2620  &3.3901  &\nodata   &\nodata   &\nodata  &different   &1   \nl
0100$+$1300  &2.3090  &\nodata   &\nodata   &similar? &different   &2   \nl
0201$+$3634  &2.4620  &$>>1$     &\nodata   &similar  &different   &3   \nl
0216$+$0803  &2.2931  &$>>1$     &similar   &\nodata  &different?  &1   \nl
             &1.7688  &\nodata   &similar   &\nodata  &\nodata     &1   \nl
0528$-$2505  &2.8110  &$>>1$     &similar   &similar  &similar?    &1   \nl
             &2.1410  &$>>1$     &similar   &\nodata  &different   &1   \nl
1202$-$0725  &4.3829  &\nodata   &\nodata   &similar  &different   &4   \nl
1331$+$1704  &1.7764  &\nodata   &\nodata   &\nodata  &different   &5   \nl
1425$+$6039  &2.8268  &\nodata   &\nodata   &similar? &different   &1   \nl
1946$+$7658  &2.8443  &\nodata   &\nodata   &similar  &different   &1   \nl
             &1.7382  &\nodata   &similar   &\nodata  &\nodata     &1   \nl
2212$-$1626  &3.6617  &\nodata   &\nodata   &similar  &different   &1   \nl
2231$-$0015  &2.0662  &\nodata   &similar   &\nodata  &\nodata     &1   \nl
2237$-$0608  &4.0803  &\nodata   &\nodata   &similar? &different   &1   \nl
\enddata

\tablerefs{(1) this paper; (2) Wolfe et al. 1994; (3) Prochaska \& Wolfe
1996; (4) Lu et al. 1996a; (5) Wolfe 1995. }

\tablenotetext{a}{This column compares the relative strength of the 
Al II $\lambda$1670 absorption with that of the Al III $\lambda\lambda$1854, 1862
absorption. A $>>1$ means the Al II absorption is much stronger than the
Al III absorption.}

\tablenotetext{b}{These columns compare the absorption profiles of low
ion lines with Al III, and with high ion lines (ie, Si IV and C IV).
The entry ``similar'' means that the two ions being compared have similar
absorption profiles in terms of the velocity distribution and relative
strength of the components. The entry ``similar?'' means the two ions
being compared are more similar than different. The entries ``different''
and ``different?'' should be interpreted in an analogous way. }

\end{planotable}

\begin{planotable}{ccrrc}
\tablewidth{38pc}
\tablecaption{Measurements of Cosmic Microwave Background Temperature}
\tablehead{
\colhead{Object}           & \colhead{redshift}    
 &\colhead{$T_{ex}^a$ (K)} & \colhead{$T_{CMB}^b$ (K)}  
 &\colhead{reference}   } 
\startdata
1331+1704      & 1.7760 & $10.4\pm0.5$  & 7.6    &1      \nl
               &        & $7.4\pm0.8$   & 7.6    &1      \nl
0528$-$2505    & 2.8110 & $<31.5^c$     & 10.4   &2      \nl
1946+7658      & 2.8443 & $<20.1$    & 10.5      &2      \nl
0636+6801      & 2.9090 & $<16.0^d$  & 10.7      &3      \nl
2212$-$1626    & 3.6617 & $<20.6$    & 12.7      &2      \nl
2237$-$0608    & 4.0803 & $<14.9$    & 13.9      &2      \nl
1202$-$0725    & 4.3829 & $<18.1$    & 14.7      &4      \nl
\enddata

\tablerefs{(1) Songaila {\it et al.\ } 1994a; (2) this paper; 
(3) Songaila {\it et al.\ } 1994b; (4) Lu {\it et al.\ } 1996a.}

\tablenotetext{a}{ Excitation temperature as derived either from $N$(C II*)/$N$(C II)
or from $N$(C I*)/$N$(C I). Upper limits are 2$\sigma$.}

\tablenotetext{b}{ Predicted temperature of the cosmic microwave background from
2.73(1+$z$).}

\tablenotetext{c}{ For the absorption complext at $v>140$ km s$^{-1}$ (see Table 8).}

\tablenotetext{d}{ Songaila {\it et al.\ } (1994b) quoted $T_{ex}<13.5$ K
using $N$(C II)=$4.6\times 10^{14}$ cm$^{-2}$ estimated from profile fitting.
Given the heavy saturation of the C II $\lambda$1334 absorption, we consider
this $N$(C II) very uncertain, and prefer to use $N$(C II)$>1.5\times 10^{14}$
cm$^{-1}$ estimated from the observed equivalent width of the C II absorption.}

\end{planotable}

\clearpage

\figcaption{Damped Ly$\alpha$ absorption in 
the $z_{damp}=3.3901$ system toward
Q 0000$-$2620. The solid curve represents the best-fit damping profile with
$N$(H I)$=2.6\times 10^{21}$ cm$^{-2}$. The dotted curves are damping profiles 
with $N$(H I)$=(2.2, 3.1)\times 10^{21}$ cm$^{-2}$. }

\figcaption{Velocity profiles of metal lines in the $z_{damp}=3.3901$
damped Ly$\alpha$ system toward Q 0000$-$2620. 
The zero velocity is fixed at $z=3.39005$. 
Absorption features marked with ``+'' signs are unrelated to the
absorption line in question. Lines that occur in the Ly$\alpha$ forest
or are blended with other absorption lines are indicated as such.}

\figcaption{Velocity profiles of metal lines in the $z_{damp}=2.2931$
damped Ly$\alpha$ system toward Q 0216+0803. 
The zero velocity is fixed at $z=2.29310$.
The absorption at $v>130$ km s$^{-1}$ in the C~IV $\lambda$1548 panel 
(marked with a ``+'' sign) is from 
C IV $\lambda$1550, and that at $v<-380$ km s$^{-1}$ in the 
C IV $\lambda$1550 panel (marked with a ``+'' sign)
is from C IV $\lambda$1548.}

\figcaption{Velocity profiles of metal lines in the $z_{damp}=1.7688$ damped
Ly$\alpha$ system toward Q 0216+0803. 
The zero velocity is fixed at $z=1.76880$.}

\figcaption{Velocity profiles of metal lines in the $z_{damp}=1.2667$ damped
Ly$\alpha$ system toward Q 0449$-$1326. 
The zero velocity is fixed at $z=1.26666$. The strongly
saturated Fe II $\lambda$2600 line is not shown due to space considerations.
Features marked  with ``+'' signs are unrelated 
to the absorption line in question.}

\figcaption{Profiles of apparent column densities, 
$N_a(v)$, for the Mn II lines in
the $z_{damp}=1.2667$ damped Ly$\alpha$ system toward Q 0449$-$1326. }

\figcaption{Velocity profiles of metal lines in the $z_{damp}=1.1743$
damped Ly$\alpha$ system toward Q 0450$-$1312. 
The zero velocity is fixed at $z=1.17428$. The
data are from Churchill \& Vogt (1997).}

\figcaption{Velocity profiles of metal lines in the $z_{damp}=0.8598$
damped Ly$\alpha$ system toward Q 0453+0356. 
The zero velocity is fixed at $z=0.85980$. The
data are from Churchill \& Vogt (1997). \label{fig. 21}}

\figcaption{Velocity profiles of metal lines in the $z_{damp}=2.8110$
damped Ly$\alpha$ system toward Q 0528$-$2505. The zero velocity is
fixed at $z=2.81100$. Features marked with ``+'' signs are unrelated
to the absorption in question. Lines that are blended with other 
absorption lines are indicated as such. }

\figcaption{Velocity profiles of metal lines in the $z_{damp}=2.1410$
damped Ly$\alpha$ system toward Q 0528$-$2505. The zero velocity is
fixed at $z=2.14100$. Absorption features marked with ``+'' signs
are unrelated to the absorption line in question. 
The Ni II $\lambda$1709 and Cr II $\lambda$2066 lines, though detected,
are not shown due to space considerations. Lines that are blended with
other absorption lines are indicated as such.}

\figcaption{Ly$\alpha$ and Ly$\beta$ absorption line profiles in the
$z_{damp}=2.8268$ damped Ly$\alpha$ system toward Q 1425+6039.
The dotted curves show the damping profiles with $z_{abs}=2.8268$,
$N$(H I)$=2\times 10^{20}$ cm$^{-2}$, and Doppler $b=10$ km s$^{-1}$.
Note the excess absorption at the red wing of the line profiles,
which comes from the absorbing subcomplex near $z_{abs}=2.83058$
($v\sim 300$ km s$^{-1}$ in figure 13).}

\figcaption{Same as figure 11, but with one additional component added
to the fit at $z_{abs}=2.83058$ with $N$(H I)$=1\times 10^{19}$ cm$^{-2}$ and
$b=10$ km s$^{-1}$.}

\figcaption{Velocity profiles of metal lines in the $z_{damp}=2.8268$
damped Ly$\alpha$ system toward Q 1425$+$6039. The zero velocity is
fixed at $z=2.82680$. Absorption features marked with ``+'' signs
are unrelated to the absorption line in question. Lines occurring in
the Ly$\alpha$ forest are indicated as such. Note the expanded vertical
scales for some panels. Essentially all absorption 
displayed in the C II* $\lambda$1335 panel is from 
the C II $\lambda$1334 line.  Only the components
at $v<120$ km s$^{-1}$ will be considered as part of the damped Ly$\alpha$
system (see main text). }

\figcaption{Damped Ly$\alpha$ absorption in the $z_{damp}=2.8443$ system
toward Q 1946+7658. The solid curve represents the best-fit damping profile
with $N$(H I)=$1.86\times 10^{20}$ cm$^{-2}$. The dotted curves
are damping profiles with $N$(H I)=$(1.62, 2.14)\times 10^{20}$ cm$^{-2}$.}

\figcaption{Velocity profiles of metal lines in the $z_{damp}=2.8443$
damped Ly$\alpha$ system toward Q 1946$+$7658. The zero velocity is
fixed at $z=2.84430$. Absorption features marked with ``+'' signs
are unrelated to the absorption line in question. Lines occurring in
the Ly$\alpha$ forest are indicated as such.}

\figcaption{Velocity profiles of metal lines in the $z_{damp}=1.7382$
damped Ly$\alpha$ system toward Q 1946$+$7658. The zero velocity is
fixed at $z=1.73820$. Absorption features marked with ``+'' signs
are unrelated to the absorption line in question. Lines blended with
other lines are indicated as such.}

\figcaption{Damped Ly$\alpha$ absorption in the $z_{damp}=3.6617$ system
toward Q 2212$-$1626. The solid curve represents the best-fit 
damping profile with $N$(H I)=$1.6\times 10^{20}$ cm$^{-2}$.
The dotted curves are damping profiles with 
$N$(H I)=(1.3, 1.9)$\times 10^{20}$ cm$^{-2}$.}

\figcaption{Velocity profiles of metal lines in the $z_{damp}=3.6617$
damped Ly$\alpha$ system toward Q 2212$-$1626. The zero velocity is
fixed at $z=3.66170$. Absorption features marked with ``+'' signs
are unrelated to the absorption line in question. Lines blended
with other absorption lines are indicated as such.}

\figcaption{Velocity profiles of metal lines in the $z_{damp}=2.0662$
damped Ly$\alpha$ system toward Q 2231$-$0015. The zero velocity is
fixed at $z=2.06615$. Absorption features marked with ``+'' signs
are unrelated to the absorption line in question.}

\figcaption{Damped Ly$\alpha$ absorption in the $z_{damp}=4.0803$
system toward Q 2237$-$0608. The solid curve represent the best-fit
damping profile with $N$(H I)$=3.0\times 10^{20}$ cm$^{-2}$. 
The dotted curves are damping profiles with
$N$(H I)=(2.3, 3.9)$\times 10^{20}$ cm$^{-2}$.}

\figcaption{Velocity profiles of metal lines in the $z_{damp}=4.0803$
damped Ly$\alpha$ system toward Q 2237$-$0608. The zero velocity is
fixed at $z=4.08026$. Absorption features marked with ``+'' signs
are unrelated to the absorption line in question.}
    
\figcaption{Age-metallicity relations. The conversion between age of
the universe and redshift is for the cosmological parameters indicated
in the figure. The solid circles are for damped Ly$\alpha$ 
galaxies in Table 16. 
The ``+''s are for Galactic disk stars in the sample of Edvardsson  et al.
(1993). }

\figcaption{Abundance ratios of selected elements against [Fe/H] for the sample
of damped Ly$\alpha$ galaxies given in Table 16. In a few cases,  except for
the [Cr/Fe] vs [Fe/H] panel, Cr abundance
has been used in place of Fe abundance when the latter is not available. }

\figcaption{Similar to fig. 23, but for diffuse ISM clouds in the Galactic disk
and halo as given in Table 17. Note that the spread in [Fe/H] in the horizontal
axis reflects  the different levels of Fe depletion in the ISM clouds rather
than intrinsic differences in the clouds' metallicity.}

\figcaption{Similar to fig. 23, but for Milky Way disk and halo stars. These
ratios reflect the chemical enrichment processes during the past history
of the Milky Way. Sources of the data are given in \S4.2.}

\figcaption{Elemental abundances as a function of condensation temperature 
for damped Ly$\alpha$ galaxies. Note that the abundances are plotted relative
to the Fe abundance (or Cr abundance, 
in case Fe measurement is not available) in 
the same system.  For Al, Si, S, Cr, Mn, Fe, Ni, and Zn, the
average  values of the actual measurements (i.e., excluding upper and 
lower limits) are plotted (solid
dots), with the vertical bars indicating the range of the measured values.
For C, N, and O, individual limits are plotted since no actual measurements
are available. Lower limits are indicated by open triangles, and upper limits
are indicated by inverted solid triangles.}

\figcaption{The three left panels show the profiles of 
apparent column density per unit velocity, $N_a(v)$, for S II, 
Si II, and Ni II in the $z_{damp}=2.8110$ damped Ly$\alpha$ 
system toward Q 0528$-$2505, based on the average of two S II lines, the 
Si II $\lambda$1808 line, and the average of six Ni II lines. The three panels
to the right give the $N_a(v)$ ratios for 
S II/Si II, S II/Ni II, and Si II/Ni II.}

\figcaption{Temperature of the cosmic microwave background as 
inferred from observations
of high-redshift quasar absorption systems. 
The dotted line is the predicted relation
of $T_{CMB}=2.73(1+z)$ from Big Bang cosmology.}

\end{document}